\documentclass[a4paper,11pt]{article}

\pdfoutput=1
\usepackage{amsfonts}
\usepackage{amsmath}
\usepackage{amssymb}
\usepackage{graphicx}
\usepackage{epstopdf}
\usepackage{epsfig}
\usepackage{color}
\usepackage{ulem}

\usepackage{jheppub}

\usepackage[T1]{fontenc}

\newcommand{\be}{\begin{equation}}
\newcommand{\ee}{\end{equation}}
\newcommand{\bea}{\begin{eqnarray}}
\newcommand{\eea}{\end{eqnarray}}
\newcommand{\ba}{\begin{array}}
\newcommand{\ea}{\end{array}}

\def \nn {\nonumber}
\newcommand{\eq}[1]{(\ref{#1})}

\newcommand{\Tr}{\mbox{Tr}}

\newcommand{\beq}{\begin{eqnarray}}
\newcommand{\eeq}{\end{eqnarray}}
\newcommand{\bes}{\begin{subequations}}
\newcommand{\ees}{\end{subequations}}

\newcommand{\singlespace}{\renewcommand{\baselinestretch}{1.0}}

\preprint{}
\title{Quantum Decoherence with Holography}

\author[a]{Shih-Hao Ho,}
\author[b]{Wei Li,}
\author[c]{Feng-Li Lin,}
\author[c]{and Bo Ning}

\affiliation[a]{Physics Division, National Center for Theoretical Sciences and Physics Department,
National Tsing-Hua University, Hsin-Chu 300, Taiwan}
\affiliation[b]{Max-Planck-Institut f\"ur Gravitationsphysik, Albert-Einstein-Institut,
Am M\"uhlenberg 1, 14476 Golm, Germany}
\affiliation[c]{Department of Physics, National Taiwan Normal University, Taipei, 116, Taiwan}

\emailAdd{shho@mx.nthu.edu.tw}
\emailAdd{wei.li@aei.mpg.de}
\emailAdd{linfengli@phy.ntnu.edu.tw}
\emailAdd{ningbo@ntnu.edu.tw} 

\abstract{

Quantum decoherence is the loss of a system's purity due to its interaction with the surrounding  environment. Via the AdS/CFT correspondence, we study how a system decoheres when its environment is  a strongly-coupled theory.
In the Feynman-Vernon formalism, we compute the influence functional holographically by relating it to the generating function of  Schwinger-Keldysh propagators
 and thereby obtain the dynamics of the system's density matrix.

We present two exactly solvable examples: (1) a straight string in a BTZ black hole and  (2) a scalar probe in AdS$_5$. We prepare an initial state that mimics Schr\"odinger's cat
and  identify different stages of its decoherence process using the time-scaling behaviors of
R\'enyi entropy. We also relate decoherence to local quantum quenches, and by comparing the time evolution behaviors of the Wigner function and R\'enyi entropy we demonstrate that the relaxation of local quantum excitations leads to the collapse of its wave-function.

}

\begin{document} 
\maketitle
\flushbottom



\section{Motivation and Summary}

Quantum decoherence is defined as the loss of `coherence' of a quantum state, and
its resulting transition to a classical state. What is its mechanism? Is it instantaneous or gradual? If latter is it possible to reduce its speed?
The development of the quantum information science during the last twenty years has not only lifted these questions from the philosophical realm to the physical one but also made them pressing issues since a realistic quantum computer requires the qubits to remain coherent long enough for their operations to complete.\singlespace\footnote{ Quantum decoherence is also crucial in resolving the information paradox of black holes and studying the primordial cosmic fluctuations during the inflation era of the Universe.}

The Environment-Induced Decoherence developed by Zurek \cite{Zurek} differs from earlier mechanisms of decoherence such as `Copenhagen Interpretation' \cite{Bohr} and `Many Worlds' \cite{many_worlds} in that it is defined purely from quantum mechanics itself: the notion of classicality, the boundary between classical and quantum, the process of decoherence can all be defined and computed within the framework of Quantum Mechanics. Adding to its theoretical elegance are numerous experimental supports such as \cite{Brune:1996zz}.

In the framework of Environment-Induced Decoherence, the quantum decoherence problem is characterized by three elements: the `system' of interest, its `environment', and the interaction between the two. Both the `system' and `environment' are quantum, together they form a closed quantum system which evolves unitarily. However, the `system' by itself is open: it starts to decohere the moment the interaction is turned on and eventually loses all its coherence and becomes a classical state.  The `environment' plays two crucial roles. First, it determines which states the `system' can decohere into. Second, it causes the decoherence to happen via its interaction with the `system'.

The natural question is then how to compute the decoherence rate for different combinations of  `system', `environment', and the interaction between the two.\singlespace{\footnote{In general, to mitigate the effect of decoherence on quantum computation one needs to implement fault-tolerant algorithms; however here we are only interested in the physical aspects of decoherence.}}   The canonical formalism to study decoherence in Environment-Induced Decoherence scheme 
is Feynman-Vernon path-integral \cite{Feynman:1963fq,Caldeira:1982iu}. Given the difficulty of decoherence problem (a non-equilibrium process of an open `system' possibly at finite temperature), most of the early studies in this formalism is limited to the toy model in which both the `system' and the `environment' are simple harmonic oscillators (SHO), for which the decoherence rate can be computed exactly \cite{Grabert:1988yt,Hu:1991di}.\singlespace{\footnote{There have also been various approximations and reductions starting from the Feynman-Vernon formalism in order to treat more non-trivial `environments' but they are more ad hoc and their validity need to be examined case-by-case (see e.g. \cite{Weiss}).}}

However, for non-trivial combinations of `system' plus `environment' with slow decoherence rates we need to extend our search beyond this toy model. The Feynman-Vernon formalism is a first-principle method that requires no assumption on the detailed nature of the `system' or the `environment', therefore is capable of dealing with generic non-trivial `environment'. This leads to the focus of the present paper:
Using Feynman-Vernon formalism, we will study quantum decoherence when the  `environment' is a non-trivial strongly-coupled quantum field theory that has a dual description in terms of a gravity theory. More precisely, we will consider conformal field theories that are dual to gravity theories living in anti de Sitter space. In this Gauge/Gravity (or AdS/CFT) correspondence, the strongly-coupled field theory can be much more easily studied in terms of its weakly-coupled gravity dual.

The motivation is two-fold. One comes from experiments:
the qubits (the `system') in the lab are usually embedded in a strongly-coupled condensed matter `environment'. For instance, a very promising realization of quantum computation is the topological quantum computation \cite{Kitaev:1997wr,Nayak:2008zza}, which utilizes the topologically-ordered phases of certain condensed-matter systems. Most of the topologically ordered phases \cite{TO} are based on the gapped states and hence are robust against ordinary environmental disturbance. However, these gapped states or the localized zero-modes such as Majorana fermions \cite{Majorana,Ho&Lin} (viewed as `system') still interact with the gapless states on the edge and might decohere via this interaction; and the gapless state should therefore be viewed as the `environment' and they are usually described by strongly-coupled 2D conformal field theories.

The other motivation is theoretical. Even beyond the class of the topological quantum computation, in general we should search among non-trivial (very likely strongly-coupled) theories for decoherence-suppressing `environments'. However, quantum decoherence with non-trivial `environment' is a difficult problem. Therefore, we could start from those with weakly-coupled gravity duals in order to gain some insight to this problem.

In Feynman-Vernon formalism, all effects of the `environment' can be packaged into a certain functional of the fields of the `system' called `influence functional'. For the holographic quantum decoherence, the key observation is that  the influence functional is nothing but the generating function of non-equilibrium Green's functions (Schwinger-Keldysh propagators)
 in the `environment' (for the case of linear coupling). This generating function can then be easily computed from the gravity side via AdS/CFT correspondence.

This  holographic quantum decoherence applies to any holographic `environment'. In this paper, we demonstrate its power with two cases of `system' plus `environment': (1) the holographic dual of a straight string in a BTZ black hole and  (2) that of a scalar probe in AdS$_5$.
In both cases we solve the dynamics of quantum decoherence exactly. Using two quantities (the negative part of Wigner function and the second order R\'enyi entropy) to characterize the quantum decoherence, we then describe the full decoherence process and further distinguish its different stages according to the scaling behavior of the second order R\'enyi entropy.  We also write a python code to numerically study more complex cases. These results not only allow us to understand quantum decoherence in more details; more importantly, with our scheme one can study holographic `environments' systematically, thus provide valuable insight for the construction of robust qubits.

   We also notice the similarity between the environment-induced decoherence and the local quantum quench. We find that the decoherence and the relaxation following the quench occur around the same time, and match the scaling behaviors (with time) of the entanglement entropy in the quantum quench process with that of the second order R\'enyi entropy in decoherence. This suggests that the relaxation following the quantum quench occurs when local excitations decohere.

   This paper is organized as follows.  In Section.\ref{setup} we review quantum decoherence and Feynman-Vernon method,  and explain how to compute the influence functional from the gravity side. In Sec.\ref{sec3} we compute the propagating function
and use it to derive the master equation for the reduced density matrix. (The detailed derivations are contained in the Appendices.) In Sec.\ref{sec4} we study Wigner function, the second order R\'enyi entropy, and relation of our setup to the local quantum quench.
Sec.\ref{sec5} examines the two exactly solvable cases.  Sec.\ref{sec6} contains a summary and discussions.

\section{Formalism for Quantum Decoherence and its Holographic Version}\label{setup}

\subsection{Quantum Decoherence}

If our world is intrinsically quantum mechanical, why does it appear to be classical most of
the time? Can the concept `classical' be defined in a purely quantum mechanical framework? How does a quantum mechanical system lose its `coherence'
and become
classical? Is the process of `decoherence', i.e. the quantum-to-classical transition, instantaneous or gradual?  If latter how can we reduce its speed in order to build a real quantum computer?

These are the questions a framework for quantum decoherence need to address. Historically, the most prominent schemes are the following two (for a review
see \cite{Zurek}).

\singlespace
\begin{enumerate}
\item Copenhagen interpretation \cite{Bohr}: Our world is divided into classical and quantum. A measuring apparatus is macroscopic and classical. A quantum mechanical system loses its coherence (i.e. wave-function collapse) the moment it is probed by a classical measuring apparatus. In this interpretation, the classicality and the boundary between the classical and the quantum cannot be defined within the framework of quantum mechanics but has to be introduced from outside.
\item Many Worlds
\cite{many_worlds}: Unlike in Copenhagen interpretation, the world does not have a priori classical subsystems and always evolves unitarily.  The decoherence is caused by selecting a particular observer or subsystem.
\end{enumerate}

The `Environment-Induced Decoherence' developed by Zurek \cite{Zurek} differs from these earlier attempts in that it does not need a deus ex machina outside quantum mechanics to come in and announce the transition from quantum to classical. All questions asked in the beginning of this section can be answered within the framework of quantum mechanics.

First of all, a closed system is always quantum. Its state evolves unitarily according to the  Schr\"odinger equation,
and would never collapse into a classical state by itself.  Now, let us divide this closed system into a subsystem (which we refer to as the `system') and its complement `environment',
and we allow the interaction (hence the flows of energy and information) between the two.
Now the `system' is an open system and due to its interaction with the `environment', in general it would not evolve unitarily --- even though the total system (`system'
plus `environment') still does. This non-unitarity, hence the decoherence of the `system', is caused by the (non-unitary) influence of the environment via a `leaking' of the quantum information from the `system' into the `environment'.

The `environment' is also quantum mechanical (unlike in Copenhagen Interpretation) and plays two roles. First, it selects from the `system''s Hilbert space a subspace of states which are stable against the disturbance from the `environment'. These states are classical states (also called pointer states) --- this is the definition of classicality.  Second, by interacting with the `system' the `environment' causes the `system' to lose its coherence between these pointer states. The process of decoherence is gradual and its speed can be computed given the `system', the `environment', and the interaction between the two. Finally we can engineer one or all of these three aspects in order to reduce the speed of decoherence --- thereby improving the robustness (against decoherence) of the quantum computer.

To discuss the quantum decoherence process, we need a language to describe the quantum and classical states in a unified way. The density matrix $\hat{\rho}$ is such one. In the basis of pointer states, the quantum coherence manifests itself in the presence of $\hat{\rho}$'s off-diagonal elements, whose  disappearance signifies the quantum decoherence process. At the end of decoherence, a quantum state becomes a classical one, whose density matrix loses all off-diagonal elements.

Therefore we need to study the evolution dynamics of the density matrix $\hat{\rho}_{\textrm{sys}}$ of the `system'. First, let us denote
the density matrix of the total system (`system' plus `environment') by $\hat{\rho}_{\textrm{tot}}(t)$.  As the total system is always quantum,  its density matrix evolves unitarily according to the Hamiltonian $H_{\textrm{tot}}$ of the total
system:
\be\label{totrho}
\hat{\rho}_{\textrm{tot}}(t) = e^{-iH_{\textrm{tot}} (t-t_i)} \hat{\rho}_{\textrm{tot}}(t_i) e^{i H_{\textrm{tot}} (t-t_i)}
\ee
where $\hat{\rho}_{\textrm{tot}}(t_i)$ is the initial density matrix of the total system at $t=t_i$  and we set
$\hbar=1$ throughout the paper. We assume the factorized initial condition, i.e. at $t=t_i$ the `system' and the `environment' are unentangled:
\be\label{initotrho}
\hat{\rho}_{\textrm{tot}}(t_i)=\hat{\rho}_{\textrm{sys}}(t_i) \otimes \hat{\rho}_{\textrm{env}}(t_i)\;.
\ee
Once $\hat{\rho}_{\textrm{tot}}(t)$ is known, the reduced density matrix of the `system' of interest is given by tracing out the degrees of freedom  of the `environment':
\be\label{redrho}
\hat{\rho}_{\textrm{sys}}(t) = \Tr_{\textrm{env}}\; \hat{\rho}_{\textrm{tot}}(t)\;.
\ee

Then we can prepare the initial `system' to be in a pure state, and the quantum decoherence process of the `system' is encoded in the way the reduced density matrix $\hat{\rho}_{\textrm{sys}}(t)$ evolves from that of a pure state into a classical one. Formally, the dynamics of the reduced density matrix can be encoded in a
Schr\"odinger-like equation, i.e.,
\be\label{master-1}
i {d\hat{\rho}_{\textrm{sys}}(t) \over dt} = [\mathbf{H}_{\textrm{sys}}, \hat{\rho}_{\textrm{sys}}(t)] + \cdots
\ee
where $\mathbf{H}_{\textrm{sys}}$ is the renormalized Hamiltonian of the `system'. The $\cdots$ terms
characterize the non-unitarity of the influence from the `environment' and is
responsible for the quantum decoherence.
In a general lab experiment, the transition of the `system' from quantum to classical happens very fast. (Indeed, in general it appears to happen instantaneously.) Now, in the framework of environment-induced decoherence, we can actually compute this decoherence time-scale by a direct study of the time-evolution of $\hat{\rho}_{\textrm{sys}}(t)$. In particular we can verify that when
the `environment' is macroscopic as in a lab experiment, and the transition is indeed very fast.

    Most of the computation in this paper is carried out in the path-integral formalism, therefore let us now look at the Lagrangian description. The Lagrangian for the total system (`system' plus `environment') consists of three parts:
\be\label{totL}
\mathcal{L}[\phi, \chi] = \mathcal{L}_{\textrm{sys}}[\phi] + \mathcal{L}_{\textrm{env}}[\chi] + \mathcal{L}_{\textrm{int}}
[\phi,\chi]\,.
\ee
The first two terms $\mathcal{L}_{\textrm{sys}}[\phi]$ and $ \mathcal{L}_{\textrm{env}}[\chi]$ define the `system' and `environment', respectively. We use $\phi$ to denote collectively the degrees of freedom of the `system', and $\chi$ those in the `environment'.
The interaction between the two is given by
$\mathcal{L}_{\textrm{int}}$. In this paper, we consider the simple case of linear coupling:
\be\label{intL}
\mathcal{L}_{\textrm{int}}[\phi,\chi] = g\; \phi\;  \mathcal{O}[\chi]
\ee
where $g$ is the coupling constant and $\mathcal{O}[\chi]$ is a given function of $\chi$.

We also assume that the number of degrees of freedom of the `system' is much smaller than that of the
`environment', such that the back-reaction of the `system' to the `environment' can be ignored during the
time scale of quantum decoherence of the `system'. This is a realistic assumption and has been adopted in the past study of quantum decoherence (see e.g. \cite{Environment,Caldeira:1982iu,Hu:1991di})

The main difference of the present paper from earlier studies is the following.
Previous studies of quantum decoherence usually consider the case in which the `system' $\phi$ is a simple harmonic oscillator (SHO) and the `environment' $\chi$
consists of a collection of independent SHOs at thermal equilibrium with temperature $T=1/\beta$. The (linear)
interaction between the two is dictated by the spectral weight $C_i$ of the SHOs in the `environment':
\be\label{spectral-SHO}
\mathcal{L}^{(\textrm{SHO})}_{\textrm{int}} = g \, \phi \sum_i C_i \chi_i\;.
\ee
This model is quadratic  therefore the influence functional and evolution dynamics of $\hat{\rho}_{\textrm{sys}}$ can be obtained exactly in this case.  Indeed, this is the simplest model to study non-equilibrium processes in the presence of environmental influence: the Feynman-Vernon formalism was developed using this model \cite{Feynman:1963fq}; and later it was used to study the quantum  Brownian motion in \cite{Caldeira:1982iu} (which derived its Langevin equation after making a suitable choice of the spectral weight $C_i$); finally it was also used to study quantum decoherence in \cite{Hu:1991di,Hu:1993vs}. See also \cite{fermion-1,fermion-2,fermion-3,NonMarkov} for the similar consideration of the fermionic version.

However, interesting physics happens when we go beyond the models with simple harmonics oscillators --- this is what we will do in this paper. Both the `system' and the `environment' have two different aspects: the theory itself (given by the Lagrangians) and its physical state (given by its time-dependent density matrix). Since the `environment' is assumed to be in a thermostatic state throughout the decoherence process (i.e. $\hat{\rho}_{\textrm{env}}=e^{-\beta \hat{H}_{\textrm{env}}}$), we will allow the `environment' Lagrangian to be arbitrary, although later in the actual computation we focus on theories with a dual gravity description so that we can use the holographic machinery to obtain results that would have been hard to compute directly in the field theory side. And we are interested in the difference  between different environment Lagrangians. On the other hand, we consider the `system' as a probe to study the decoherence effects of different `environments'; therefore we will choose a simple `system': a canonical scalar with Lagrangian
\be\label{Lsys}
\mathcal{L}_{\textrm{sys}}[\phi] =  -{1\over 2} (\partial_{\mu} \phi)^2 - {1\over 2} {\Omega}^2 \phi^2\;.
\ee
The influence  functional can be regarded as the probe's effective action, which is obtained after the  `environmental' degrees of freedom $\chi$ are integrated out.

\subsection{Feynman-Vernon and Schwinger-Keldysh}

  In the coupled total system given by the Lagrangian \eq{totL}, we are only interested in the evolution dynamics of the `system', but not in the detailed dynamics of the `environment'. Therefore the degrees of freedom of the `environment' should be integrated out and its whole influence on the `system' packaged into one or a few quantities. For the generic time-dependent problem at hand, this task requires the path-integral formalism devised by Feynman and Vernon in \cite{Feynman:1963fq}.

In this subsection, we review the Feynman-Vernon (FV) formalism. The main point of FV formalism is in rewriting the evolution \eq{totrho} in the path-integral representation, and thereby integrating out the `environmental' degrees of freedom $\chi$ to produce an `influence functional' that contains all the `environmental' effect on the `system'. As the evolution
of density matrix involves both forward ($e^{-iH_{\textrm{tot}}t}$) and backward ($e^{iH_{\textrm{tot}}t}$)
propagators, the path integral should be formulated on an ordered closed-time path (Keldysh contour $\mathcal{K}$), i.e., from $t=t_i$ to $t_f$ and then back \cite{Schwinger:1960qe,Keldysh:1964ud}. The ordinary causal Green's function is replaced by a $2\times 2$ matrix of Green's functions (a.k.a. Schwinger-Keldysh propagators) to account for the two branches in the Keldysh contour.
The influence functional is precisely the generating function of these Schwinger-Keldysh propagators \cite{Su:1987pi}.

\begin{figure}[h]
\begin{center}
\includegraphics[scale=0.35]{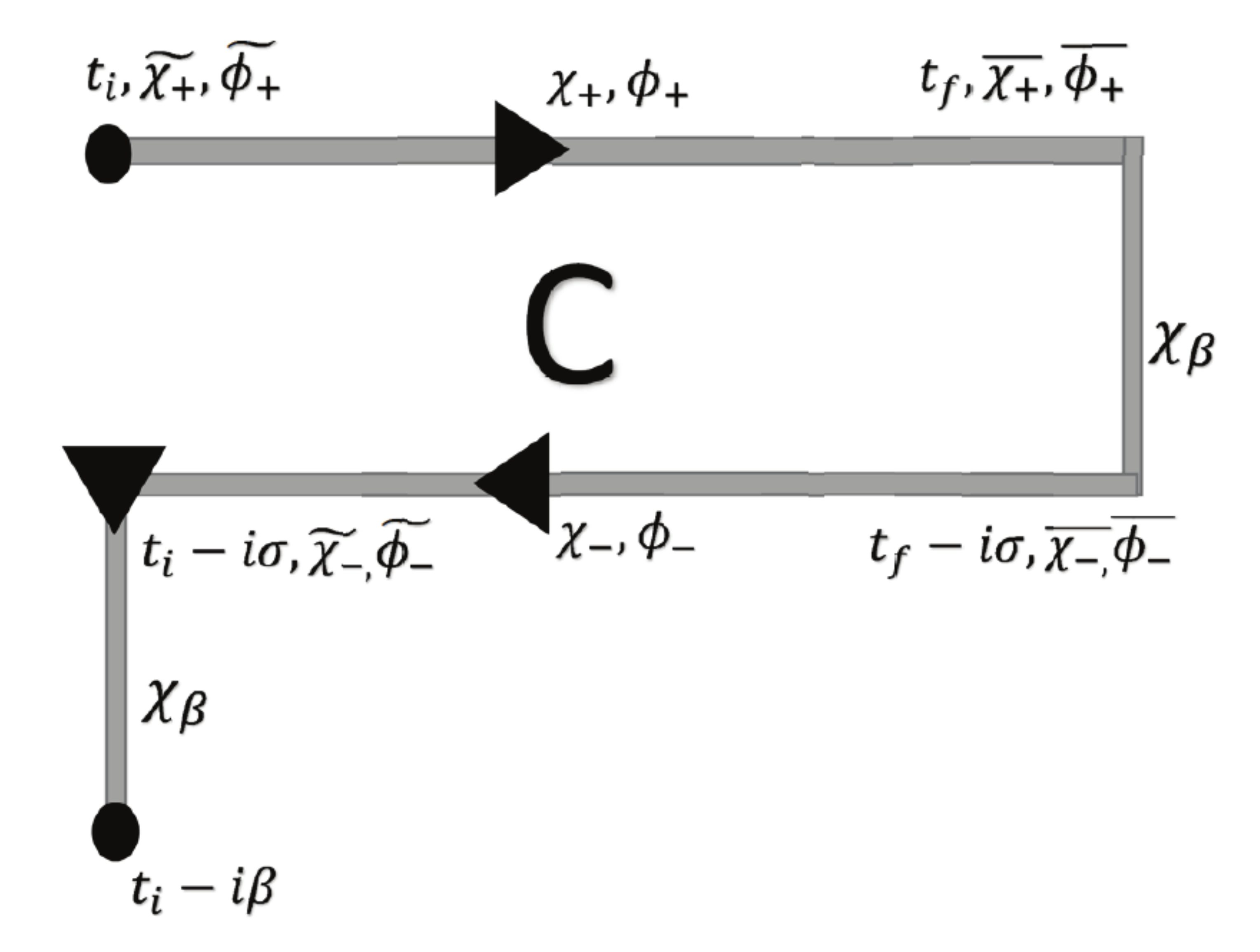}
\end{center}
\singlespace
\caption{The thermo-Keldysh contour $\mathcal{C}$.
The two horizontal segments represent the closed-time path (i.e. Keldysh contour $\mathcal{K}$), which is the contour for a generic `environment' density matrix  $\hat{\rho}_{\textrm{env}}$.
When the `environment' is thermal with temperature $\frac{1}{\beta}$, the average of the `environment' can be accounted for by the path-integral along the additional  vertical segments. 
}
\label{Keldysh contour}
\end{figure}

This formalism, including the connection between Feynman-Vernon and Schwinger-Keldysh, is valid for any initial density matrix $\hat{\rho}_{\textrm{env}}$ of the `environment'. For our present case of thermostatic environment with $\hat{\rho}_{\textrm{env}}=e^{-\beta \hat{H}_{\textrm{env}}}$, the effect of the thermo-average can also be represented as a path-integral, but along the imaginary time direction from $t_i$ to $t_i-i\beta$. The thermo-Keldysh contour $\mathcal{C}$ is along the path
\begin{equation}\label{fullcontour}
\mathcal{C}:\qquad t_i \quad \xrightarrow{\quad 1\quad} \quad t_f \quad \xrightarrow{\quad 2\quad} \quad t_f-i \sigma \quad \xrightarrow{\quad 3\quad} \quad t_i-i\sigma \quad \xrightarrow{\quad 4\quad} \quad t_i-i\beta\;.
\end{equation}
In the original thermo-Keldysh contour, $\sigma$ is chosen to be $0$. Later it was shown in \cite{Matsumoto:1982ry} that $\sigma$ can actually be chosen arbitrarily since the `environment' is thermostatic. To compare with the bulk computation, the symmetric choice $\sigma=\frac{\beta}{2}$ is the most convenient one \cite{Herzog:2002pc}. Therefore in this paper we will use the thermo-Keldysh contour $\mathcal{C}$ with $\sigma=\frac{\beta}{2}$, as shown in Fig. \ref{Keldysh contour}.

Now let us use the method of Feynman-Vernon to compute the reduced density matrix $\hat{\rho}_{\textrm{sys}}$ at the final time $t_f$.
Take an arbitrary element $\langle \bar{\phi}_+|\hat{\rho}_{\textrm{sys}}(t_f)|\bar{\phi}_-\rangle$. It is given by the total density matrix $\hat{\rho}_{\textrm{tot}}(t_f)$ via \eq{redrho}:
\begin{equation}\label{redrho1}
\langle \bar{\phi}_+|\hat{\rho}_{\textrm{sys}}(t_f)|\bar{\phi}_-\rangle =  \int d\bar{\chi}\; \langle \bar{\phi}_+,\bar{\chi}|\hat{\rho}_{\textrm{tot}}(t_f)|\bar{\phi}_-,\bar{\chi}\rangle
\end{equation}
where we use the bar to label the final values (at $t=t_f$) of fields.
Since we know that the total density matrix $\hat{\rho}_{\textrm{tot}}$ evolves by \eq{totrho}, after inserting  complete sets of fields at $t=t_i$ we can expand \eq{redrho1} into
\begin{equation}
\begin{aligned}
&\langle \bar{\phi}_+|\hat{\rho}_{\textrm{sys}}(t_f)|\bar{\phi}_-\rangle
 = \int d\bar{\chi}d\tilde{\chi}_+ d\tilde{\chi}_-  d\tilde{\phi}_+  d\tilde{\phi}_- \;
 \\ \label{rdm-1}
  &\langle \bar{\phi}_+, \bar{\chi} | e^{-iH_{\textrm{tot}}(t_f-t_i)}|\tilde{\phi}_+,\tilde{\chi}_+\rangle
  \cdot
  \langle \tilde{\phi}_+,\tilde{\chi}_+| \hat{\rho}_{\textrm{sys}}(t_i)\otimes \hat{\rho}_{\textrm{env}}
(t_i)|\tilde{\phi}_-,\tilde{\chi}_-\rangle
\cdot
 \langle \tilde{\phi}_-,\tilde{\chi}_-|e^{i H_{\textrm{tot}}
(t_f-t_i)}|\bar{\phi}_-,\bar{\chi}\rangle
\end{aligned}
\end{equation}
where we used tilde to label the initial values (at $t=t_i$) of fields. Note that the initial density matrix of the `environment' $\hat{\rho}_{\textrm{env}}(t_i)$ can be arbitrary.

The integrand of \eq{rdm-1} consists of three terms. The first and third are forward and backward  propagators, respectively, which can be rewritten in the path-integral representation as
\begin{equation}\label{fb}
\begin{aligned}
 \langle \bar{\phi}_+, \bar{\chi}_+| e^{- iH_{\textrm{tot}}(t_f-t_i)}|\tilde{\phi}_+,\tilde{\chi}_+\rangle &=
\int_{\tilde{\phi}_+}^{\bar{\phi}_+} \mathcal{D}\phi_+ \int_{\tilde{\chi}_+}^{\bar{\chi}_+}
\mathcal{D} \chi_+ \; e^{i \int_{t_i}^{t_f} dt \; \mathcal{L}[\phi_+,\chi_+] }\;, \\
 \langle \tilde{\phi}_-, \tilde{\chi}_-| e^{iH_{\textrm{tot}}(t_f-t_i)}|\bar{\phi}_-,\bar{\chi}_-\rangle &=
\int_{\tilde{\phi}_-}^{\bar{\phi}_-}  \mathcal{D}\phi_- \int_{\tilde{\chi}_-}^{\bar{\chi}_-}
\mathcal{D} \chi_- \; e^{-i \int_{t_i}^{t_f} dt \; \mathcal{L}[\phi_-,\chi_-] }
\end{aligned}
\end{equation}
where $\bar{\chi}_{+}=\bar{\chi}_{-}=\bar{\chi}$ and we have used the subscript $+/-$ to denote fields living on forward/backward time path.

Now let us extract all the information of the `environment' field $\chi$ from the r.h.s of \eq{rdm-1}. Plugging \eq{fb} back into \eq{rdm-1} and recalling that the Lagrangian in \eq{fb} is given by \eq{totL}, we can package all the information of the `environment' field $\chi$  into an influence functional $\mathcal{F}$ defined as:
\begin{equation}\label{FV}
\begin{aligned}
\mathcal{F}[\phi_+,\phi_-] &\equiv \int d\bar{\chi}\; d\tilde{\chi}_{+}\; d\tilde{\chi}_{-}\; \langle
\tilde{\chi}_+|\hat{\rho}_{\textrm{env}}(t_i) |\tilde{\chi}_-\rangle
\\
& \cdot \int_{\tilde{\chi}_+}^{\bar{\chi}}  \mathcal{D} \chi_+ \int_{\tilde{\chi}_-}
^{\bar{\chi}}\mathcal{D} \chi_- e^{-i\int_{t_i}^{t_f} dt (\mathcal{L}_{\textrm{env}}[\chi_+]-\mathcal{L}
_{\textrm{env}}[\chi_-]+\mathcal{L}_{\textrm{int}}[\phi_+,\chi_+]-\mathcal{L}_{\textrm{int}}[\phi_-,\chi_-])} \;.
\end{aligned}
\end{equation}
Once the influence functional is known, \eq{rdm-1} is given by
\be\label{rho-dy}
\langle \bar{\phi}_+|\hat{\rho}_{\textrm{sys}}(t_f)|\bar{\phi}_-\rangle = \int  d\tilde{\phi}_+  d
\tilde{\phi}_-  J\left[\bar{\phi}_+, \bar{\phi}_-; t_f|\tilde{\phi}_+,\tilde{\phi}_-; t_i\right] \langle
\tilde{\phi}_+ | \hat{\rho}_{\textrm{sys}}(t_{i}) |\tilde{\phi}_- \rangle
\ee
where $J$ is the propagating function that dictates the evolution of $\hat{\rho}_{\textrm{sys}}$ from $t=t_i$ to $t_f$, and it depends on the `system' Lagrangian and `environmental' influence functional via:
\be\label{Jrho}
J\left[\bar{\phi}_+, \bar{\phi}_-; t_f|\tilde{\phi}_+,\tilde{\phi}_-; t_i\right] \equiv \int_{\tilde{\phi}_+}
^{\bar{\phi}_+} \mathcal{D}\phi_+ \int_{\tilde{\phi}_-}^{\bar{\phi}_-}  \mathcal{D}\phi_-\; e^{i
\int_{t_i}^{t_f} dt \; (\mathcal{L}_{\textrm{sys}}[\phi_+] -\mathcal{L}_{\textrm{sys}}[\phi_-])}  \mathcal{F}[\phi_
+,\phi_-]\;.
\ee
Therefore the most crucial task is to compute the influence functional $\mathcal{F}$.

First, \eq{FV} can be written concisely as
\begin{equation}\label{Ftrace}
\mathcal{F}[\phi_{+},\phi_{-}]=\langle \mathcal{T}_{\mathcal{K}}e^{i\int_{\mathcal{K}}\mathcal{L}_{\textrm{int}}[\phi,\chi]}\rangle_{\textrm{env}}
\end{equation}
where the correlator is averaged w.r.t. the initial density matrix of the `environment':
\begin{equation}\label{average}
\langle\dots  \rangle_{\textrm{env}} \equiv \Tr_{\textrm{env}}\left[\hat{\rho}_{\textrm{env}}\dots\right]
\end{equation}
and $\mathcal{K}$ denotes the closed-time path (i.e. the two horizontal segments in the contour $\mathcal{C}$ of Fig. \ref{Keldysh contour}.), $\mathcal{T}_{\mathcal{K}}$ is the path-ordering operator. For a generic interaction $\mathcal{L}_{\textrm{int}}$, the influence functional $\mathcal{F}$ is difficult to compute. However, when the interaction $\mathcal{L}_{\textrm{int}}$ is linear in $\phi$ as given in \eq{intL}, \eq{Ftrace} becomes
\begin{equation}\label{Ftrace-1}
\mathcal{F}[\phi_{+},\phi_{-}]=\langle \mathcal{T}_{\mathcal{K}}e^{i g \int_{\mathcal{K}}\phi \,\mathcal{O}[\chi]}\rangle_{\textrm{env}}\;.
\end{equation}
If we consider the `system' field $\phi$ as the source of the `environment' field $\chi$ (or more precisely of $\mathcal{O}[\chi]$), the influence functional $\mathcal{F}$ with the linear coupling \eq{Ftrace-1} is precisely the generating function of the real-time Green's functions at finite temperature (a.k.a. Schwinger-Keldysh propagators), which are defined as
\be
G_{ss'}(1,2) \equiv -i \langle \mathcal{T}_{\mathcal{K}}\mathcal{O}_{s}(1)\mathcal{O}_{s'}(2)  \rangle_{\textrm{env}}
\ee
where $\mathcal{T}_{\mathcal{K}}$ is the path-ordering operator along the closed-time path $\mathcal{K}$, $s,s'=\pm$, and $\mathcal{O}_{s}(1)\equiv \mathcal{O}[\chi_{s}(t_1,\vec{x}_1)]$ \cite{Su:1987pi}. Namely
\begin{equation}\label{FVtoSK}
G_{ss'}(1,2)=-i\frac{\delta \ln \mathcal{F}[\phi_{+},\phi_{-}]}{\delta \phi_{s}(1)\, \delta \phi_{s'}(2)} \big\vert_{\phi=0}\;.
\end{equation}

Given the influence functional, the Schwinger-Keldysh Green's functions can be obtained via
\eq{FVtoSK}. However, for the decoherence problem at hand, the question is the inverse: How to compute the influence functional $\mathcal{F}$ once we know the Schwinger-Keldysh Green's function $G_{ss'}(1,2)$ of the `environment'? This question can be answered when the coupling between `system' and `environment' is weak (i.e. $g\ll 1$), where we can approximate by keeping only the quadratic coupling term (in the exponent of $\mathcal{F}$) and dropping all higher order terms:\footnote{It was explicitly
shown in \cite{Boyanovsky:2004dj} that the leading term of the influence functional yields the
generating function for the Schwinger-Keldysh Green's functions of $\mathcal{O}[\chi]$ in a
thermal reservoir.}
\be\label{FV-G1}
-i \ln \mathcal{F} [\phi_+,\phi_-] \approx -{g^2\over 2} \int d^{d+1}x \int d^{d+1}x' \;  \sum_{s,s'}
\textrm{sgn}(ss') \phi_s(x) G_{ss'}(x-x') \phi_{s'}(x')\;.
\ee
In fact, when the `environment' is composed of a collection of simple harmonic oscillators, the coupling is necessarily quadratic (in $\phi$), and \eq{FV-G1} is actually the exact answer.

To summarize, the strategy of studying the decoherence (in the FV formalism) is to first solve the Schwinger-Keldysh Green's function of the `environment', then using \eq{FV-G1} we can immediately obtain the influence functional $\mathcal{F}$, which in turn gives the propagating function $J$ via \eq{Jrho}, and then it is straightforward to study the evolution of $\hat{\rho}_{\textrm{sys}}$ using  \eq{rho-dy}.

Before we proceed, let us first simplify \eq{FV-G1} further. First of all, out of the four Green's functions $G_{ss'}$ only three are independent, since by definition
\begin{equation}
G_{++}+G_{--}=G_{-+}+G_{+-}\;.
\end{equation}
The remaining three linear combinations of $G_{ss'}$ contain two types of information on the `environment'. The advanced and retarded Green's functions
 $\{G_{\textrm{A}},G_{\textrm{R}}\}$
encode the dispersion relations of the `environment', whereas the symmetric Green's function $G_{\textrm{sym}}$
describes the actual state of the `environment'.\footnote{We remind that $\{G_{\textrm{A}},G_{\textrm{R}}, G_{\textrm{sym}}\}$ are defined as follows
\begin{equation}
iG_{\textrm{A}}=\theta(t_2-t_1) \langle [\mathcal{O}(2),\mathcal{O}(1)]  \rangle_{\textrm{env}}\,, \quad iG_{\textrm{R}}=\theta(t_1-t_2) \langle [\mathcal{O}(1),\mathcal{O}(2)]  \rangle_{\textrm{env}}\,, \quad G_{\textrm{sym}}=\frac{1}{2} \langle \{ \mathcal{O}(1),\mathcal{O}(2)\} \rangle_{\textrm{env}}\,.
\end{equation}
and they are related to the Schwinger-Keldysh Green's functions via:
\be
G_{\textrm{A}}=G_{++}-G_{-+}\;,\quad G_{\textrm{R}}=G_{++}-G_{+-}\;,\quad G_{\textrm{sym}}={i\over 2}(G_{++}+G_{--})\;.
\ee}
 Therefore it is more convenient to switch to a basis for the matrix Green's function such that its four components are manifestly $\{G_{\textrm{A}},G_{\textrm{R}},G_{\textrm{sym}},0\}$.
Accordingly, instead of $\{\phi_{+},\phi_{-}\}$, we should switch to the basis $\{\Sigma, \Delta\}$ in `system' \cite{Keldysh:1964ud,KCChou}:
\begin{equation}
\Sigma=\frac{\phi_{+}+\phi_{-}}{2}\;, \qquad \qquad \Delta=\phi_{+}-\phi_{-}\;.
\end{equation}
The field $\Delta$ is the difference (i.e. relative) between the two fields $\phi_+$ and $\phi_{-}$ (from the two branches of the contour), and $\Sigma$ is the average --- therefore $\{\Sigma, \Delta\}$ is called the `ra' basis, in which the influence functional can be written as \be\label{FV-final}
-i \ln \mathcal{F} = - g^2 \int_{t_i}^{t_f} d\tau  \int_{t_i}^{t_f} d\tau' \int d^d x\; d^d x'
\; [ \Delta(x) G_{\textrm{R}}(x-x') \Sigma(x')-{i\over 2} \Delta(x) G_{\textrm{sym}}(x-x') \Delta(x') ]
\ee
where we have used the relation
\begin{equation}\label{RA}
G_{\textrm{A}}(1,2)=G_{\textrm{R}}(2,1)\;.
\end{equation}

So far, the derivation applies to generic $\hat{\rho}_{\textrm{env}}(t_i)$. Indeed, as long as we can obtain the `environment''s Green's function, we can study the decoherence in this `environment' using \eq{FV-final} no matter how exotic the `environment' is --- even if the `environment' itself is in non-equilibrium (with interesting examples being de Sitter space and non-thermal `environment').

 Now let's restrict to the most realistic and simplest situation when the `environment' is in a thermo-equilibrium with temperature $\frac{1}{\beta}$, i.e. the density matrix $\hat{\rho}_{\textrm{env}}=e^{-\beta \hat{H}_{\textrm{env}}}$.  In such cases, the trace in (\ref{average}) reduces to a thermal average and can be accounted for by a path-integral from $t_i$ to $t_i -i\beta$. Namely, the Keldysh contour $\mathcal{K}$ (the two horizontal segments in Fig. \ref{Keldysh contour}) is supplemented by the vertical segment to give the full thermo-Keldysh contour $\mathcal{C}$ defined in (\ref{fullcontour}) and represented in Fig. \ref{Keldysh contour}.
The fields that appear in the definition of Schwinger-Keldysh propagators now become
$\mathcal{O}_{\pm}(1)\equiv \mathcal{O}[\chi_{\pm}(t_1,\vec{x}_1)]$ with
\begin{equation}
\chi_{+}(t)=e^{i H_{\textrm{env}}t}\chi e^{-i H_{\textrm{env}}t}\;, \qquad \chi_{-}(t)=e^{i H_{\textrm{env}}(t-i\sigma)}\chi e^{-i H_{\textrm{env}}(t-i\sigma)}\;.
\end{equation}

When the `environment' is in a thermo-equilibrium,\footnote{Otherwise, the condition (\ref{g3}) and (\ref{KMS-w}) are not valid. However, in this paper we will restrict to the thermostatic environment at all time as the probe is small and local compared to the environment. For discussions of quantum decoherence in non-thermostatic environments, see \cite{fermion-1,fermion-2,NonMarkov}.} its Green's functions satisfies another constraint, namely the Kubo-Martin-Schwinger (KMS) condition \cite{Martin:1959jp}:
\begin{equation}
G_{+-}(t-i\beta,\vec{x})=G_{-+}(t,\vec{x}) \label{g3}
\end{equation}
which gives rise to the condition
\be\label{KMS-w}
G_{\textrm{sym}}(\omega) = - [1+2 n(\omega)] \textrm{Im} G_{\textrm{R}}(\omega)
\ee
in the frequency domain, where $n(\omega)= {1\over e^{\beta \omega}-1}$ is the thermal distribution of the
`environment'. Eq. \eq{KMS-w} relates the symmetric Green's function to the imaginary part of the
retarded Green's function, which is the spectral function for the $\mathcal{O}[\chi]$ excitations.

To summarize, when the `environment' is thermostatic, all its dynamical information can be encoded in the retarded Green's function $G_{\textrm{R}}$. Note that there is no quadratic term for $\Sigma$ in \eq{FV-final} due to the condition \eq{RA}.
Moreover, the quadratic form of \eq{FV-final} ensures that one can further solve the propagating
function for $\hat{\rho}_{\textrm{sys}}$, i.e. \eq{Jrho} in a closed form as long as $S_{\textrm{sys}}[\phi]$ is
also quadratic. This implies that the dynamics of $\hat{\rho}_{\textrm{sys}}$ can be
accurately determined  once $G_{\textrm{R}}$ is known.

We consider the relation \eq{FV-final} (or together with its higher order term) the most essential element in the study of quantum decoherence as it connects the Feynman-Vernon (the theoretical framework of environment-induced decoherence) to Schwinger-Keldysh (the machinery that allows us to capture and compute the influence from the `environment'). Its interpretation is the following. Once the `system' starts to interact with the `environment', it is under the influence of the `environmental noise'. The basic properties of this noise, such as its energy and lifetime, is characterized by the
 `environment''s retarded Green's function $G_{\textrm{R}}$ (in terms of its poles and zeros), from which one can extract the corresponding transport coefficients of the thermal reservoir by, for instance, the Kubo
formula. (This step is the standard application of Schwinger-Keldysh formalism.) Then from the viewpoint of the `system', the transport phenomena carries away its
quantum information and causes its (gradual) decoherence.

At this point, one might wonder why this natural connection  hasn't been very visible in the literature of quantum decoherence. There are two reasons. In the usual discussion of quantum decoherence, the `environment' is modeled by a collection of SHOs for simplicity,\footnote{A notable exception is the spin-bath model studied in \cite{SpinBath}. Modeling localized modes such as nuclear spins and defects in the environments, a spin-bath is intrinsically strongly-coupled to the `system' therefore cannot be mapped to an oscillator bath: whereas the back-reaction from the `system' to the oscillator-bath is usually negligible, the coupling between the `system' and the spin-bath is so strong that the dynamics of spin-bath is slaved by the `system' \cite{SpinBath}.
It would be interesting to holographically model the decoherence due to a spin-bath.} and its spectral function (i.e. the spectral weight $C_i$ in \eq{spectral-SHO}) can then be chosen at will and put in by hand (instead of being determined from first principle QFT computation). Furthermore, the simplicity of the SHO model makes the machinery like \eq{FV-final} unnecessary since in this case the influence functional can be easily obtained in closed-form. It is this  simplicity that has delayed the application of Feynman-Vernon formalism to more general
`environments' and therefore the study of quantum decoherence in more physical and
interesting situations.

\subsection{Holographic influence functional}\label{holographic GR}

The relation \eq{FV-final} is central to Environment-induced decoherence and applies to any `environment'; now let us apply it to cases beyond the toy model with SHO `environment'.
A crucial step is to evaluate the retarded
Green's function $G_{\textrm{R}}$ of $\mathcal{O}[{\chi}]$ (the operator  in the `environment' that couples to the `system' field $\phi$).  After obtaining the retarded Green's function, we can plug it into \eq{FV-final} to
calculate the influence functional from which one can explore the behaviors of quantum
decoherence. The retarded Green's function is in general difficult to evaluate except for the free
theory. This is part of the reason that the quantum decoherence for the interacting
`environment'
is much less studied.

    However, besides the free theory, there exists a class of strongly-coupled theories whose retarded Green's functions can be obtained in practice. These are the theories with dual gravity descriptions. The most developed class of Gauge/Gravity duality is the AdS/CFT correspondence, in which the (non-gravitational) field theory is a $d$-dimensional conformal field theory and its gravity dual lives in a $(d+1)$-dimensional asymptotically AdS space.  The
zero-temperature (Euclidean) two-point functions of these CFTs have power-law behaviors; whereas  the finite-temperature real-time Green's functions cannot be determined by the conformal symmetry alone, but can be evaluated through its gravity dual.

    In the present context, our probe field $\phi$ plays the role of the external source
coupled to the operator $\mathcal{O}[\chi]$. In the gravity dual, $\phi$ is the boundary
value of a massive elementary field in the asymptotically AdS space, with its mass determined by
the conformal dimension of operator $\mathcal{O}[\chi]$.  Then, the generating function for the
correlation functions of operator $\mathcal{O}[\chi]$, i.e., the influence functional, is
determined by the on-shell action of the corresponding massive field \cite{GKPW}:
\be\label{AdSCFT for IF}
\mathcal{F}^{(h)}[\phi]:= \big{\langle} e^{i g \int_{\mathcal{C}} d^{d+1}x \; \phi\;
\mathcal{O}[\chi]} \big{\rangle}_{\chi} = e^{i S_{\textrm{bulk}}[\Phi|_{\phi}]}
\ee
where $\Phi|_{\phi}$ is the on-shell bulk scalar with $\phi$ being its value at AdS boundary, and
$S_{\textrm{bulk}}$ is the bulk action of $\Phi$. The superscript $h$ denotes the ``holographic".

   The prescription of calculating the on-shell bulk action to derive the Euclidean two-point
functions was proposed in \cite{GKPW}: one imposes Dirichlet boundary condition at the AdS boundary and needs to choose appropriate
boundary action when evaluating the on-shell
action.  For evaluating the Lorentzian correlation functions such as the retarded Green's
function, especially at finite temperature (corresponding to a black hole in asymptotically AdS space), the
choice of the boundary condition at the black hole horizon and the corresponding boundary
action are more subtle. The subtlety arises because there is a coordinate singularity at the
horizon, thus one should choose the appropriate boundary condition for $\Phi|_{\phi}$ at horizon
to have a smooth solution. This choice corresponds to the thermal vacuum for the dual
CFT. To compute the retarded Green's function, one should choose the in-falling condition for the on-shell $\Phi|_{\phi}
$ at the black hole horizon, and the result is \cite{Son:2002sd,Iqbal:2008by}:
\be\label{holoRetard}
G^{(h)}_{\textrm{R}}(x) = - \lim_{r \rightarrow \infty} {\Pi(x,r)|_{\Phi|{\phi}} \over \Phi|_{\phi}
(x,r)}
\ee
where $x$ is the transverse coordinate of the AdS space and $r$ is the radial one with the AdS
boundary at $r \rightarrow \infty$. Here $\Pi|_{\Phi|{\phi}}$  is the conjugate momentum of
$\Phi$ with respect to the on-shell bulk action.

  In fact, the denominator of \eq{holoRetard}  is the leading term in the large $r$ expansion of
$\Phi|_{\phi}$, and the numerator is the sub-leading one.  After scaling away the $r$ factors in
both the numerator and denominator of \eq{holoRetard}, it can be seen as  a linear response
theory for the dual CFT: $
G^{(h)}_{\textrm{R}} = { \langle \mathcal{O} \rangle|_{\phi} \over \phi}$
if one identifies $\Pi|_{\Phi|{\phi}}$ as $\langle \mathcal{O} \rangle|_{\phi}$ after some
appropriate  rescaling.\footnote{For a massive canonical bulk scalar $\Phi$ in a $(d+1)$-dimensional AdS
space, $\phi(x)=\lim_{r\rightarrow \infty} \Phi_{\phi}(x,r)$ and $\langle \mathcal{O}
\rangle|_{\phi}=\lim_{r\rightarrow \infty}r^{\Delta-d}\Pi(x,r)|_{\Phi|{\phi}}$ with $\Delta$
the conformal dimension of $\mathcal{O}$.}

   It was shown in \cite{Herzog:2002pc} that the on-shell action for a bulk scalar in the AdS
black hole background yields the same form as the r.h.s. of \eq{FV-G1}.  The forward and
backward paths in the Schwinger-Keldysh formalism correspond to the time path in a causal patch
and the one in its mirror image of the Penrose diagram for an eternal AdS black hole.
\footnote{This kind of identification was first proposed in \cite{Israel:1976ur} for
Schwarzschild black hole, and later generalized to AdS one in \cite{Maldacena:2001kr}.}
Combining this result with \eq{AdSCFT for IF}, we summarize that the holographic influence
functional is given as follows:
\be\label{holo-fv-if}
-i \ln \mathcal{F}^{(h)} = - g^2 \int_{t_i}^{t_f} d\tau  \int_{t_i}^{t_f} d\tau' \int d^d x\;
d^d x' \; [ \Delta(x) G^{(h)}_{\textrm{R}}(x-x') \Sigma(x')-{i \over 2} \Delta(x) G^{(h)}_{\textrm{sym}}(x-x')
\Delta(x') ]
\ee
where $G^{(h)}_{\textrm{R}}$ is evaluated as in \eq{holoRetard}, and $G^{(h)}_{\textrm{sym}}$ is related to
$G^{(h)}_{\textrm{R}}$ by the KMS condition \eq{KMS-w}. This is the holographic version of \eq{FV-final}. Again,
the quadratic form of \eq{holo-fv-if} is guaranteed by the weak coupling $g$ so that the higher
order correlation terms are suppressed.

  Here, two remarks are in order:
\begin{enumerate}
\item Although one can obtain the holographic influence functional by treating the `environment' as a holographic CFT, one cannot naturally incorporate the kinetic term for the probe field in
this context as it is treated as an external source. Thus, we need to put in probe's kinetic
term such as \eq{Lsys} by hand. It would be more satisfactory if the dynamics of the probe field
could also be embedded in the bulk holographically.
\item The holographic CFTs are usually gauge theories therefore the operator $\mathcal{O}[\chi]$
should be a gauge invariant observable. Hence the probe can only decohere through a particular
channel of physical observable to which it couples. The probe and its decoherence behavior are
characterized by the spectral density of $\mathcal{O}[\chi]$ encoded in the retarded Green's
function. This is in contrast to the case of SHO `environment' in which the probe SHO couples
directly to all the `environmental' degrees of freedom and the probe and its
decoherence behavior are characterized by the spectral weight $C_i$ in \eq{spectral-SHO}.  This
difference is illustrated in Fig. \ref{ways}.
\begin{figure}[h]
\begin{center}
\includegraphics[scale=0.35]{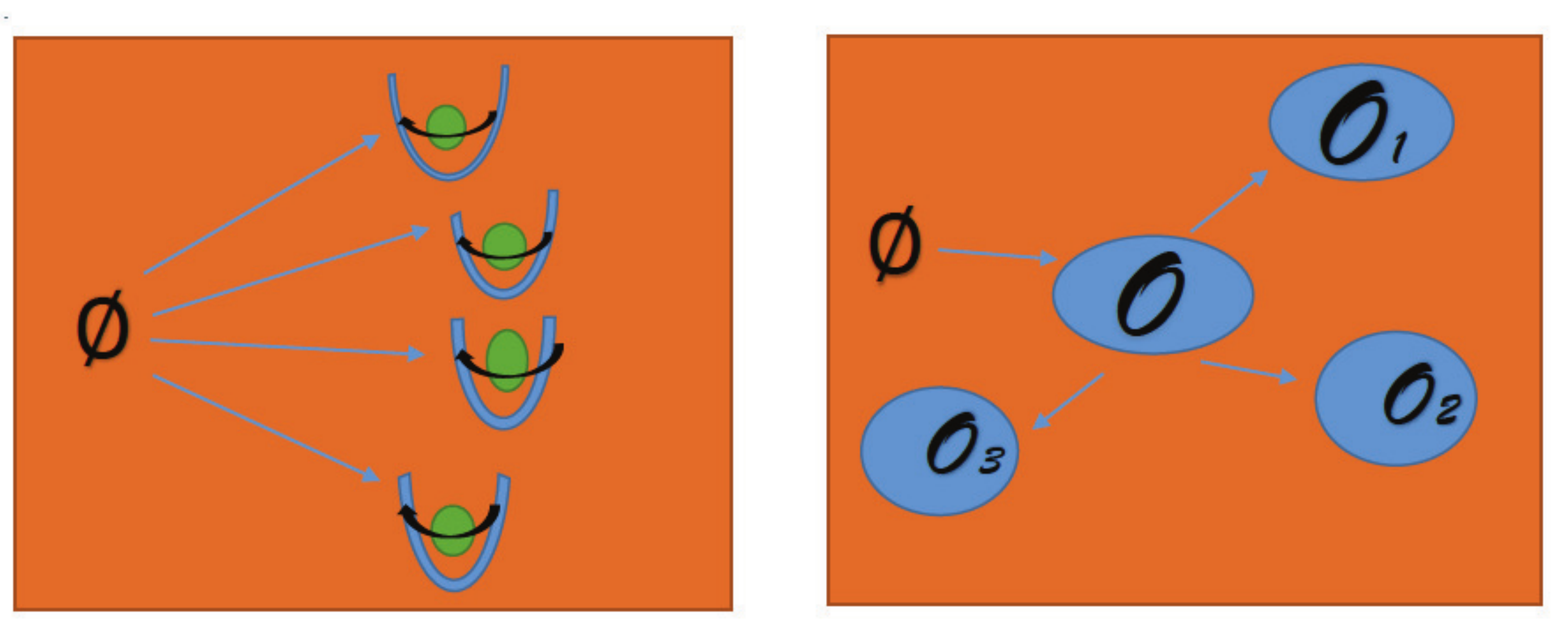}
\end{center}
\singlespace\caption{Different decoherence mechanisms for the probe $\phi$. Left: $\phi$ in thermal reservoir of simple harmonic oscillators. Right: $\phi$ in holographic thermal reservoir.}
\label{ways}
\end{figure}
\end{enumerate}

\section{Propagating function}\label{sec3}

As shown in the previous section, the evolution of the `system''s reduced density matrix $\hat{\rho}_{\textrm{sys}}$ is given by an integral transform \eq{rho-dy}, whose kernel is the propagating function $J$ given by \eq{Jrho}. The propagating function $J$ is determined by two inputs via the path-integral \eq{Jrho}: (1) the data of the `system' (i.e. its Lagrangian
and its initial and final states $\{\tilde{\phi}_{\pm},\bar{\phi}_{\pm}\}$); (2) the effect from the `environment' (i.e. the influence functional $\mathcal{F}$, which can be expressed in terms of Schwinger-Keldysh propagators of the `environment').  In this section, we first combine the two inputs to rewrite the propagating function directly in terms of the `environment''s Green's function plus the `system''s data. This then allows us to derive the master equation of $\hat{\rho}_{\textrm{sys}}$.

\subsection{Langevin equation from propagating function}\label{brownian}

The propagating function $J$ given in \eq{Jrho} can also be interpreted as the path-integral of the `system' on the Keldysh contour $\mathcal{K}$:
\be\label{Jrhophi}
J\left[\bar{\phi}_{+}, \bar{\phi}_{-}; t_f | \tilde{\phi}_{+},\tilde{\phi}_{-}; t_i\right] =
\int_{\tilde{\phi}_+}^{\bar{\phi}_+} \mathcal{D}\phi_+ \int _{\tilde{\phi}_-}^{\bar{\phi}_-}
\mathcal{D}\phi_-  \; e^{iS_{\textrm{eff}}[\phi_+,\phi_-]}
\ee
in which the action of the `system' is modified into $S_{\textrm{eff}}$ due to the influence of the `environment':
\be\label{Seffphi}
S_{\textrm{eff}}[\phi_{+},\phi_{-}] \equiv \int_{t_i}^{t_f} dt \; (\mathcal{L}_{\textrm{sys}}[\phi_+] -\mathcal{L}_{\textrm{sys}}[\phi_-])-i\ln
\mathcal{F}[\phi_+,\phi_-]\;.
\ee
Now we switch to the `ra' basis which is more suitable when studying the non-equilibrium process in terms of `environment''s Green's functions, and set $t_i=0$ and $t_f=t$,  with which $J$ becomes a partition function of $\{\Sigma, \Delta\}$:
\be\label{Jrho-ra}
J\left[\bar{\Sigma}, \bar{\Delta}; t | \tilde{\Sigma},\tilde{\Delta}; 0\right] =
\int_{\tilde{\Sigma}}^{\bar{\Sigma}} \mathcal{D}\Sigma \int _{\tilde{\Delta}}^{\bar{\Delta}}
\mathcal{D}\Delta  \; e^{iS_{\textrm{eff}}[\Sigma,\Delta]}
\ee
with action
\begin{equation}
\begin{aligned}
S_{\textrm{eff}}[\Sigma,\Delta] =&  \int_{0}^{t} d\tau [\dot{\Sigma}(\tau)  \dot{\Delta} (\tau) - \Omega^2\,  \Sigma(\tau) \Delta(\tau)]
 \\ &- g^2 \int_{0}^{t} d\tau  \int_{0}^{t} d\tau' [\Delta(\tau)G_{\textrm{R}}(\tau-\tau')
 \Sigma(\tau')-{i\over 2} \Delta(\tau) G_{\textrm{sym}}(\tau-\tau') \Delta(\tau') ] \;.\label{seff-main}
\end{aligned}
\end{equation}
where we have used the `system' Lagrangian \eq{Lsys} and the expression of $\mathcal{F}$ in terms of Schwinger-Keldysh propagators \eq{FV-final}, and $\dot{\Sigma}\equiv \frac{\partial \Sigma}{\partial \tau}$, and we suppress the spatial dependence for simplicity.

Before we proceed to evaluate the propagating function $J$ using \eq{Jrho-ra} and \eq{seff-main}, let's first check that it has the correct semi-classical limit.  In the semi-classical approximation,  $J$ should give the Langevin equation for the quantum Brownian motion as shown earlier in  \cite{Caldeira:1982iu,deBoer:2008gu,Son:2009vu}. Now we review the derivation. Viewing $J$ as the partition function of $\{\Sigma,\Delta\}$, and recall that the `average' field $\Sigma$ is slow and heavy whereas the `relative' field $\Delta$ is fast and light, we can integrate out the light field $\Delta$ and obtain the classical trajectory of the heavy field $\Sigma$.

First, via a Legendre transform, the $\Delta^2$ term in the r.h.s. of \eq{seff-main} can be rewritten as
\be\label{response}
e^{- {1\over 2}g^2 \int_{0}^{t} d\tau
\int_{0}^{t} d\tau' \Delta(\tau) G_{\textrm{sym}}(\tau-\tau') \Delta(\tau')}
= \int \mathcal{D} \xi \; e^{i\int_{0}^{t} d\tau  \Delta(\tau) \xi(\tau)
- {1\over 2 g^2}\int_{0}^{t} d\tau  \int_{0}^{t} d\tau'  \xi(\tau) G_{\textrm{sym}}
^{-1}(\tau-\tau') \xi(\tau')}\;.
\ee
Namely the fast field $\Delta$ can be regarded as the response field of a Gaussian random force $\xi$ which has the correlation
\begin{equation}\label{noise}
\langle \xi(\tau)\xi(\tau')\rangle_{\textrm{env}} = g^2 \; G_{\textrm{sym}}(\tau-\tau')\;.
\end{equation}
Intuitively, integrating out the `relative' (light and fast) mode $\Delta$ should provide random kicks to the `center-of-mass' (heavy and slow) mode $\Sigma$. Therefore we plug \eq{response} into \eq{Jrho-ra} and \eq{seff-main} and integrate out $\Delta$ to obtain
\be
J= \int \mathcal{D}\xi  e^{- {1\over 2}\int_{0}^{t} d\tau  \int_{0}^{t} d\tau'  \xi(\tau)
G_{\textrm{sym}}^{-1}(\tau-\tau') \xi(\tau')}
\int  \mathcal{D}\Sigma \delta\big[\ddot{\Sigma}(\tau)
+  \Omega^2 \Sigma(\tau) + g^2 \int^{\tau} d\tau' G_{\textrm{R}}(\tau-\tau')\Sigma(\tau')-\xi(\tau)\big]\;.
\ee
The argument of the delta-function gives the classical equation obeyed by the center-of-mass field $\Sigma$:
\be
\ddot{\Sigma}(\tau) +  \Omega^2 \Sigma(\tau) + g^2 \int^{\tau} d\tau' G_{\textrm{R}}(\tau-\tau')\Sigma(\tau')
=\xi(\tau)
\ee
which is precisely the  Langevin equation, where $\Sigma$ is the slow field and $\xi(\tau)$ the noise that satisfies \eq{noise}. The fluctuation-dissipation theorem for this quantum Brownian motion is nothing but the KMS condition \eq{KMS-w}.

\subsection{Evaluating the propagating function and master equation}

   Unlike the discussion in the previous subsection, to fully characterize the dynamics of
$\hat{\rho}_{\textrm{sys}}$ and examine the quantum decoherence of the probe, one should carry out
explicitly the Gaussian path integral of \eq{Jrho-ra}.  Although the calculation has been done, for
example in \cite{Hu:1991di}, we review it in Appendix \ref{app-a}
for completeness and to fix the notation. Besides, there are some issues about the normalization
of $\hat{\rho}_{\textrm{sys}}$ which were not considered explicitly in  \cite{Hu:1991di}, thus we carry
out the explicit calculation in Appendix \ref{app-normal}.

Since in the decoherence process, we are mainly interested in the time-dependence of $\hat{\rho}_{\textrm{sys}}$, from now on we will focus on the case in which the `system' field has no spatial dependence. For example, this is the case when the `system' is a particle and the $\phi$ field its coordinate, or if we only consider one particular momentum mode of $\phi$.
The result of $J$ is then summarized here:
\be\label{Jrho-exp}
J[\bar{\Sigma},\bar{\Delta},t|\tilde{\Sigma},\tilde{\Delta},0]
={1\over 2\pi |h(t)|} e^{i \; \mathcal{W}_{\textrm{eff}}[\bar{\Sigma},\bar{\Delta},\tilde{\Sigma},\tilde{\Delta};t]}
\ee
where
\be
\mathcal{W}_{\textrm{eff}}\equiv \left(\dot{f}_0(t) \tilde{\Sigma}+\dot{f}_1(t)\bar{\Sigma}\right)
\bar{\Delta}- \left(\dot{f}_0(0) \tilde{\Sigma}+\dot{f}_1(0)\bar{\Sigma}\right)\tilde{\Delta}-
{1\over 2}\left(a_{00}\tilde{\Delta}^2 + (a_{01}+a_{10})\tilde{\Delta}\bar{\Delta}
+a_{11}\bar{\Delta}^2\right)
\ee
and
\be\label{aij-main}
a_{ij}(t)\equiv g^2 \int_0^t d\tau \int_0^t d\tau' g_i(\tau) G_{\textrm{sym}}(\tau-\tau') g_j(\tau')\;.
\ee
For simplicity, we set $t_i=0$ and $t_f=t$ from now on. The functions $\{h,f_0,f_1,g_0,g_1\}$
are defined as follows. First, $h(\tau)$ is the central one based on which all others are defined; and it is the solution of the initial value problem:
\be\label{heom-main}
\ddot{h}(\tau)+\Omega^2 h (\tau) + g^2 \int_0^\tau d\tau' G_{\textrm{R}}(\tau-\tau') h(\tau')=0 \qquad \textrm{with} \quad h(0)=0\,,\quad \dot{h}(0)=1\;
\ee
which can be solved via Laplace transform:\footnote{In practice, this is not necessarily easier than directly solving \eq{heom-main} numerically unless when $G_{\textrm{R}}(\tau)$ can be obtained analytically.}
\begin{equation}\label{hinverseLap}
h(\tau)=\mathcal{L}^{-1}\left[\hat{h}(s)\right] \qquad \qquad  \textrm{with}\qquad \hat{h}(s)=\frac{1}{s^2+\Omega^2+g^2 \hat{G}_{\textrm{R}}(s)}\;.
\end{equation}
Then $\{f_0,f_1,g_0,g_1\}$ are defined in terms of $h$ via:
\begin{equation}\label{fg-def}
f_0(\tau)=-\frac{\dot{h}(t)}{h(t)}h(\tau)+\dot{h}(\tau) \qquad \qquad f_1(\tau)=\frac{1}{h(t)}h(\tau)
\end{equation}
and
\be
g_0(\tau)=f_1(t-\tau), \qquad g_1(\tau)=f_0(t-\tau)\;.
\ee
Note that $\{f_{0,1},g_{0,1}\}$ are actually functions of both $\tau$ and $t$, with $t$ entering as the boundary point. For details of the derivation above see Appendix \ref{app-a}.

With the above explicit form of the propagating function,  we can now derive $\hat{\rho}_{\textrm{sys}}$'s master equation (the evolution equation of a density matrix) starting from \eq{rho-dy}. In the `ra' basis, the l.h.s. of \eq{rho-dy} is a function of $\{\bar{\Sigma},\bar{\Delta}\}$, and we compute its time-derivation. Since the derivation is rather long, we leave the details in Appendix \ref{mastereq} and here only write down the final result:
\bea\label{rhomaster}
i {\partial \rho_{\textrm{sys}}(\bar{\Sigma},\bar{\Delta}, t) \over \partial t}&=& [-{\partial^2 \over
\partial \bar{\Sigma} \partial \bar{\Delta}} +\Omega_{ren}^2(t) \bar{\Sigma} \bar{\Delta}]
\rho_{\textrm{sys}}(\bar{\Sigma},\bar{\Delta},t)
\nn \\  \nn &&+ [ i \gamma_1(t) \bar{\Delta}{\partial \over \partial \bar{\Delta}}+\gamma_2(t)
\bar{\Delta} {\partial \over \partial \bar{\Sigma}} + i \gamma_3(t) \bar{\Delta}^2 ]  \rho_{\textrm{sys}}
(\bar{\Sigma},\bar{\Delta},t) \\ \label{master}
&&+ i \gamma_4(t) {\partial^2 \over \partial \bar{\Sigma}^2}  \rho_{\textrm{sys}}(\bar{\Sigma},
\bar{\Delta},t)
\eea
(for detailed expression of $\Omega_{ren}$ and $\gamma_i$ please refer to Appendix  \ref{mastereq}).

Let us now explain the master equation \eq{rhomaster}. In the first line, $\Omega_{ten}(t)$ is the renormalized frequency, and this line is the renormalized Liouville term which preserves the unitarity. The terms in the second line are responsible for the quantum decoherence. $\gamma_i(t)$ with
$i=1,\dots,4$ are functions of time in generic situations; in the special case where all $\gamma$'s are time-independent, the evolution
dynamics is Markovian. Note that the last term is absent in \cite{Hu:1991di}, however it can be re-absorbed into the kinetic term of the Hamiltonian via a renormalization of the mass. The difference could be due to the different routes in the derivation of \eq{master}: We obtained \eq{master} by directly taking time derivative on \eq{Jrho-exp}, whereas in \cite{Hu:1991di} the counterpart of \eq{master} was derived via a perturbation analysis of the path integral expression of \eq{Jrho}.

\section{Characterization of  Quantum Decoherence}\label{sec4}

After deriving the explicit form of the master equation \eq{rhomaster}, we could now solve it to obtain the density matrix $\hat{\rho}_{\textrm{sys}}(t)$ for a given initial state. Alternatively, since we have the explicit form of the propagating function given by \eq{Jrho-exp}, we can also directly compute $\hat{\rho}_{\textrm{sys}}(t)$ via \eq{rho-dy}. Once we obtain $\hat{\rho}_{\textrm{sys}}(t)$, how can we use it to characterize a decoherence process?

Recall that the `environment' is not only responsible for destroying the coherence in the `system': first and foremost it selects from the Hilbert space of the `system' a subspace of states that are most stable against the `environmental' disturbance; these states are called pointer states, and become classical states at the classical limit $\hbar \rightarrow 0$ \cite{Zurek}.

In the basis of these pointer states, the quantum coherence  is encoded in the off-diagonal elements of the density matrix of the `system', and the Environment-Induced decoherence is then the process of the gradual decrease (to zero) of these off-diagonal elements.

However, for our present `system' of a scalar $\phi$, it is not feasible to directly look at the explicit elements of $\hat{\rho}_{\textrm{sys}}$, for the following two reasons. First, since $\phi$ takes continuous values, $\hat{\rho}_{\textrm{sys}}$ is an infinite-dimensional matrix. Second, ``decoherence is the disappearance of $\hat{\rho}_{\textrm{sys}}$'s off-diagonal elements" is a basis-dependent statement; however, the pointer states are not exactly stable under the `environmental' disturbance (since we do not take $\hbar \rightarrow 0$ limit) therefore the pointer basis of the density matrix is not invariant during the decoherence process.

To solve the first problem, we can coarse-grain the `system' --- i.e. to prepare its initial state to resemble a simple finite-dimensional system, even a qubit (the quintessence of Schr\"odinger's cat). In this section, we will first explain how to prepare (an approximation of) Schr\"odinger's cat with the scalar $\phi$ in our `system'.

The second problem can be solved in two ways. The first basis-independent information on the degree of `coherence' can be found in the Wigner function $W(\Sigma,p,t)$ (the Fourier transform w.r.t. the fast variable $\Delta=\phi_{+}-\phi_{-}$) of a density matrix $\rho(\Sigma, \Delta,t)$.  $W(x,p,t)$ is the quantum version of the distribution function $f(x, p, t)$ in the phase
space $\{x, p\}$; and unlike $f(x,p,t)$, $W(x,p,t)$ is not always positive-definite, but will become so once the system completely decoheres and becomes classical. Therefore, we can unambiguously use the disappearance of $W(r,p,t)$'s negative parts to characterize the decoherence.

 Another way to solve the basis-dependence problem is to look at the entanglement entropy
or the R\'enyi entropy since they are scalars therefore are independent of the basis choice. In this paper, we will choose the second order R\'enyi entropy (which is the logarithm of the `purity'). As we will see, the Wigner function and the second order R\'enyi entropy are both very effective in characterizing decoherence.

Finally, we will  show in this section that our setup to study quantum decoherence is almost parallel with the one for the local quantum quench \cite{quench1}. Thus, it is interesting to
establish the connection (if there exists any) between quantum decoherence and the local quantum quench.

\subsection{Preparation of initial state}

A prototype for the study of quantum decoherence is  Schr\"odinger's cat. The initial pure state of the cat is in a superposition of the two pointer states $|\textrm{Dead}\rangle$ and $|\textrm{Alive}\rangle$:\footnote{For the cat, the states $|\textrm{Dead}\rangle$ and $|\textrm{Alive}\rangle$ are stable against further `environmental' disturbance: once the cat becomes dead/alive, it remains dead/alive. In addition, $\langle\textrm{Dead}|\textrm{Alive}\rangle=0$ since we take $\hbar \rightarrow 0$ limit for the cat.} $|\textrm{Cat}\rangle =\frac{1}{\sqrt{2}}\left(|\textrm{Dead}\rangle +|\textrm{Alive}\rangle\right)$. After decoherence, the cat has a classical probability of being $50\%$ $|\textrm{Dead}\rangle$  and $50\%$$| \textrm{Alive}\rangle$ (after she comes into contact with the environment for long enough but before we open the window to peek). The initial and final states of the density matrix in the $|\textrm{Dead}\rangle /|\textrm{Alive}\rangle$ basis is:
\begin{equation}
\hat{\rho}_{\textrm{sys}}(0)=
\begin{pmatrix}
\frac{1}{2} & \frac{1}{2}\\
\frac{1}{2} & \frac{1}{2}
\end{pmatrix}
\qquad \qquad \longrightarrow \qquad \qquad
\hat{\rho}_{\textrm{sys}}(t_f)=
\begin{pmatrix}
\frac{1}{2} & 0\\
0 & \frac{1}{2}
\end{pmatrix}\;.
\end{equation}

The simplest lab realization of Schr\"odinger's cat is a qubit: $|\textrm{Dead}\rangle\equiv|\uparrow\rangle $ and $|\textrm{Alive}\rangle\equiv |\downarrow\rangle$.
However, our `system' is a canonical scalar $\phi$ defined by \eq{Lsys}, how do we prepare a $|\textrm{Dead Cat}\rangle$ and  an $|\textrm{Alive Cat}\rangle$ with $\phi$? First of all, we need to use pointer states from the Hilbert space of $\phi$, namely states that are least perturbed by the `environment' and therefore resemble classical states.
There are various definitions for pointer states; and for a generic `environment', locating the pointer states is a non-trivial problem \cite{zurek1993,Zurek}. Intuitively, the pointer states in the present case should be
Gaussian wave-packets in $\phi$ because they are  closest to the classical point since $\Delta \phi \Delta p_{\phi}=\frac{\hbar}{2}$ --- the minimal value for a quantum state.

Therefore we describe the two states $|\textrm{Dead}\rangle$ and $|\textrm{Alive}\rangle$ by  two Gaussian wave-packets wavefunctions of width $\sigma$ that are centered at $\phi=\pm \phi_0$:
\begin{equation}
|\textrm{Dead}\rangle \equiv  \Psi_1(\phi)={1\over \sqrt{\mathcal{N}/2}}
e^{-\frac{(\phi-\phi_0)^2}{2\sigma^2}}\;, \qquad \qquad |\textrm{Alive}\rangle \equiv  \Psi_2(\phi)={1\over \sqrt{\mathcal{N}/2}}
e^{-\frac
{(\phi+\phi_0)^2}{2\sigma^2}}\;.
\end{equation}
The initial state and the density matrix of the cat is then
\beq \label{TG01}
|\textrm{Cat} \rangle\equiv\Psi(\tilde{\phi}) =\frac{1}{\sqrt{2}}\left( \Psi_1(\tilde{\phi})+\Psi_2(\tilde{\phi})\right)\;, \qquad \qquad \hat{\rho}_{\textrm{sys}}(0)=|\Psi(\tilde{\phi})\rangle\langle \Psi(\tilde{\phi})| \;.
\eeq
Note that  in this representation  $\langle\textrm{Dead}|\textrm{Alive}\rangle=0$ only when $\phi_0 \gg \sigma$, therefore we define the normalization factor by $\Tr\hat{\rho}_{\textrm{sys}}=1$ and get $\mathcal{N}=  \sqrt{2\pi} \sigma (1+  e^{-{\phi_0^2 \over \sigma^2}})$.
The elements of $\hat{\rho}_{\textrm{sys}}(0)$ are shown in the left panel of
Fig. \ref{rho-Wigner}.
 The two `diagonal' peaks represent the two wave-packets centered at $\phi=\pm\phi_0$ (i.e. the high probability of the scalar $\phi$ being near $\pm \phi_{0}$)
:
\begin{equation}\label{certain}
\phi_{+}=\phi_{-}=\pm \phi_{0} \qquad \qquad \Longrightarrow \qquad \qquad (\Sigma,\Delta)=(\pm \phi_{0} ,0)
\end{equation} which corresponds to the $|\textrm{Dead}\rangle\langle \textrm{Dead}|$ and $|\textrm{Alive}\rangle\langle \textrm{Alive}|$ elements.  The  two `off-diagonal' peaks signify the interference of these two wave-packets:
\begin{equation}\label{uncertain}
\phi_{+}=-\phi_{-}=\pm \phi_{0} \qquad \qquad \Longrightarrow \qquad \qquad (\Sigma,\Delta)=(0  ,\pm 2 \phi_0)
\end{equation}
 which correspond to the $|\textrm{Dead}\rangle\langle \textrm{Alive}|$ and $|\textrm{Alive}\rangle\langle \textrm{Dead}|$ elements.

\subsection{Wigner function: quantum distribution function of phase-space}

One effective way to study quantum decoherence is to go to $\rho_{\textrm{sys}}$'s phase-space description furnished by the Wigner function. So far we have been working with the density matrix $\hat{\rho}_{\textrm{sys}}$ in the position space: the element $\rho(\bar{\Sigma}, \bar{\Delta},t)$ characterizes the overlap between two points (in position space) that are centered in $\bar{\Sigma}=\frac{\bar{\phi}_{+}+\bar{\phi}_{-}}{2}$ and separated by $\bar{\Delta}=\bar{\phi}_{+}-\bar{\phi}_{-}$. Intuitively, when $\rho(\bar{\Sigma}, \bar{\Delta}\neq0,t)\neq 0$, the nonzero $\bar{\Delta}$ signifies the uncertainty in position space, therefore should be conjugate to the momentum of the (average) position $\bar{\Sigma}$. This intuition is borne out by the Wigner function which is defined as the Fourier transform (w.r.t. the fast variable $\bar{\Delta}$) of $\rho_{\textrm{sys}}(\bar{\Sigma}, \bar{\Delta},t)$:
\be
W(\bar{\Sigma},p,t)\equiv {1\over 2\pi} \int_{-\infty}^{\infty} d\bar{\Delta} \; e^{-i p
\bar{\Delta}}\, \rho_{\textrm{sys}}
(\bar{\Sigma}, \bar{\Delta},t) \; .
\ee
The Wigner function $W(\bar{\Sigma},p,t)$ is the quantum version of the distribution function $f(\bar{\Sigma}, p, t)$ in the phase
space $\{\bar{\Sigma}, p\}$ and reduces to $f(\bar{\Sigma},p,t)$ in the classical limit. However, unlike its classical counterpart $f(\bar{\Sigma},p,t)$, $W(\bar{\Sigma},p,t)$ is not always positive-definite --- a consequence of the quantum uncertainty. Thus, in terms of the Wigner function, one signature of the
quantum-to-classical transition is the disappearance of its negative part. After this process is complete, the Wigner function becomes the probability distribution function of the classical states.

\begin{center}
\begin{figure}[bp]
\begin{tabular}{ll}
\begin{minipage}{70mm}
\begin{center}
\unitlength=1mm
\resizebox{!}{4.5cm}{
   \includegraphics{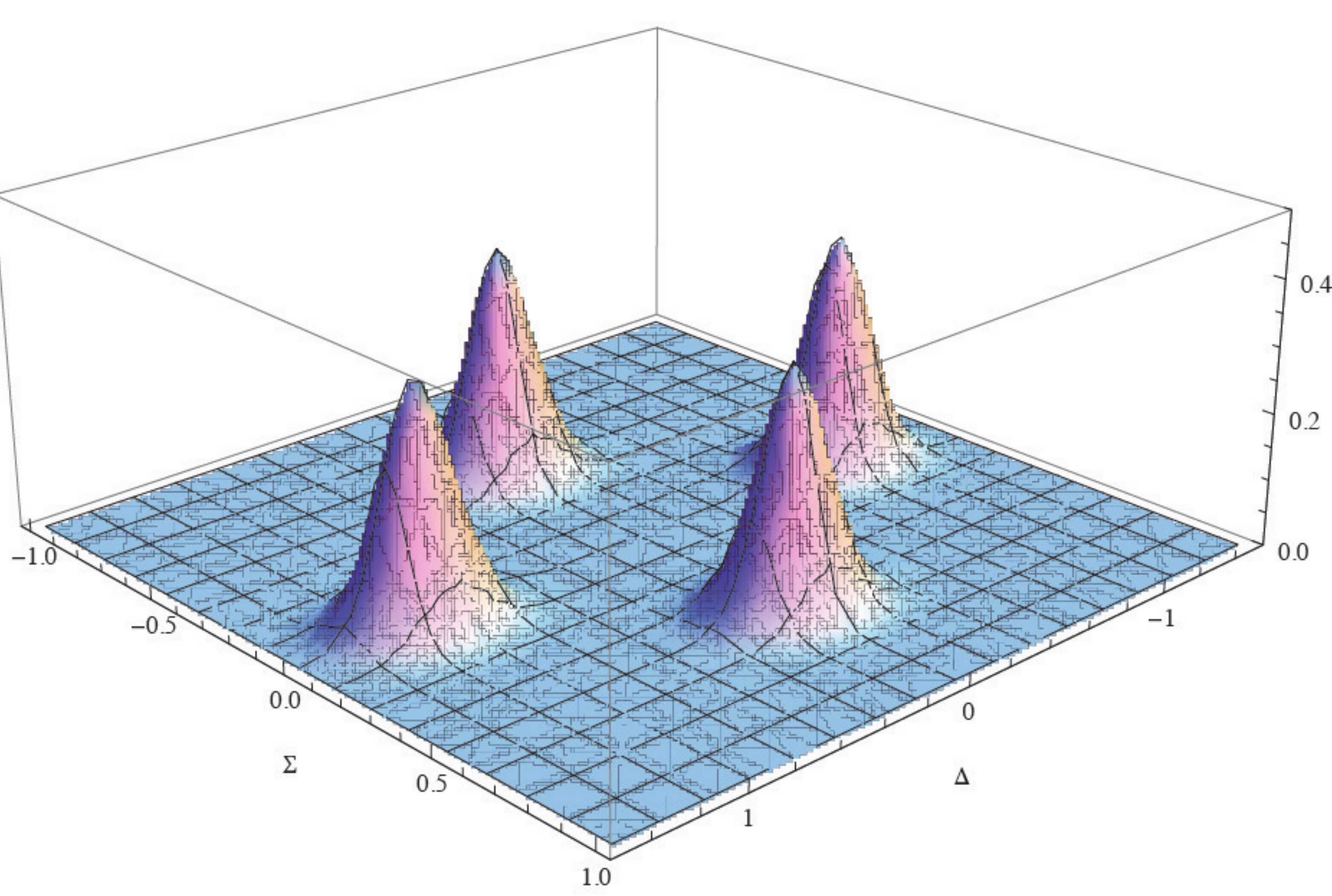}
                  }
\end{center}
\end{minipage}
&
\begin{minipage}{70mm}
\begin{center}
\unitlength=1mm
\resizebox{!}{4.5cm}{
   \includegraphics{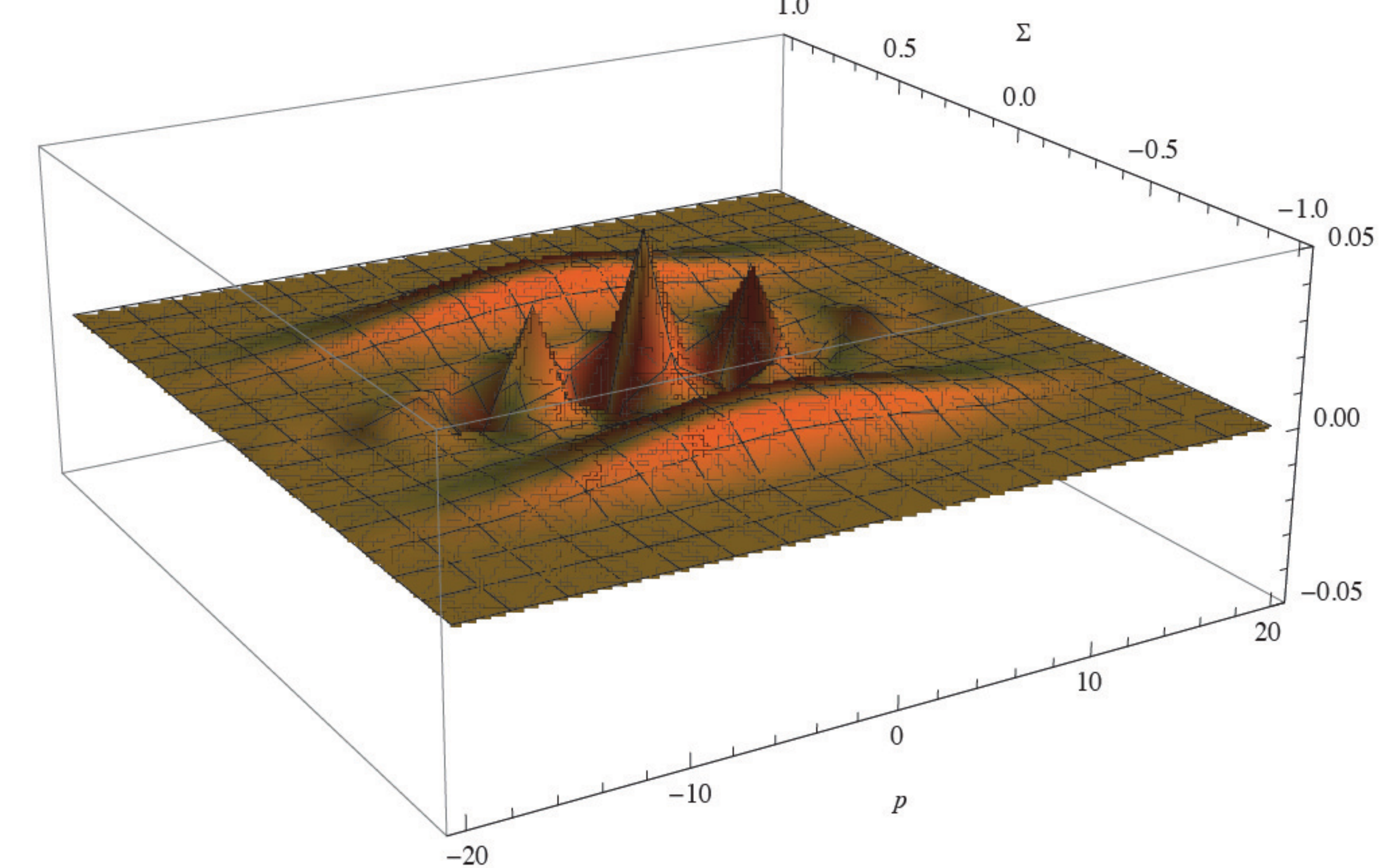}
                  }
\end{center}
\end{minipage}

\end{tabular}
\singlespace\caption{The initial state of a superposition of two Gaussian wave-packets (defined in Eq.\eq{TG01}) with
$\sigma=0.1$ and $\phi_0=0.5$. Left: Density matrix $\rho(\Sigma, \Delta,t=0)$; Right: Wigner function $W(\Sigma, p,t=0)$.}
\label{rho-Wigner}
\end{figure}
\end{center}

Let us now compute the Wigner function for the cat made of our scalar $\phi$ --- the superposition of two Gaussian wave-packets defined in \eq{TG01}. In the right panel of Fig. \ref{rho-Wigner} we plot the Wigner function of the cat at the initial time. This should be compared with $\hat{\rho}_{\textrm{sys}}(0)$ in the left panel. The two diagonal peaks of $\hat{\rho}_{\textrm{sys}}(0)$ defined in \eq{certain} represent the two wave-packets, therefore in the Wigner function they correspond to the two ridges that are centered in $\Sigma=\pm\phi_{0}$ and extend along $p$. The two off-diagonal peaks of $\hat{\rho}_{\textrm{sys}}(0)$ defined in \eq{uncertain} correspond to the interference of the two wave-packets, therefore in the Wigner function they give rises to the interference pattern along the $\Sigma=0$ line (note its negative parts).


The cat starts to interact with the `environment' and decohere after $t\geq 0$. To see this, we first
compute $\hat{\rho}_{\textrm{sys}}(t)$ starting with $\hat{\rho}_{\textrm{sys}}(0)$ given in \eq{TG01}, using the convolution \eq{rho-dy} and the propagating function \eq{Jrho-exp}. Then we Fourier transform to obtain its Wigner function.  The result has three terms:
\be	\label{TG-08}
W(\bar{\Sigma}, p ,t)  = W_1+W_2+W_c
\ee
where
\bes \label{TG-09}
\beq
\label{TG09-1}
W_1(\bar{\Sigma}, p ,t) && =  \frac{  \sigma L_{\Delta}}{4 \mathcal{N} |h(t)| \sqrt{\pi C}}
\left [ e^{\frac{\phi_0^2}{\sigma^2 C} \left( \frac{1}{4\sigma^2} +\frac{1}{2}a_{00}\right) -
\frac{L_{\Delta}^2}{4} \left(\frac{1}{L_c^2}\bar{\Sigma} + p - \frac{1}{l_1}\right)^2 - \frac{1}
{L_{\Sigma}^2}\bar{\Sigma}^2 - \frac{1}{l_2}\bar{\Sigma}}\right]e^{- {\phi_0^2 \over \sigma^2}}
\\
\label{TG09-2}
W_2 (\bar{\Sigma}, p ,t)&& =  \frac{  \sigma L_{\Delta}}{4 \mathcal{N} |h(t)| \sqrt{\pi C}}
\left [ e^{\frac{\phi_0^2}{\sigma^2 C} \left( \frac{1}{4\sigma^2} +\frac{1}{2}a_{00}\right) -
\frac{L_{\Delta}^2}{4} \left(\frac{1}{L_c^2}\bar{\Sigma} + p + \frac{1}{l_1}\right)^2 - \frac{1}
{L_{\Sigma}^2}\bar{\Sigma}^2 +\frac{1}{l_2}\bar{\Sigma}}\right]e^{- {\phi_0^2 \over \sigma^2}}
\eeq
\ees
correspond to the two individual Gaussian wave-packets and
\beq
\label{TG09-3}
W_c (\bar{\Sigma}, p ,t) &&  = \mathcal{A}_w(t)\, \cos \left[\left( \frac{L_{\Delta}^2}{2L_c^2 l_4}+\frac{1}{l_3}
\right) \bar{\Sigma}+\frac{L_{\Delta}^2}{2l_4 }p \right]  \;   e^{ (-\frac{L_{\Delta}^2}{4L_c^4} -
\frac{1}{L_{\Sigma}^2})\bar{\Sigma}^2 -\frac{L_{\Delta}^2}{2L_c^2} \bar{\Sigma} p -
\frac{L_{\Delta}^2}{4} p^2} \eeq
with
\be\label{Nw}
\mathcal{A}_w(t) \equiv \frac{  \sigma L_{\Delta}}{4 \mathcal{N} |h(t)| \sqrt{\pi C}}\;
e^{\frac{\phi_0^2}{4C \sigma^4} +\frac{L_{\Delta}^2}{4 l_4^2} - \frac{\phi_0^2}{\sigma^2}}\;
\ee
gives the interference between the two wave-packets. The parameters in \eq{TG-09} and \eq{TG09-3}  are defined as follows:
\bes \label{TG07}
\be
\label{TG07-4}
C = \frac{\sigma^2}{4} \dot{f}_0(0)^2+\frac{1}{4\sigma^2}+\frac{1}{2} a_{00}(t)\;,
\ee\beq
\label{TG07-1}
L_{\Sigma}^{-2} && = \frac{1}{4C} \dot{f}_1(0)^2\;, \\
\label{TG07-2}
L_c^{-2} && =  \frac{1}{4C} \left(  \sigma^2 \dot{f}_0(t)\dot{f}_0(0) \dot{f}_1(0) -
\dot{f}_1(0) (a_{10}+a_{01}) -  4C \dot{f}_1(t) \right)\;, \\
\label{TG07-3}
L_{\Delta}^{-2} && =  \frac{1}{4C} \left(\sigma^2 C \dot{f}_0(t)^2 - \frac{\sigma^4}{4}
\dot{f}_0(t)^2 \dot{f}_0(0)^2 -(\frac{a_{10}+a_{01}}{2})^2+\frac{ \sigma^2 \dot{f}_0(t)
\dot{f}_0(0) (a_{10}+a_{01})}{2}  \right)   \nonumber \\
&&+\frac{1}{2} a_{11}\;,
\eeq
and
\beq \label{TG07-5}
l_1^{-1} && =\frac{\phi_0}{2C} \left( \dot{f}_0(0) \frac{a_{10}+a_{01}}{2} + \dot{f}_0(t)
\left( {1\over 2\sigma^2} +a_{00} \right)  \right)\;, \\
\label{TG07-6}
l_2^{-1} && =\frac{\phi_0}{2C}\dot{f}_0(0) \dot{f}_1(0)\;, \\
\label{TG07-7}
l_3^{-1} && = \frac{\phi_0}{2C \sigma^2} \dot{f}_1(0)\;, \\
\label{TG07-8}
l_4^{-1} && = -\frac{\phi_0}{4C \sigma^2} \left( a_{10} +a_{01} - \sigma^2 \dot{f}_0(t)
\dot{f}_0(0)  \right)\;.
\eeq
\ees
Let us focus on the $W_c$ term which characterizes the interference between the two wave-packets. It has three factors: an enveloping amplitude $\mathcal{A}_w(t)$ that deceases over time to zero, a cosine factor that produces the oscillation both in $\Sigma$ and in $p$ and therefore is responsible for the non-positive-definiteness of the Wigner function, and finally an exponential factor that encode the shapes and positions of the two wave-packets.
To study the decoherence process, we only need to watch how the amplitude $\mathcal{A}_w(t)$ evolves with time.

\subsection{Purity/R\'enyi entropy}

    One way to characterize the purity of a quantum state is to examine its entanglement
entropy. The entanglement entropy is zero for pure states and reaches its maximum for the
completely mixed state.  As quantum decoherence is a process from a pure state to a mixed
state, we may use the entanglement entropy (or the R\'enyi entropy) to characterize it.  Then, one may wonder how the 
entanglement entropy/R\'enyi entropy behaves when the decoherence happens.

     One hint to answer this question is the analogue with the thermalization, for which the
thermal entropy saturates when reaching the thermal equilibrium. In a thermal environment, quantum decoherence could be closely related to the thermalization, and part of the entanglement
entropy could be related to the thermal one. Thus, the entanglement entropy might saturate when
the pure state decoheres in a similar way to the saturation of the thermal entropy when reaching
the thermal equilibrium.

    There are many possible quantities qualified to be the entanglement entropy, which needs to be positive definite
and concave. They can all be derived from the reduced density matrix. The most commonly used one
is the von Neumann entropy: $S_{vN}=-\Tr \; \hat{\rho}_{\textrm{sys}} \log \hat{\rho}_{\textrm{sys}}$. The other
ones are called the R\'enyi entropy of order $\alpha$:
\beq \label{TG-13}
S_{\alpha} = \frac{1}{1-\alpha}  \mbox{log} \mbox{ Tr}\; \hat{\rho}_{\textrm{sys}}^{\alpha}
\eeq
where $\alpha$ is a positive real number. It can be shown that $S_{vN}=\lim_{\alpha
\rightarrow 1} S_{\alpha}$.

   For simplicity, we will only calculate the R\'enyi entropy of second order, which is related to
$\mathcal{P}:=\Tr \hat{\rho}^2_{\textrm{sys}}$, named as ``purity" in the community of quantum information sciences. To compare with the results from the phase-space view point discussed in
the previous subsection, here we also consider the two Gaussian wave-packets for the initial
state, i.e.,  \eq{TG01}. Using the explicit form \eq{Jrho-exp} of $\rho_{\textrm{sys}}(t)$, after lengthy calculations we arrive at
\beq \label{TG-12}
 \mathcal{P} && = {A(t)^2 \pi^3\sigma^2 L_{\Delta}L_{\Sigma} \over C \mathcal{N}^2}
e^{\frac{\phi_0^2(1-4C\sigma^2)}{2C\sigma^4}}  \left(e^{-\frac{L_{\Sigma}^2}{2\ell_3^2}}
+e^{\frac{L_{\Delta}^2}{2\ell_4^2}}+e^{-\frac{L_{\Delta}^2}{2\ell_1^2}+\frac{\phi_0^2 a_{00}}{C
\sigma^2}}+e^{\frac{L_{\Sigma}^2}{2\ell_2^2}+\frac{\phi_0^2 a_{00}}{C\sigma^2}} \right)
\nonumber \\
 && +  8 {A(t)^2 \pi^3\sigma^2 L_{\Delta}L_{\Sigma}\over C \mathcal{N}^2} e^{\frac{1}{8}
\left( -\frac{L_{\Delta}^2}{l_1^2}+\frac{L_{\Delta}^2}{l_4^2}+\frac{L_{\Sigma}^2}{l_2^2}+
\frac{L_{\Sigma}^2}{l_3^2}+ \frac{4 \phi_0^2 (1+a_{00}-4C)}{C\sigma^2}    \right)}  \mbox{cos}
[\frac{1}{4} \left( \frac{L_{\Delta}^2}{l_1 l_4} - \frac{L_{\Sigma}^2}{l_2 l_3} \right) ]
\eeq
where the parameters used here are already defined in \eq{TG07}.

\subsection{Relation to local quantum quench}

    The entanglement entropy has been used to characterize the quantum quench process
\cite{quench,quench2}.  The setup for the quantum quench is to bring the total system to a
highly excited state and let it evolve. There are different ways to create such a setup. The
simplest one is to tune a parameter of the Hamiltonian homogeneously such that the original ground
state turns into non-eigenstate of the new Hamiltonian suddenly.  This is the so-called global
quantum quench.

\begin{figure}[h]
\begin{center}
\includegraphics[scale=0.35]{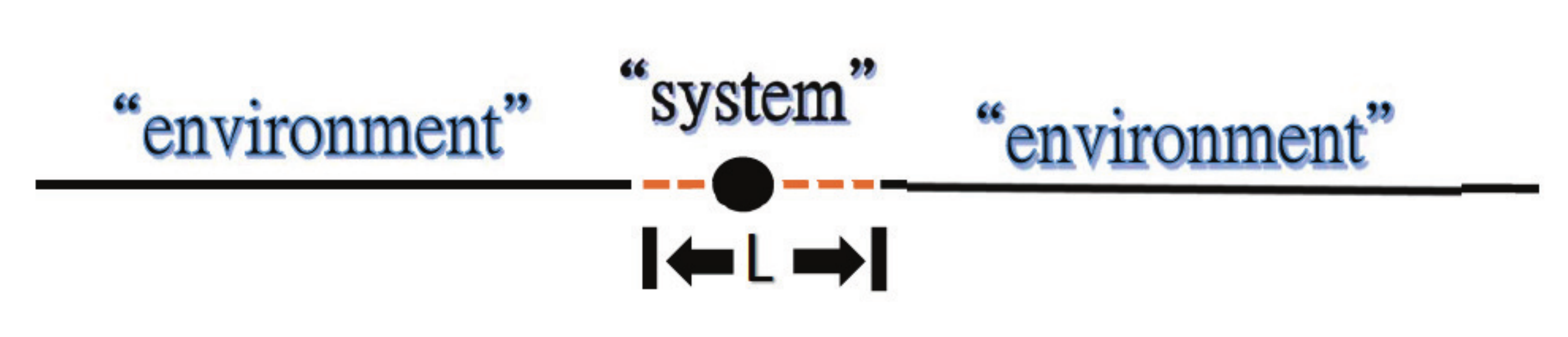}
\end{center}
\singlespace\caption{Our setup for quantum decoherence can be also thought as the one for local quantum
quench in $(1+1)$ dimensions.}
\label{quench}
\end{figure}

    The other is the local quantum quench: Excite the total system locally by either
tuning a parameter of the Hamiltonian inhomogeneously or simply creating local excitation such
as quasi-particles or qubits. This setup is the same as ours for quantum decoherence if we treat
the local excitations (unentangled with the `environment' when created) as the `system' and
the rest of the unperturbed region as the `environment', see Fig. \ref{quench}.  More
specifically, the initial state of two Gaussian wave-packets given in \eq{TG01} can be thought
as the local excitations of some effective linear size $L$ which should be determined by the
profile of the wave-packets, i.e., by $\phi_0$ and $\sigma$.  We will try to determine $L$ later.

     One way to characterize the quantum quench is to study the time evolution of the
entanglement entropy for a region of size $L$ enclosing the initial local excitations or the
`system'. By the causal motions of the entangled quasi-particle one expects a cross-over to
a saturated or mildly growing regime for the growth of the entanglement entropy. The detailed
time evolution behaviors of the (holographic) entanglement entropy for the local quantum quench
have been studied extensively \cite{quench1,quench2,holo-quench,Nozaki:2013vta,quench}: the
initial growing rate is quadratic in $t$, and then turns to the so-called linear $t$ ``Tsunami"
regime \cite{quench} before dropping suddenly to the mildly growing regime.   For the case of
quantum quench in $(1+1)$-dimensional CFT, the entanglement entropy for the a region of
size $L$ \cite{quench1,quench},
\be\label{eeL}
S_L(t)  \sim  \begin{cases} t^2, &  t \approx 0\;; \\
 t, &  t < L\;; \\
 \mbox{const. or}\quad \ln t, & t > L\;.
\end{cases}
\ee
For $t>L$, $S_L$ either saturates (for global quantum quench) or mildly grows as $\ln t$ (for
local quantum quench). Thus, $t_Q \approx L$ here can be regarded as the relaxation time scale (for the probe to relax to its classical state).

    As we can calculate the reduced density matrix for the `system', from it we can then
evaluate the entanglement entropy for our setup and compare the relaxation time scale $t_Q$ with the
decoherence time scale $t_D$. For simplicity, we will consider the R\'enyi entropy of 2nd order,
i.e., $S_2 = -\log \mathcal{P}$ see \eq{TG-13} to extract $t_Q$ for our decoherence setup. If
these two time scales are comparable in order of magnitude, then the quantum decoherence is
closely related to the quantum quench. Though this could be naively expected, it is still worthwhile
to check this connection directly, especially since there are other environmental influences such as
temperature which are usually turned off in the consideration of quantum quench. We will find
that indeed these two time scales are of the same order. Moreover,  our approach yields
similar behaviors to \eq{eeL}.

\section{Holographic Quantum Decoherence: Case Studies}\label{sec5}

To recapitulate, our `system' is a scalar $\phi$ whose initial state is a superposition of two Gaussian wave-packets (simulating Schr\"odinger's cat); the `environment' is a thermal CFT with a gravity dual. The decoherence process is captured by the time-evolution of the enveloping function $\mathcal{A}_w(t)$ of the Wigner function's interference term and  the second order R\'enyi entropy $S_2(t)$. The effect from the environment is encoded in the influence functional which is given by `environment''s Green's function. On the one hand, the `environment' is non-trivial (in fact strongly interacting), on the other hand, having a gravity dual allows its Green's function to be computed holographically. Once we obtain the Green's function, we can compute $\mathcal{A}_w(t)$ via \eq{Nw} and the second order R\'enyi entropy $S_2(t)$ via \eq{TG-12}, and then use them to quantify the decoherence process.

We will use two examples to illustrate the above. Since the problem is technically rather demanding, we present here only simple cases for which analytical computation can be carried out as much as possible (in fact up to the point of computing $h(\tau)$) and the rest is done numerically.
Then, to  study generic holographic `environment', we also composed a python program \cite{python} which can compute numerically $h(\tau)$ and then $\mathcal{A}_w(t)$ and  $S_2(t)$ for any holographic `environment' once its Green's functions are given.

\subsection{Two Cases}

The two cases we choose are: (1) a static particle coupled to the `environment' of a (1+1)-dimensional holographic CFT at finite temperature; (2) a scalar zero mode coupled to a (3+1)-dimensional holographic CFT at zero temperature. Their retarded Green's functions were computed holographically and have closed forms, which allow the function $h(t)$ to be obtained analytically as well.
\subsubsection{Straight string in BTZ black hole}

Let us first look at a static particle in a $(1+1)$-dimensional CFT at finite temperature. The
finite temperature CFT is holographically dual to the ($(2+1)$-dimensional) BTZ black
hole, and the point particle is dual to a straight string connecting the black hole horizon and the
AdS$_3$ boundary.\footnote{This setup was also used in \cite{deBoer:2008gu,Son:2009vu} to  study the
holographic quantum Brownian motion (which has been briefly reviewed in Sec. \ref{brownian}). The horizon-end of the string receives random kicks from the Hawking radiation near the horizon, and these random kicks then propagate along the string and cause the Brownian motion of the boundary-end of the string.}
Now we summarize the procedure to compute the  retarded Green's function holographically.

The metric of the BTZ black hole is
\be
ds^2 = -\frac{r^2 - r^2_H}{\ell^2} dt^2 + \frac{r^2}{\ell^2} dX^2 + \frac{\ell^2}{r^2 - r^2_H}
dr^2
\ee
with $\ell$ the AdS radius and $r_H$ the radius of the horizon.  Its Hawking
temperature is
\be
T = \frac{r_H}{2 \pi \ell^2}\,.
\ee
Let us consider a static string configuration connecting the black hole horizon and the AdS$_3$
boundary. In the static gauge this configuration is given by $X(t,r)=0$. Now consider the
perturbation of $X$ along this static string, the equation of motion of this perturbation $
\delta X(t,r)$ can be obtained from the Nambu-Goto action of the probe string
\cite{deBoer:2008gu}:
\be
\left[ -\partial_t^2 + \frac{r^2 - r_H^2}{r^2 \ell^4} \partial_r
\left( r^2 (r^2 - r_H^2) \partial_r \right)  \right] \delta X(t,r) = 0\;.
\ee
The solution with the in-falling boundary condition at the black hole horizon is
\be
\delta X(t,r) = e^{-i\,\omega t} \,  X_{\omega}(r) \,,\qquad \textrm{with}\qquad
X_{\omega}(r) \propto \frac{r - i\, \omega \ell^2}{r} \left( \frac{r - r_H}{r + r_H} \right)^{-
\frac{i\, \omega \ell^2}{2\, r_H}}\,.
\ee
The conjugate momentum of $X_{\omega}(r)$ is
\be
\Pi_{\omega}(r) = - T_s \frac{r^2 (r^2 - r_H^2)}{\ell^4} \,\partial_r   X_{\omega}(r)
\ee
where $T_s={1\over \ell_s^2}$ is the string tension. Then using the prescription given by \eq{holoRetard}, we obtain the retarded Green's function:
\bea \label{Green-straight}
G_{\textrm{R}}(\omega,r_c) &=& - \lim_{r \to r_c} \frac{\Pi_{\omega}(r)}{X_{\omega}(r)}
= - T_s \frac{\omega^2 r_c^2 \ell^2 + i\, \omega\, r_c r_H^2}{r_c \ell^2 - i\, \omega \ell^4} \nn\\
\label{GR-string}
&\approx& T_s  r_c \left(-1+\frac{r_H^2}{r_c^2} \right) \omega^2 -i T_s
\left( \ell^2 \omega^3 +\frac{r_H^2}{\ell^2} \omega \right)
\eea
with $r_c$ ($\approx \infty$) the UV-cutoff.

    The real part of \eq{GR-string} is UV-divergent but can be considered as the renormalization of
the probe's mass, and only the imaginary part of \eq{GR-string} drives the quantum decoherence.
However, its form has
unreasonable high frequency behaviors and we need to regularize by introducing some window function, which we choose to be the commonly used
Lorentzian function (with
width $\Gamma_w$).\footnote{These unreasonable UV behaviors are
termed ohmic ($\sim\omega$) or super-ohmic ($\sim\omega^{\delta}$ with $\delta>1$) in the literature, see e.g.\cite{Hu:1991di} which also used the window function to regularize the UV-behavior.}  With the real part dropped and after regularization, the retarded Green's function \eq{GR-string} becomes:
\be \label{str-GR-cutoff}
G_{\textrm{R}}(\omega) = - i N_{st}^2 \left( \omega^3 + 4\pi^2 T^2 \omega \right) \frac{\Gamma_w^2}
{\Gamma_w^2+ \omega^2}
\ee
where $N_{st}\equiv\ell \sqrt{T_s}=\ell/\ell_s$ is the number of degrees of freedom of the dual CFT. Then the symmetric Green's function $G_{\textrm{sym}}$ can be obtained from \eq{str-GR-cutoff} via \eq{KMS-w}.

    With the closed form of $G_{\textrm{R}}$ given by \eq{str-GR-cutoff} we can solve for $h(\tau)$ via the inverse Laplace transform \eq{hinverseLap}. First, the Laplace transform of $G_{\textrm{R}}(\tau)$ is
\be \label{str-GR-lap}
\hat{G}_{\textrm{R}}(s)= \pi N_{st}^2 \Gamma_w^2 \left( 4 \pi^2 T^2 - \Gamma_w^2   \right) \frac{1}{\Gamma_w +s}
\;.
\ee
We then compute $h(\tau)$ using  \eq{hinverseLap}. (Recall that once $h(\tau)$ is known, all other functions can be immediately expressed in terms of simple functions of $h(\tau)$ and its derivatives.)
Let us consider the case given for which $\hat{h}(s)$ has two complex poles at $s=s_1\pm i s_2$
and one real pole at $s=s_3$, where $s_i$'s are real functions of $T$ and $\Gamma$. However,
their detailed expressions are not very illuminating and will be omitted.  Then the inverse Laplace transform of $\hat{h}(s)$ gives $h(\tau)$:
\be \label{str-ht}
h(\tau) = \frac{s_2 (s_3 +\Gamma_w)(e^{s_3 \tau} - e^{s_1 \tau}   \cos s_2 \tau) +
(s_1^2+s_2^2-s_3 \Gamma_w +s_1 (\Gamma_w -s_3)) e^{s_1 \tau} \sin s_2 \tau }{s_2 (s_1^2+s_2^2 -2
s_1 s_3 + s_3^2)}\;.
\ee
Knowing $h(\tau)$, we can then first compute $f_i$'s  and $g_i$'s by
\eq{fg-def}, and then use them to calculate the Wigner function and the second order R\'enyi entropy.

\subsubsection{Scalar probe in AdS$_5$ spacetime}

   Most of the earlier studies of quantum decoherence assumed that the `environment' is a
thermal reservoir. This might cause the misconception that
the thermal fluctuation of the `environment' is necessary for the quantum decoherence of the `system'. However, as we are about to see, decoherence can happen even in zero temperature.

To show this, we choose the `environment' to be a $(3+1)$-dimensional CFT at zero temperature and the `system' a scalar operator. The total setup is holographically dual to a scalar in AdS$_5$ spacetime.
Following the procedure described in Sec.
\ref{holographic GR}, one can compute holographically the retarded Green's function $G_{\textrm{R}}$ of an
operator $\mathcal{O}$ with conformal dimension $\Delta_{\mathcal{O}}$. This was done in \cite{Son:2002sd}, and we summarize the results here:
\begin{equation}\label{S01}
G_{\textrm{R}}(\omega)=\begin{cases}
\frac{N_{sc}^2 \Gamma(3-\Delta_{\mathcal{O}}) \epsilon^{2(\Delta_{\mathcal{O}}-4)}}
{8 \pi^2 \Gamma(\Delta_{\mathcal{O}}-2) 2^{2\Delta_{\mathcal{O}}-5}} \;
(q^2)^{\Delta_{\mathcal{O}}-2}\; [\; \cos \pi \Delta_{\mathcal{O}} - i\;  \textrm{sgn}(\omega)
\sin \pi\Delta_{\mathcal{O}}\; ]& 2<\Delta_{\mathcal{O}}\notin \mathbb{N}\\
\frac{N_{sc}^2 \epsilon^{2(\Delta_{\mathcal{O}}-4)}}{8\pi^2
( \Delta_{\mathcal{O}}-3)!^2 2^{2\Delta_{\mathcal{O}}-5}}\; (q^2)^{\Delta_{\mathcal{O}}-2}\; [\;
\ln q^2 - i\;\pi\; \textrm{sgn}(\omega)\;]& 2\leq\Delta_{\mathcal{O}}\in \mathbb{N}
\end{cases}
\end{equation}
where $N_{sc}^2$ is the number of degrees of freedom of the dual CFT, and $\epsilon \approx 0$
is the UV cutoff.

For simplicity, we only consider the zero-mode (with
$q \equiv \sqrt{\omega^2-\vec{k}^2}=\omega$) of $\phi$, namely we assume the `system' is homogenous in space.  The problem is then effectively $(1+1)$-dimensional, which can be compared directly with the straight string case.
In contrast to the static string case, the non-zero and non-renormalizable real part of the retarded Green's function is absent here.

    In order to solve $h(\tau)$ analytically by the Laplace transform \eq{hinverseLap}, we
only consider the integer and half-integer $\Delta_{\mathcal{O}}$ cases. Moreover, as $G_{\textrm{R}}$ contains the `super-ohmic' factor $(q^2)^{\Delta_{\mathcal{O}}-2}$ we
will again regularize by introducing the Lorentzian window function of width $\Gamma_w$.
The Laplace transform of $G_{\textrm{R}}(\tau)$ is the following:
\begin{equation}\label{S04}
\hat{G}_{\textrm{R}}(s)=\begin{cases}
- \frac{N_{sc}^2 \Gamma(3-\Delta_{\mathcal{O}}) \epsilon^{2(\Delta_{\mathcal{O}}-4)}}
{8 \pi \Gamma(\Delta_{\mathcal{O}}-2) 2^{2\Delta_{\mathcal{O}}-5}} \;  \Gamma_w^{2n+2}
\frac{1}{s+\Gamma_w}& 2<\Delta_{\mathcal{O}}\in \mathbb{N}+\frac{1}{2}\\
\frac{N_{sc}^2 \epsilon^{2(\Delta_{\mathcal{O}}-4)}}
{8\pi^2 ( \Delta_{\mathcal{O}}-3)!^2 2^{2\Delta_{\mathcal{O}}-5}} (-1)^{\Delta_{\mathcal{O}}}
\Gamma_w^{2\Delta_{\mathcal{O}}-3} \left( \pi \ln\Gamma_w^2 \ \frac{1}{s+\Gamma_w} + \frac{1}{s-
\Gamma_w} - \frac{1}{s+\Gamma_w} \right)& 2\leq\Delta_{\mathcal{O}}\in \mathbb{N}
\end{cases}\;.
\end{equation}

Then $h(t)$ can be computed by inverse Laplace transform \eq{hinverseLap}. For large $N_{sc}$ and $\Gamma_w$,  $\hat{h}(s)$ has two conjugated poles at $s=s_1\pm is_2$ and two real poles at $s=s_3,s_4$, therefore $h(t)$ takes the following form:
\beq \label{S06}
h(t) & =& \big[ s_2(s_2^2+(s_1-s_3)^2)(s_2^2+(s_1-s_4)^2)(s_3-s_4) \big]^{-1}  \nonumber\\
&\cdot& \big[  s_2 (s_3^2-\Gamma_w^2) (s_2^2 +(s_1-s_4)^2) e^{s_3 t}  + s_2 (s_2^2+(s_1-s_3)^2)
(\Gamma_w^2-s_4^2) e^{s_4 t} \nonumber \\
&+&  \big(s_1^4-s_1 (s_3+s_4)(s_1^2-s_2^2+\Gamma_w^2)+(s_2^2+\Gamma_w^2)(s_2^2-s_3s_4)
+s_1^2(2s_2^2-\Gamma_w^2+s_3 s_4)\big)  (s_3 -s_4) e^{s_1 t}\sin s_2 t \nonumber \\
&-&  \big(s_1^2(s_3+s_4)+(s_2^2+\Gamma_w^2)(s_3+s_4)-2s_1(\Gamma_w^2+s_3s_4)\big) s_2(s_3-s_4)
e^{s_1 t}   \cos s_2 t\;  \big]\;.
\eeq
Again, we omit the detailed expressions for $s_i$'s. In the numerical calculation, we
set $N_{sc}=10$ and the UV-cutoff $\epsilon \approx 1/\Gamma_w$.

\bigskip
\subsection{Results and remarks}
\begin{center}
\begin{figure}[tbp]
\begin{tabular}{ll}
\begin{minipage}{70mm}
\begin{center}
\unitlength=1mm
\resizebox{!}{3cm}{
   \includegraphics{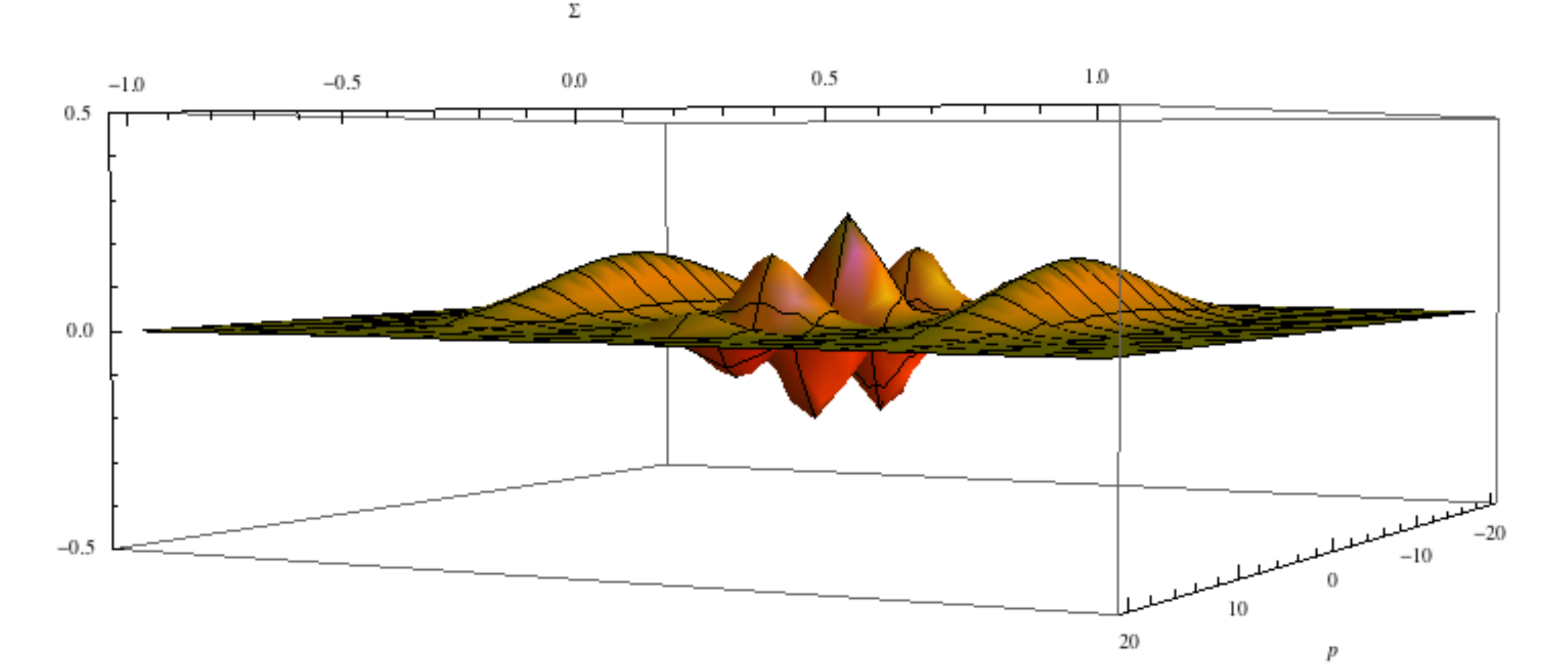}
                  }
\end{center}
\end{minipage}
&
\begin{minipage}{70mm}
\begin{center}
\unitlength=1mm
\resizebox{!}{3cm}{
   \includegraphics{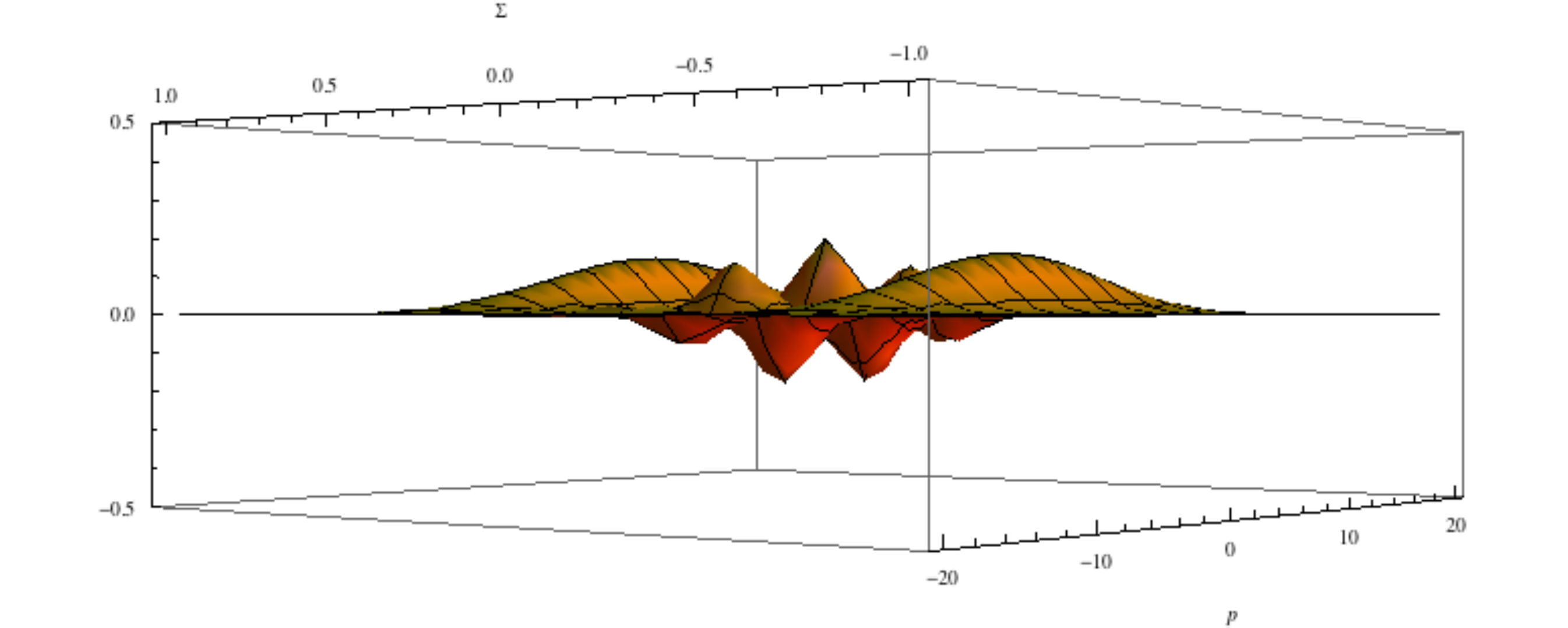}
                  }
\end{center}
\end{minipage} \\

\begin{minipage}{70mm}
\begin{center}
\unitlength=1mm
\resizebox{!}{3cm}{
   \includegraphics{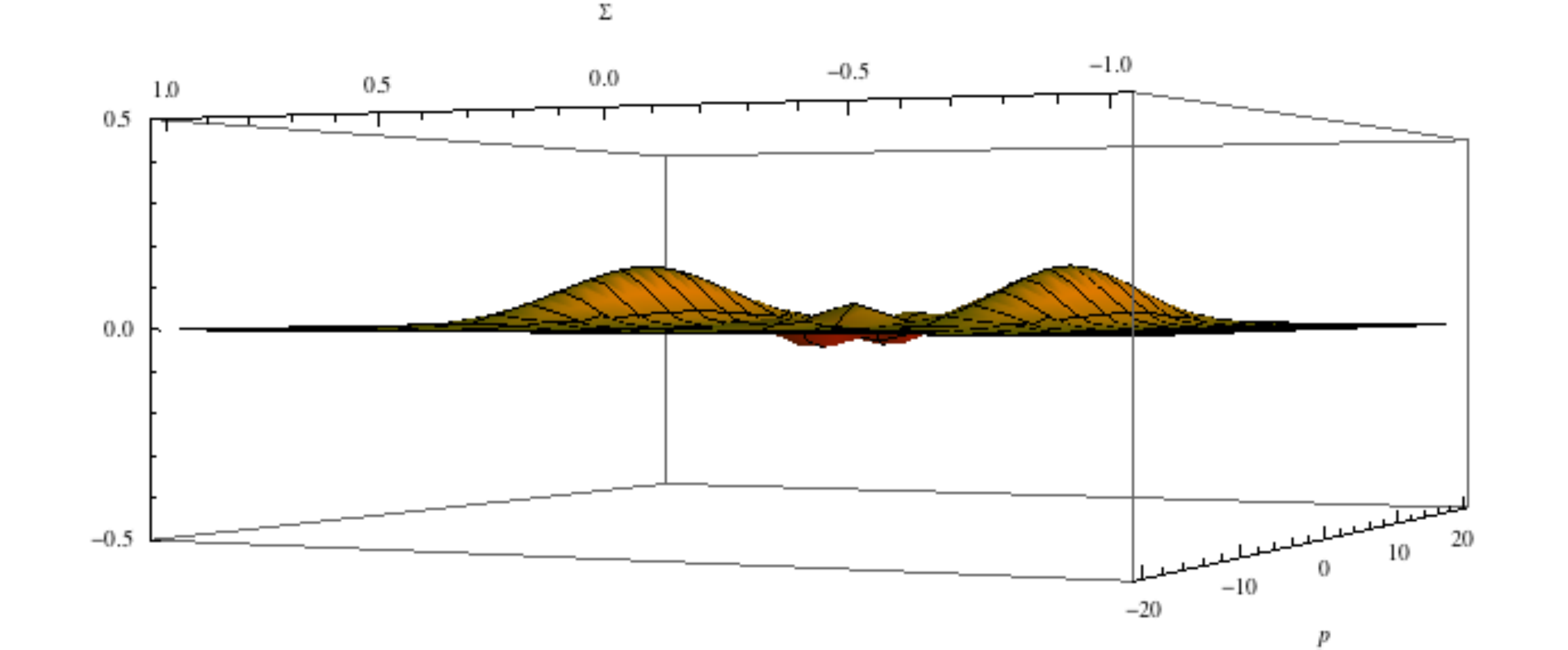}
                  }
\end{center}
\end{minipage}
&
\begin{minipage}{70mm}
\begin{center}
\unitlength=1mm
\resizebox{!}{3cm}{
   \includegraphics{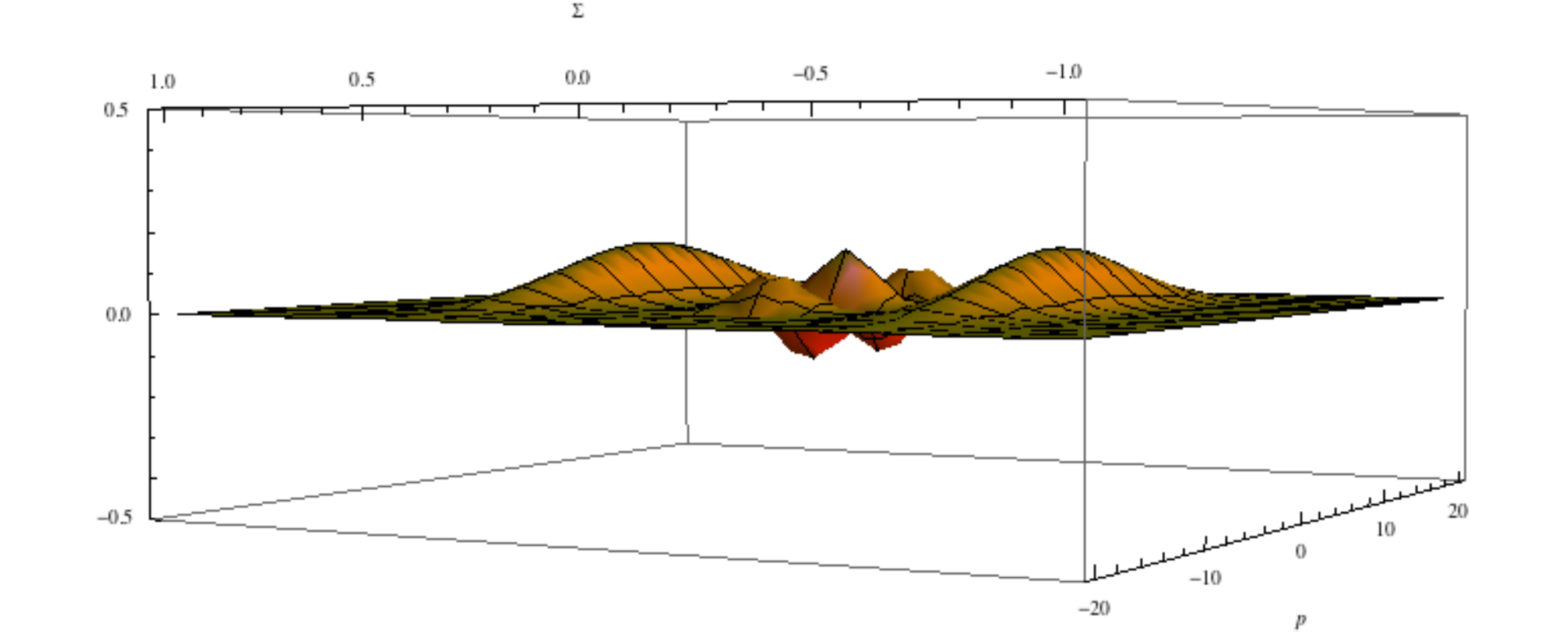}
                  }
\end{center}
\end{minipage}

\end{tabular}
\singlespace\caption{The evolution of the Wigner function $W(\Sigma,p,t)$ for the two Gaussian wave-packets made of the scalar $\phi$ with
$\Delta_{\mathcal{O}}=3$. The evolution goes clockwise as
$t=10$, $t=15$, $t=20$ and $t=30$ units respectively.}
\label{Wigner history}
\end{figure}
\end{center}

   Based on the the expressions of $h(t)$ in the previous subsection, we can now calculate the enveloping function $\mathcal{A}_w(t)$ via \eq{Nw} and the second order R\'enyi entropy $S_2(t)=-\log \mathcal{P}$ via \eq{TG-12}, and use them to characterize the decoherence of the superposition of two Gaussian wave-packets.

   The task requires long numerical calculations. Before we present the detailed results below, let us first show (in Fig. \ref{Wigner history}) four snap shots in the time-evolution of the Wigner function. Note the decay of its negative part (which characterizes the quantumness of the Wigner function).
In the scenario of the environment-induced decoherence, the coherence decay exponentially.
Therefore, we first need to define precisely what we mean by decoherence. Since the decay is exponential, only the order of magnitude of $t_D$ is relevant. We will declare that the `system' has almost decohered once the enveloping function $\mathcal{A}_w(t)$ decreases to the $1\%$ of its initial value and define the decoherence time $t_{\textrm{D}}$ by: \footnote{The decoherence time $t_D$ is roughly of the same order as the half-life scale of  $\mathcal{A}_w(t)$ since $\mathcal{A}_w(t)$ is exponentially decaying. Here defining the `almost decoherence' by (\ref{decotime}) is simply for convenience. }
\begin{equation} \label{decotime}
\textrm{decoherence time } t_{\textrm{D}}: \qquad \frac{\mathcal{A}_w(t_{\textrm{D}})}{\mathcal{A}_w(0) }= 0.01\;.
\end{equation}
Note that the time-dependence of the R\'enyi entropy $S_2$ exhibits a crossover around $t\approx t_D$.

From our computation we extract two important results. First, for the decoherence time $t_{\textrm{D}}$, we obtain its dependence on the temperature $T$ of the `environment' and on the conformal dimension $\Delta_{\mathcal{O}}$ of the field $\mathcal{O}$ coupled to the `system'. Second, we obtain the growth rate of $S_2(t)$, using which we can define different stages of decoherence. We then show that these stages are nicely matched with those in the holographic quantum quench discussed recently in
\cite{quench,holo-quench,Nozaki:2013vta}.

     Note that in all the numerical plots presented below, we set the width of the
window function $\Gamma_w=10$, the coupling $g=1$ and the constants $N_{st}=1$ and
$N_{sc}=10$.\footnote{Note that once we have taken the `weak coupling limit' by keeping only the quadratic terms in $\ln \mathcal{F}$ (and reaching (\ref{FV-G1})), we are free to absorb the coupling constant $g$ into the scalar fields, i.e. we can simply set $g=1$ for convenience.}
The width $\Gamma_w$ also serves as a reference inverse time scale.

\bigskip
\subsubsection{String probe}

\begin{center}
\begin{figure}[tbp]
\begin{tabular}{ll}
\begin{minipage}{70mm}
\begin{center}
\unitlength=1mm
\resizebox{!}{4.2cm}{
   \includegraphics{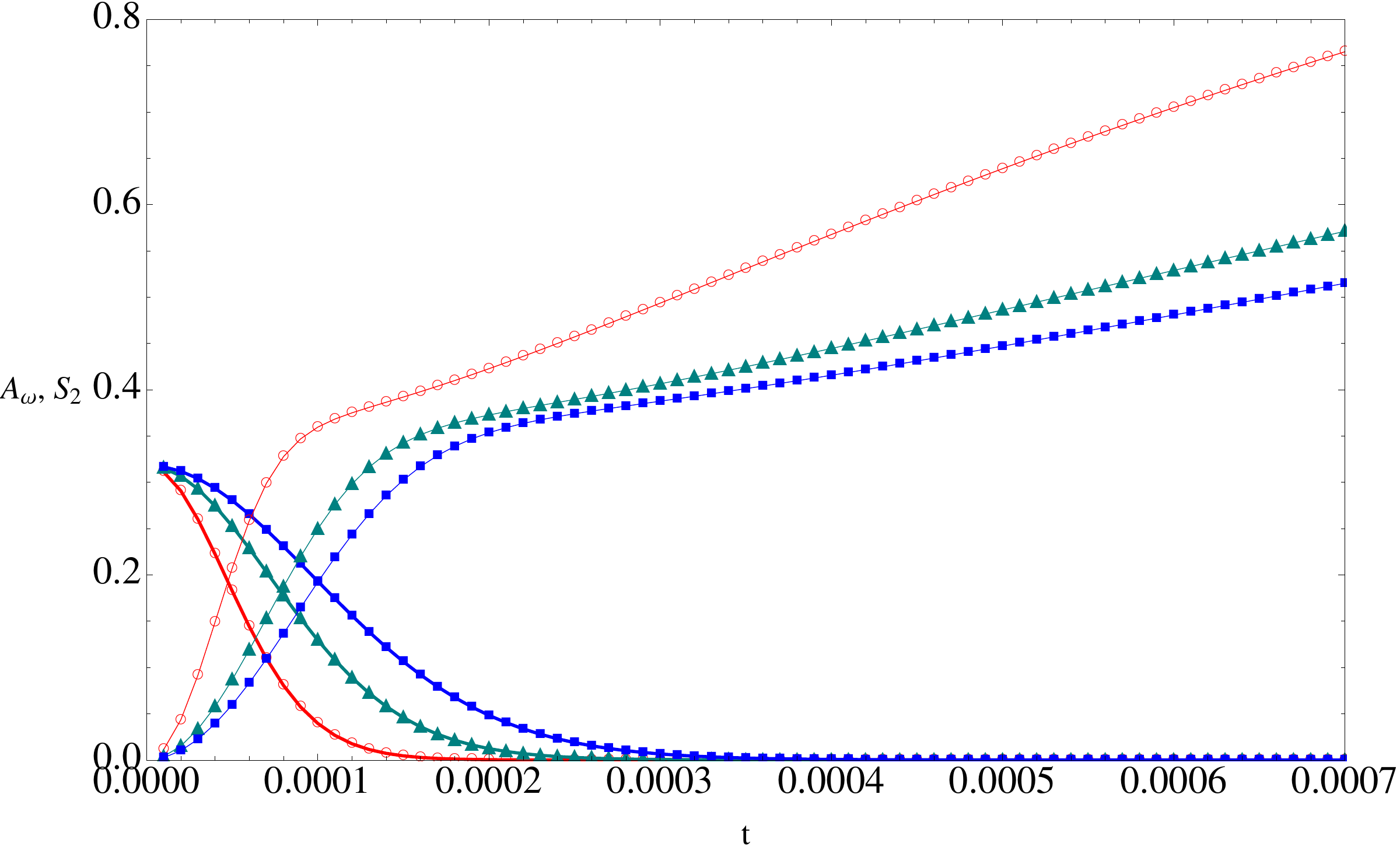}
                  }
\end{center}
\end{minipage}
&
\begin{minipage}{70mm}
\begin{center}
\unitlength=1mm
\resizebox{!}{4.2cm}{
   \includegraphics{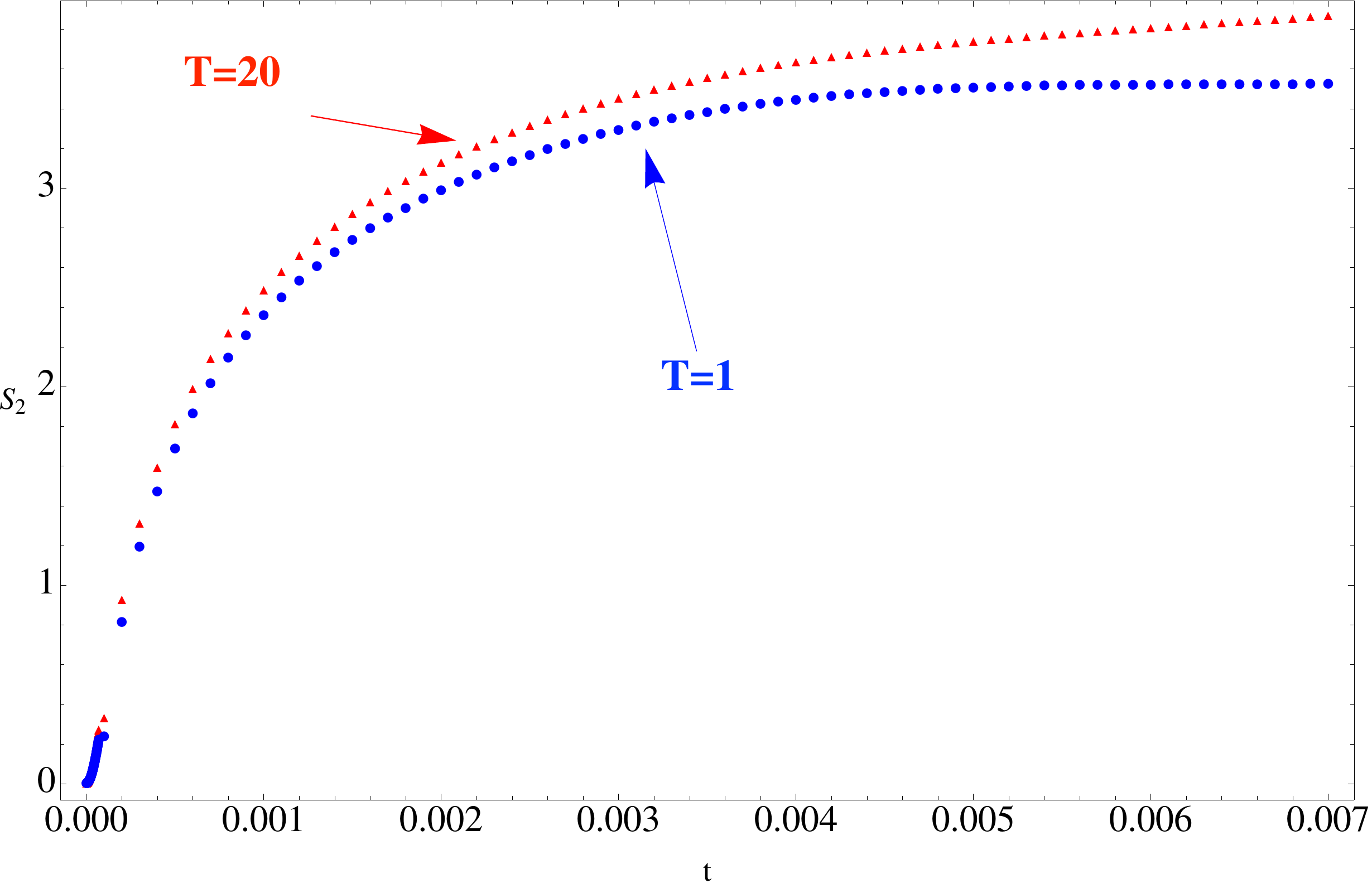}
                  }
\end{center}
\end{minipage}

\end{tabular}
\singlespace\caption{Time evolutions of the enveloping function $\mathcal{A}_w$ and the second order R\'enyi entropy $S_2$ for  a string probe in BTZ background.
The parameters chosen for \eq{Nw}, \eq{TG-12} and \eq{TG-13} are $(\phi_0, \sigma, N_{st}, \Omega, \Gamma_{w})=(05, 0.1, 1,1,1,10)$.  Left: $\mathcal{A}_w$ (thick lines) and $S_2$ (thin lines) for $T$=50 (circle), 30 (triangle) and 0.1 (square).  Note that we  have rescaled $S_2$ by $1/2$.  Right: The long-term time behavior of $S_2$ for $T=$20 and 1, which will be used for the fitting in Fig. \ref{S2-string}.}
\label{string_deco}
\end{figure}
\end{center}

For the case in which the `system' plus `environment' are the holographic dual of a string probe in a BTZ background, we present in Fig. \ref{string_deco} the time-evolution of the $\mathcal{A}_w(t)$ and $S_2(t)$ at different temperatures. We can see that $\mathcal{A}_w$ almost decays away at $t=t_D$, and around this time scale $S_2$ goes through a crossover. As
discussed, this implies that the quantum quench, which is characterized by this crossover behavior, is
closely related to the quantum decoherence.

\begin{center}
\begin{figure}[tbp]
\begin{tabular}{ll}
\begin{minipage}{70mm}
\begin{center}
\unitlength=1mm
\resizebox{!}{4.2cm}{
   \includegraphics{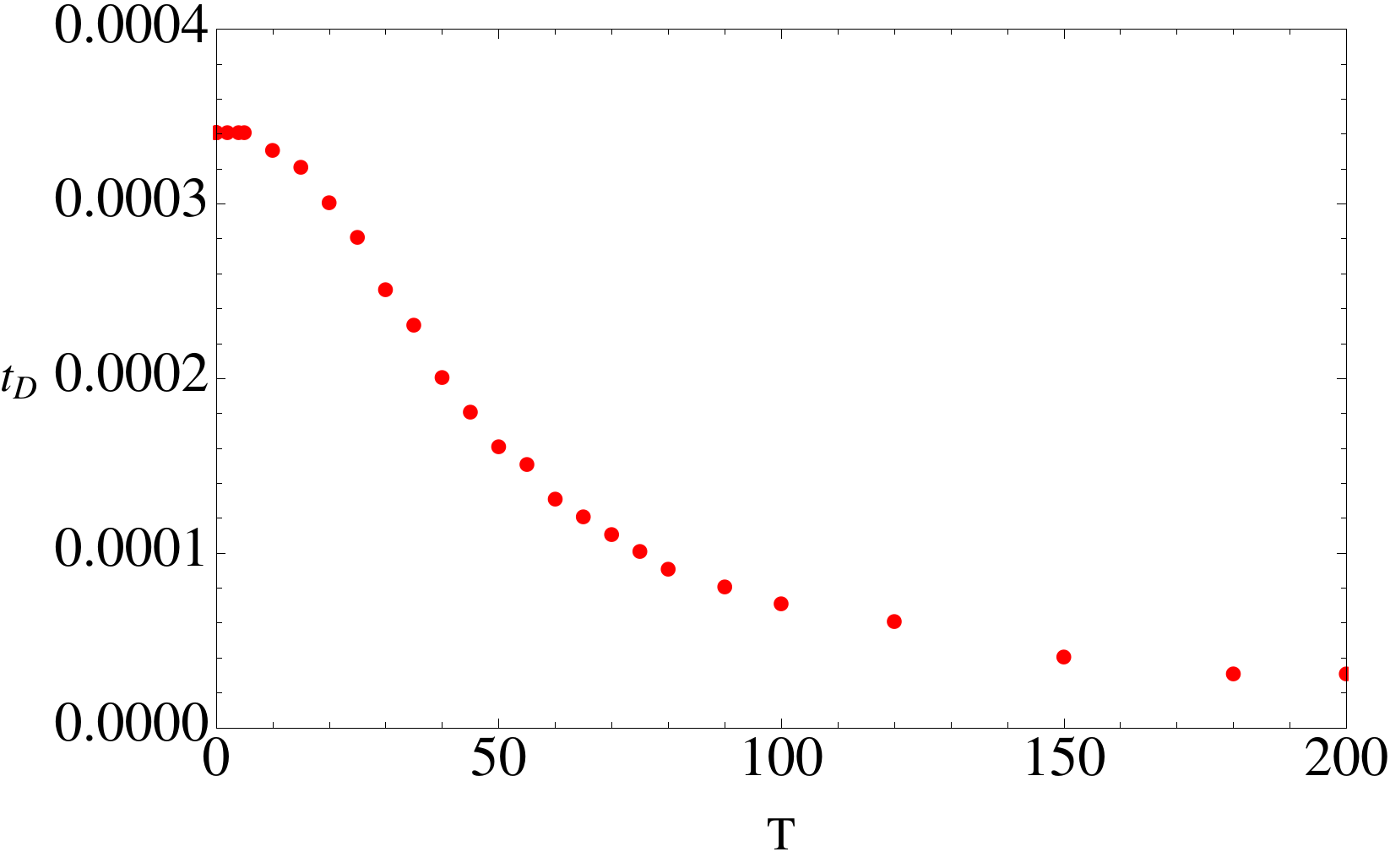}
                  }
\end{center}
\end{minipage}
&
\begin{minipage}{70mm}
\begin{center}
\unitlength=1mm
\resizebox{!}{4.2cm}{
   \includegraphics{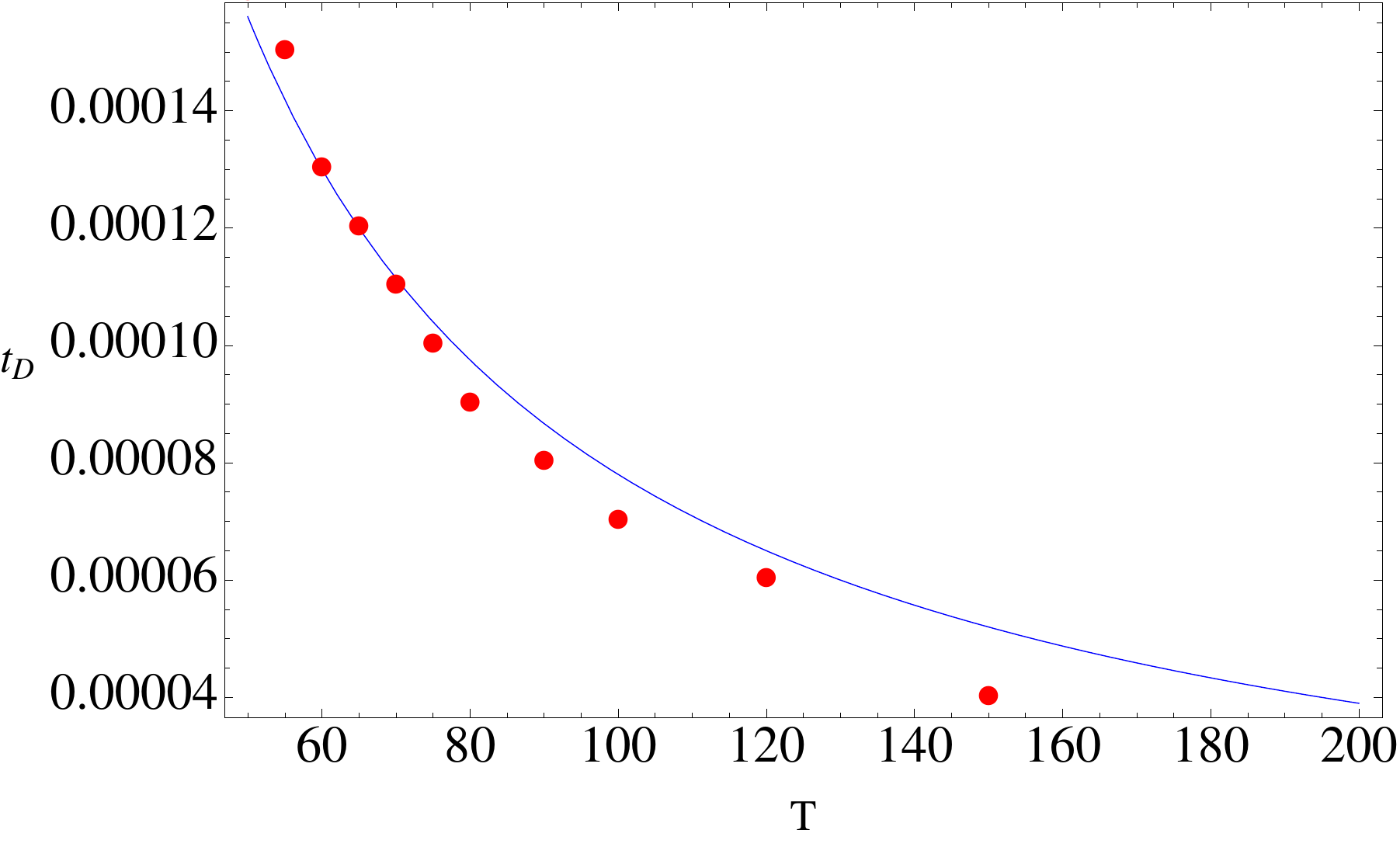}
                  }
\end{center}
\end{minipage}

\end{tabular}
\singlespace\caption{Left: Decoherence time $t_{\textrm{D}}$ v.s.  temperature $T$. Right:  Zoom-in of high -$T$ regime ($T>50$) with fitting function (blue line) $t={0.0078 \over T}$. The `soft mass gap' $\Gamma_{w}=10$.}
\label{tDvsT-string}
\end{figure}
\end{center}

Next we study the temperature ($T$) dependence of the decoherence time $t_D$. The results are shown in Fig. \ref{tDvsT-string}. First, let us look at the left panel which covers the entire $0<T<200$ region.  We see $t_{\textrm{D}}$ decreases as $T$ increases, namely, the hotter the `environment' the faster the `system' decoheres --- consistent with the intuitive picture of decoherence.

Next, we look at the high-$T$ and low-$T$ regimes separately. A conformal field theory has no intrinsic scale which we can use to define high-$T$ or low-$T$ regime. However, when we regularize the Green's function with a window function, we introduce a length scale into the system in term of its width $\Gamma_w$; therefore we can define high-$T$ or low-$T$ w.r.t. $\Gamma_w$.

In the low-$T$ ($T<\Gamma_w$) regime, we can see the decoherence time $t_{\textrm{D}}$ is almost independent of $T$. The reason is the following. The window function of width $\Gamma_w$ introduces a `soft mass gap' into the CFT. When $T<\Gamma_w$ the excitation is suppressed by the mass gap and the decoherence process resembles the one at zero-temperature therefore is insensitive to the actual temperature.

For the high-$T$ regime (see right panel of Fig. \ref{tDvsT-string}) we discover a nice scaling behavior
\be\label{scaling}
t_D \sim {1\over T}.
\ee
by fitting the data points.
This scaling behavior is the same as the one extracted directly from the master equation \eq{rhomaster} for the
simple toy model in \cite{Zurek}. While not surprising, this is the first time this scaling behavior is obtained when the `environment' is a non-trivial CFT.
\begin{center}
\begin{figure}[tbp]
\begin{tabular}{ll}
\begin{minipage}{70mm}
\begin{center}
\unitlength=1mm
\resizebox{!}{4.2cm}{
   \includegraphics{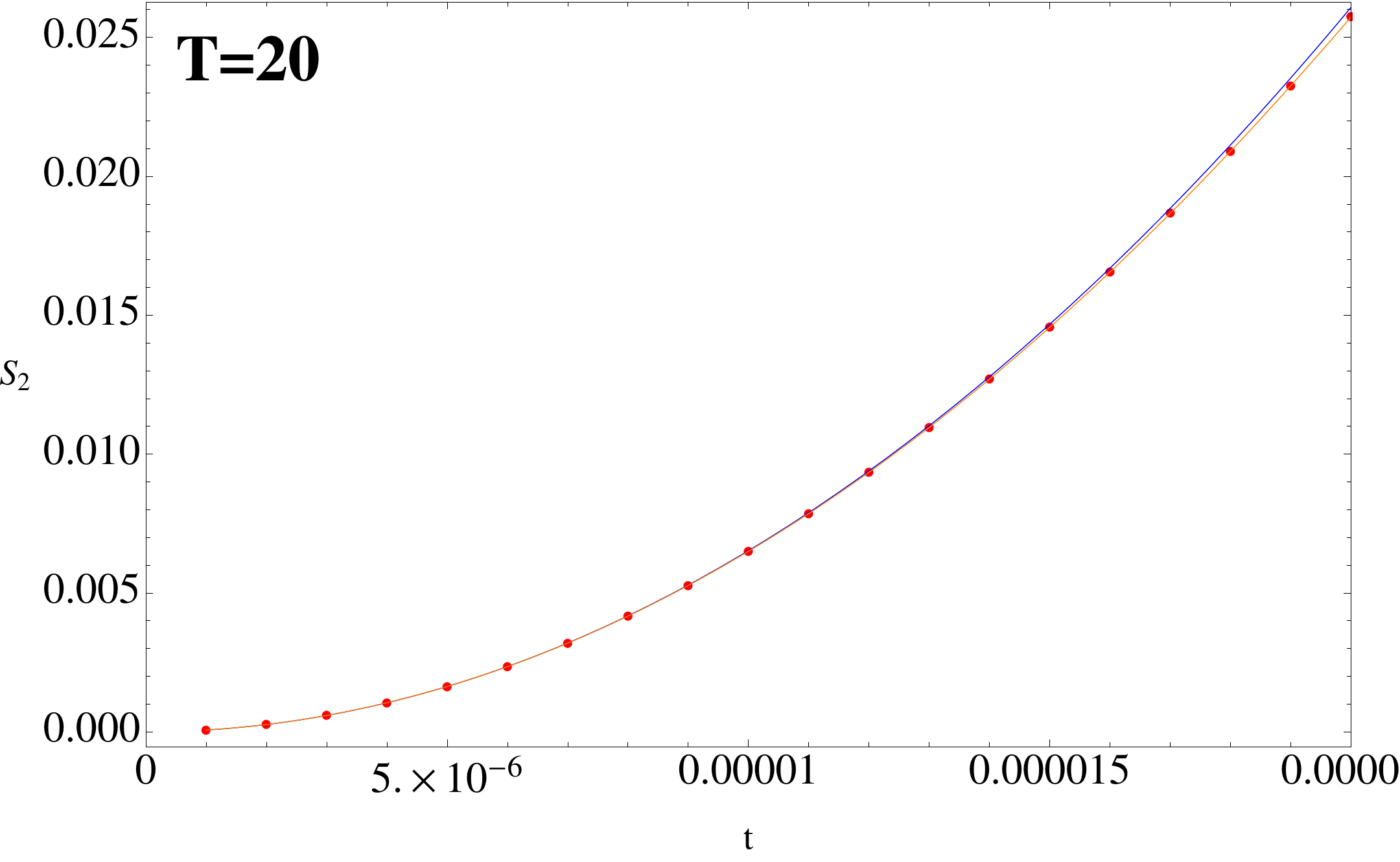}
                  }
\end{center}
\end{minipage}
&
\begin{minipage}{70mm}
\begin{center}
\unitlength=1mm
\resizebox{!}{4.2cm}{
   \includegraphics{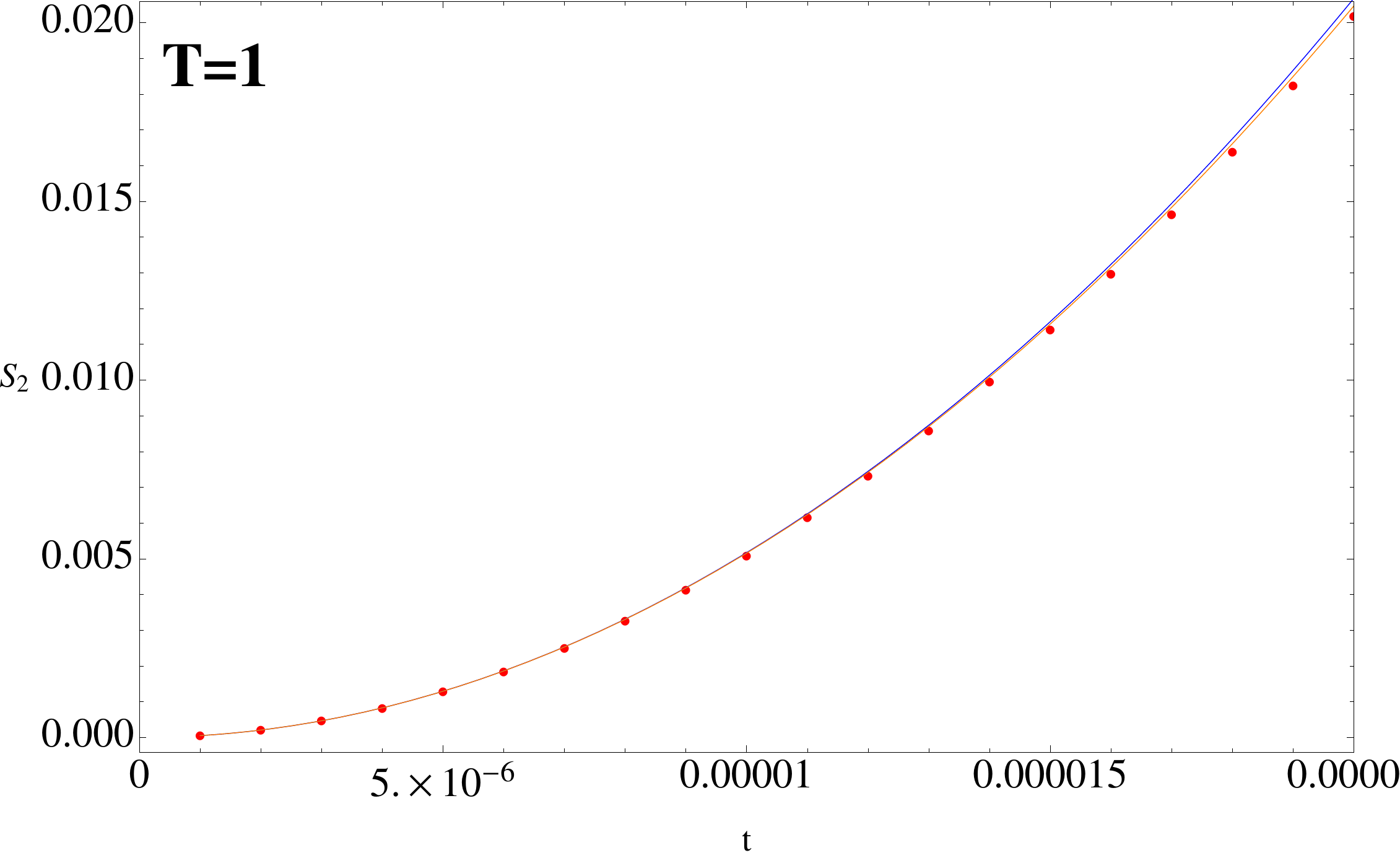}
                  }
\end{center}
\end{minipage} \\

\begin{minipage}{70mm}
\begin{center}
\unitlength=1mm
\resizebox{!}{4.2cm}{
   \includegraphics{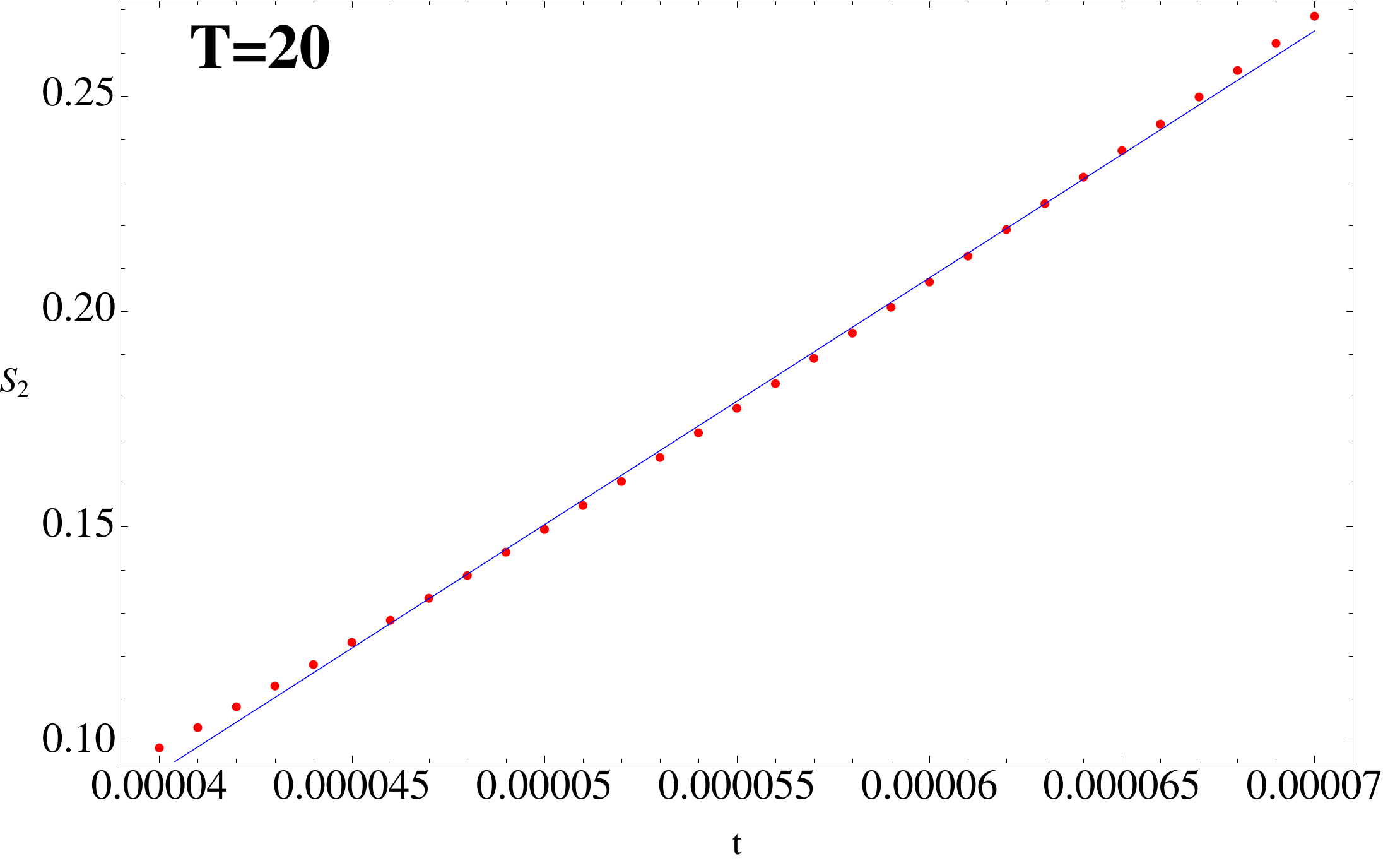}
                  }
\end{center}
\end{minipage}
&
\begin{minipage}{70mm}
\begin{center}
\unitlength=1mm
\resizebox{!}{4.2cm}{
   \includegraphics{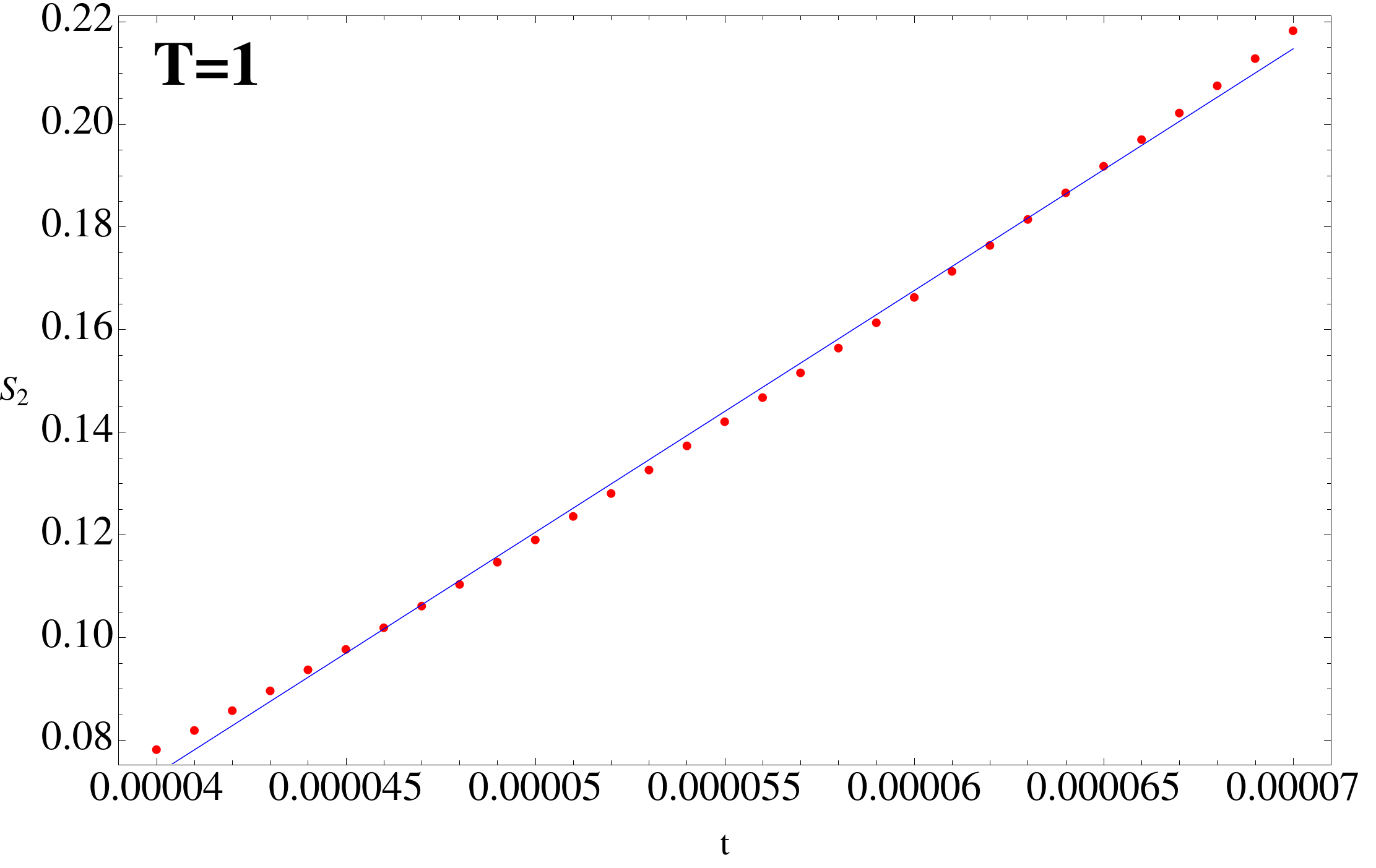}
                  }
\end{center}
\end{minipage} \\

\begin{minipage}{70mm}
\begin{center}
\unitlength=1mm
\resizebox{!}{4.2cm}{
   \includegraphics{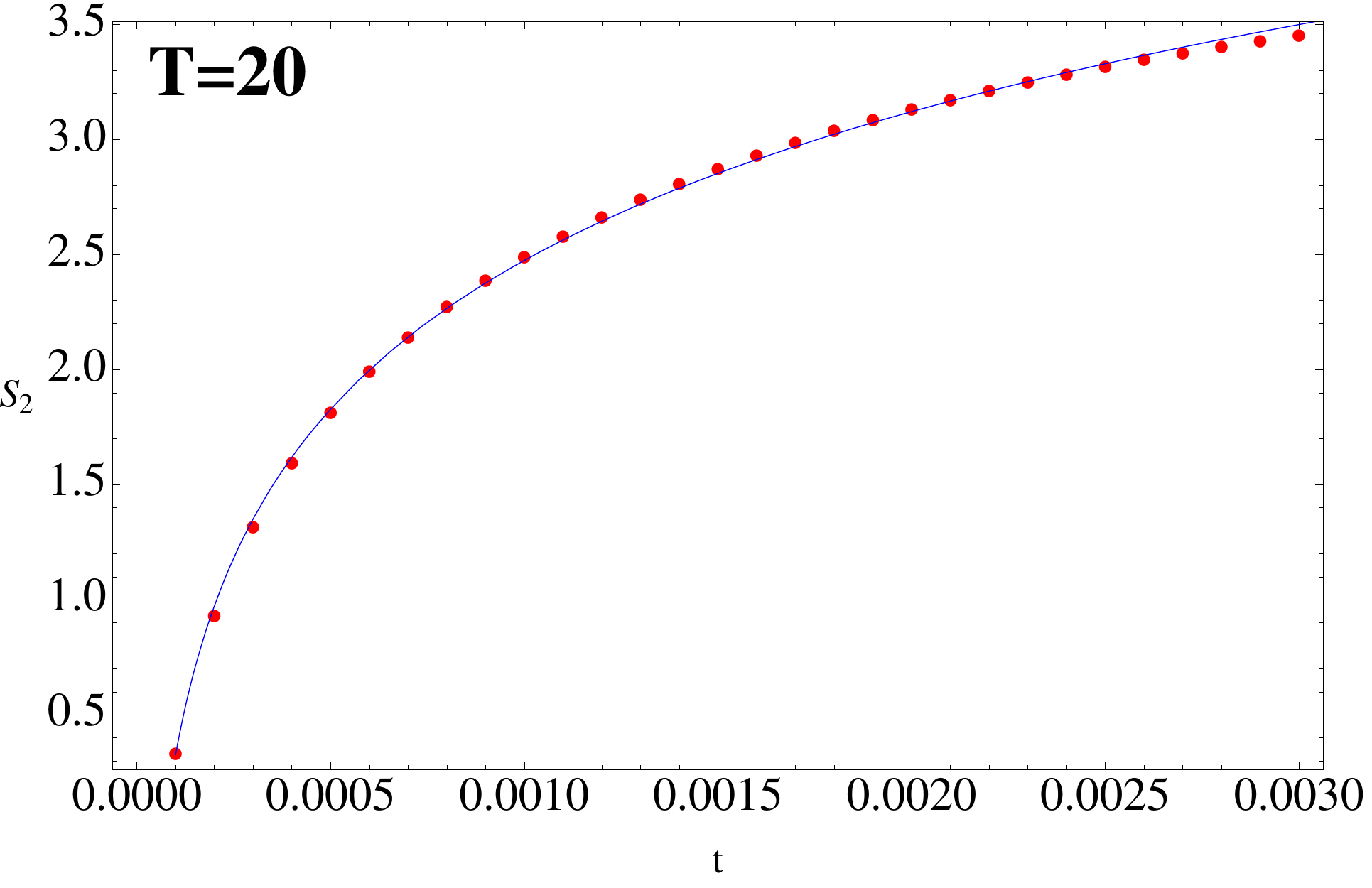}
                  }
\end{center}
\end{minipage}
&
\begin{minipage}{70mm}
\begin{center}
\unitlength=1mm
\resizebox{!}{4.2cm}{
   \includegraphics{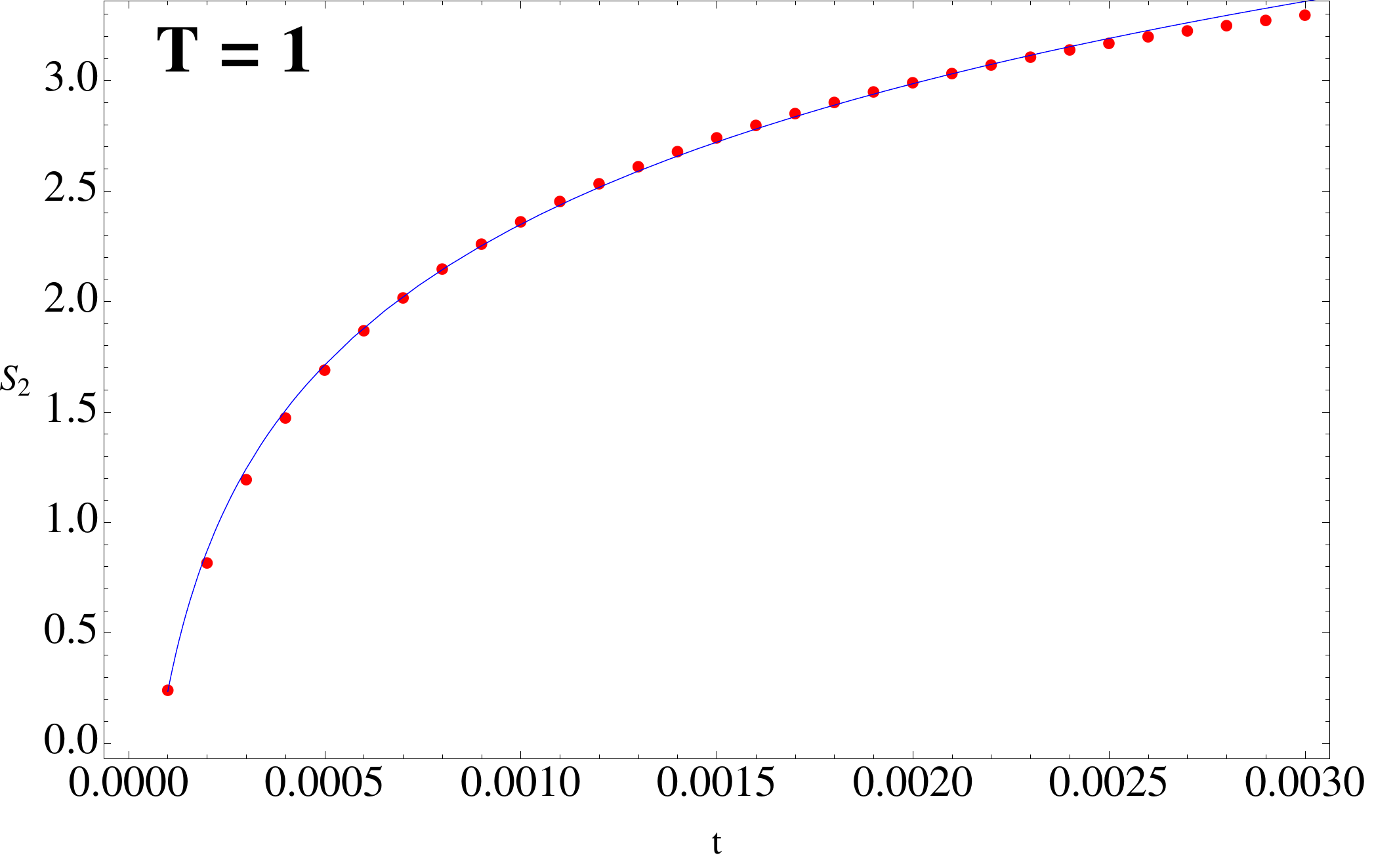}
                  }
\end{center}
\end{minipage}

\end{tabular}
\singlespace\caption{Scaling behaviors of $S_2$ at different stages of the growth shown in
the right panel of Fig. \ref{string_deco}. The three rows from up to down corresponds to the three stages defined in the main text. Initial: The solid
(orange) lines corresponds to $S_2 \sim  t^2 + \mathcal{O}(t^4)$ and fit better than the solid (blue) lines that have only $t^2$ term.
Intermediate: $S_2 \sim t + \mathcal{O}(1)$. Late: $S_2 \sim \ln t + \mathcal{O}(1)$.}
\label{S2-string}
\end{figure}
\end{center}

Finally, let us look at the time-evolution of the second order R\'enyi entropy $S_2$ of the `system'. In Fig. \ref{string_deco} we can already see that $S_2$ has different scaling behaviors at different regions of $t$. This suggests that the decoherence process happens in stages. With the fitting tool `formulize' \cite{formulize}, we fit the data and divide the decoherence process into the following four different stages according to the scaling behavior of $S_2(t)$:
\be\label{S2vst}
S_2(t)  \sim
\begin{cases}
C_0 \; t^2+\mathcal{O}(t^4)\;, \qquad &  t < \frac{t_{\textrm{D}}}{10} \qquad (\textrm{initial})\;; \\
 C_{b.c.} \; t+\mathcal{O}(1)\;, \qquad &  \frac{t_{\textrm{D}}}{10}\lesssim  t \lesssim  t_{\textrm{D}}\qquad (\textrm{intermediate})\;;\\
 C_{a.c.} \ln  t+\mathcal{O}(1)\;,  \qquad &  t_{\textrm{D}}\lesssim t \lesssim  10t_{\textrm{D}} \qquad (\textrm{late})\;;\\
\textrm{slower than } C_1 \ln t 
 \;, \qquad &   10t_{\textrm{D}}<t  \qquad (\textrm{final})\;.
\end{cases}
\ee
where $C_0$, $C_{b.c.}$, $C_{a.c.}$, $C_1(<C_{a.c.})$ are all positive and increase monotonically with $T$. The data (together with their fittings) of the first three stages are plotted in Fig. \ref{S2-string}. We also compare the high-$T$ and low-$T$ cases and find that the scaling behavior of $S_2(t)$ is insensitive to the temperature.
The intermediate stage is about one order
of magnitude longer than the initial stage, which agrees with the corresponding result for the quantum quench in \cite{quench}.  After the crossover until the end of the late stage (at $\sim 10 t_D$), $S_2$ grows as $\ln t$, in agreement with the results in \cite{quench1,holo-quench}. During the final stage, we see the $\log{t}$-growth of the previous stage slows down, although our present limit with numerical computation (due to accumulated numerical errors that affect particularly the large-$t$ region) prevents us from determining the long-term trend of $S_2(t)$.\footnote{In particular, $S_2$ acquires  a negative linear-$t$ term to offset the positive $\ln{t}$ term. However, instead of the eventual saturation that we (albeit maybe naively) expect, we see a $C_2 t-C_3 \ln{t}$ growth at the very end (using $20$ additional data points extending the right figure of  Fig. \ref{string_deco}). The two terms $C_2 t$ and $C_3 \ln{t}$ are roughly equal in magnitude  and have opposite signs therefore we cannot predict the long-term trend based on this very end; and although this could be a mirage caused by the large-$t$ numerical errors, we mention this as an issue that deserves future study.}
Matching the first three stages of \eq{S2vst} with \eq{eeL} we conclude that the relaxation following a local quantum quench occurs when the local excitations decohere.  

\begin{center}
\begin{figure}[tbp]
\begin{tabular}{ll}
\begin{minipage}{70mm}
\begin{center}
\unitlength=1mm
\resizebox{!}{4.2cm}{
   \includegraphics{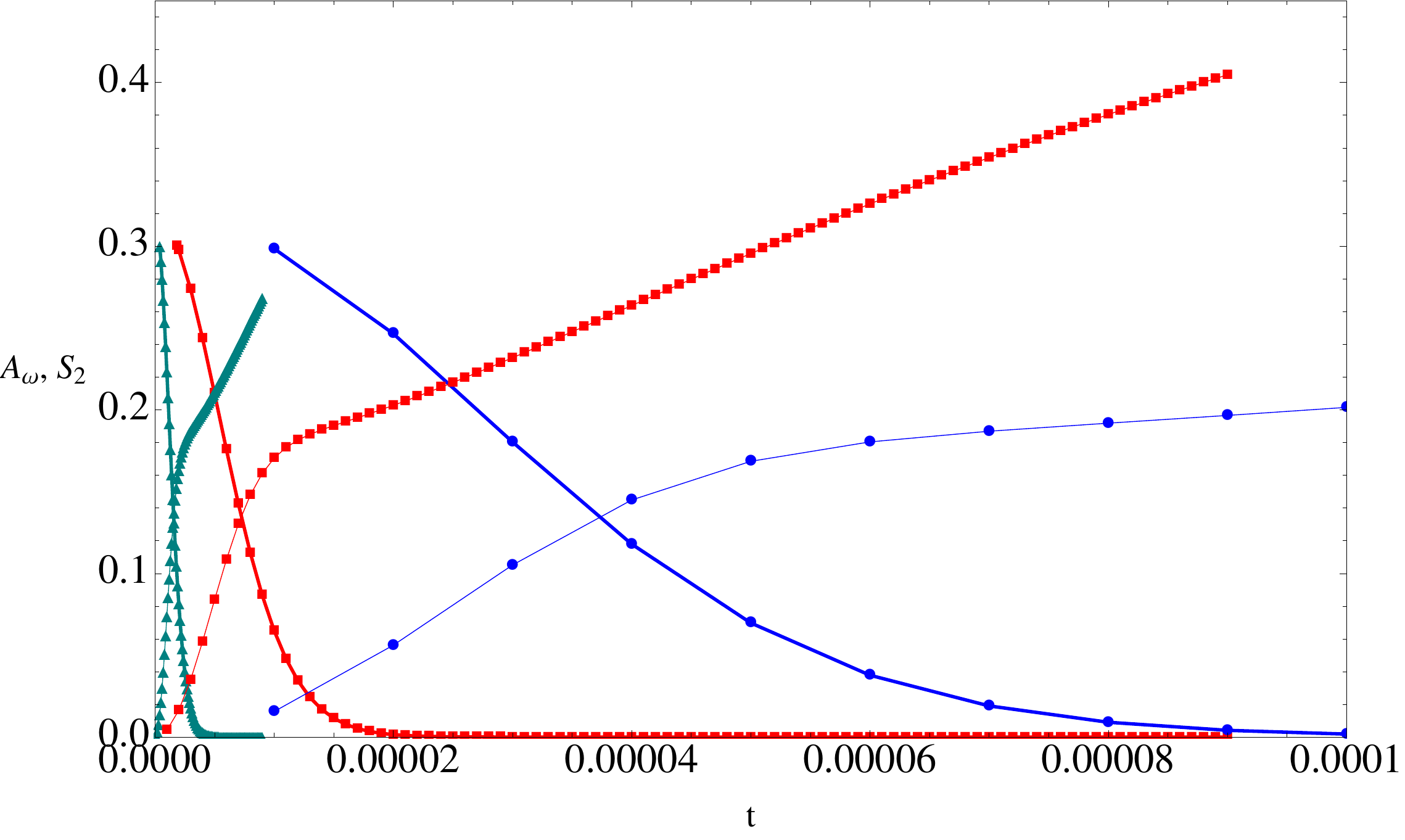}
                  }
\end{center}
\end{minipage}
&
\begin{minipage}{70mm}
\begin{center}
\unitlength=1mm
\resizebox{!}{4.2cm}{
   \includegraphics{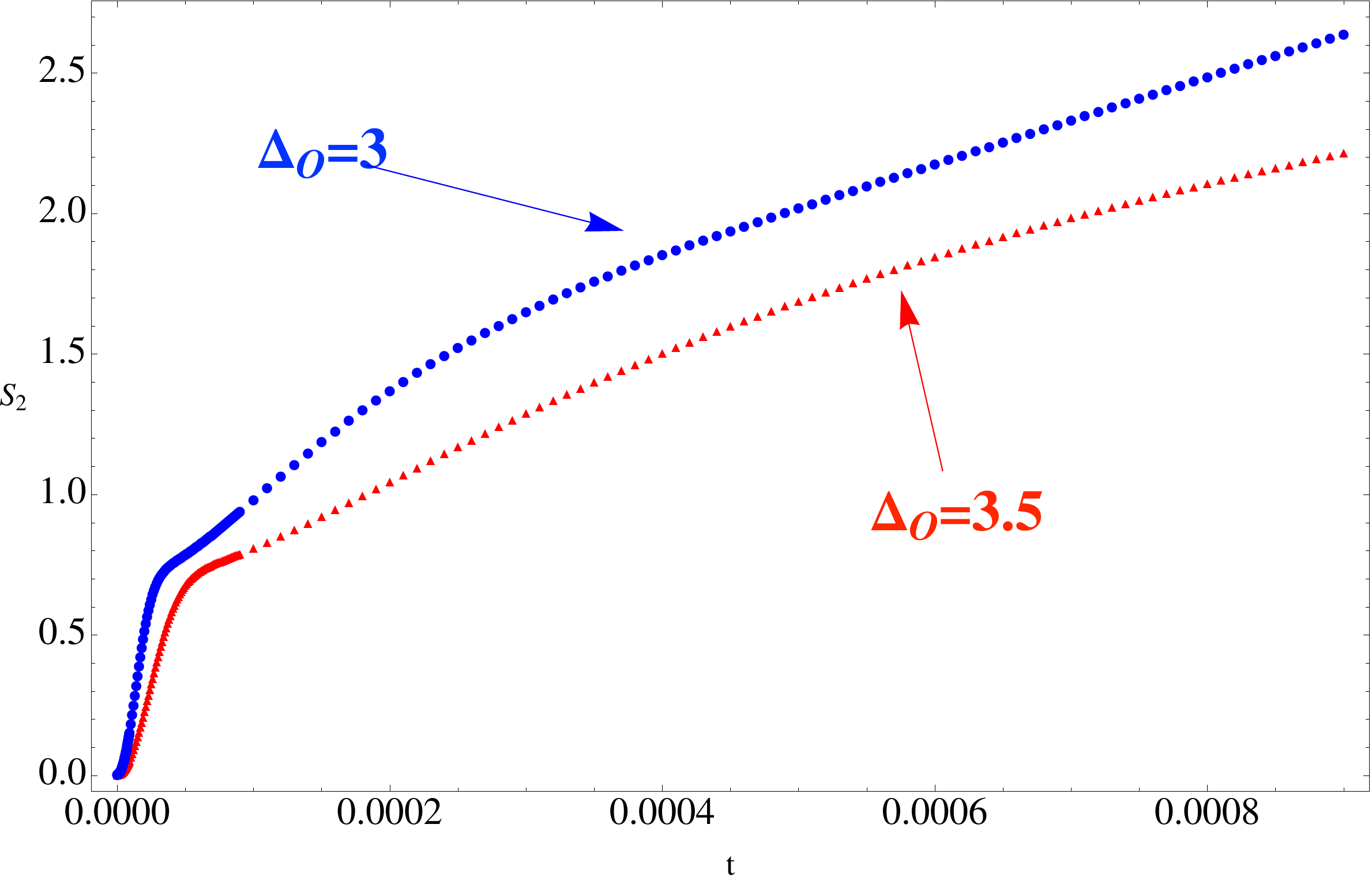}
                  }
\end{center}
\end{minipage}

\end{tabular}
\singlespace\caption{ Time evolutions of the enveloping function $\mathcal{A}_w$ and the second order R\'enyi entropy $S_2$ for a scalar probe (zero-mode only) in AdS$_5$ space-time.
The parameters chosen for \eq{Nw}, \eq{TG-12} and \eq{TG-13} are $(\phi_0, \sigma, N_{sc}, \Omega, \Gamma_{w})=(05, 0.1, 10,1,1,10)$.  Left: $\mathcal{A}_w$ (thick lines) and $S_2$ (thin lines) for $\Delta_{\mathcal{O}}=$4.5 (triangle), 4 (circle) and 3.5 (square).  Note we  have rescaled $S_2$ by $1/2$.  Right: The long-term behaviors of $S_2$ for $\Delta_{\mathcal{O}}=$3.5 and $3$, which will be used for the fitting in Fig. \ref{S2-scalar}. Note that we have rescaled the time for $\Delta_{\mathcal{O}}=3$ case by $1/10$ in order to fit it in the same plot with  $\Delta_{\mathcal{O}}=3.5$ case.
}
\label{scalar_deco}
\end{figure}
\end{center}
\begin{figure}[h]
\begin{center}
\includegraphics[scale=0.5]{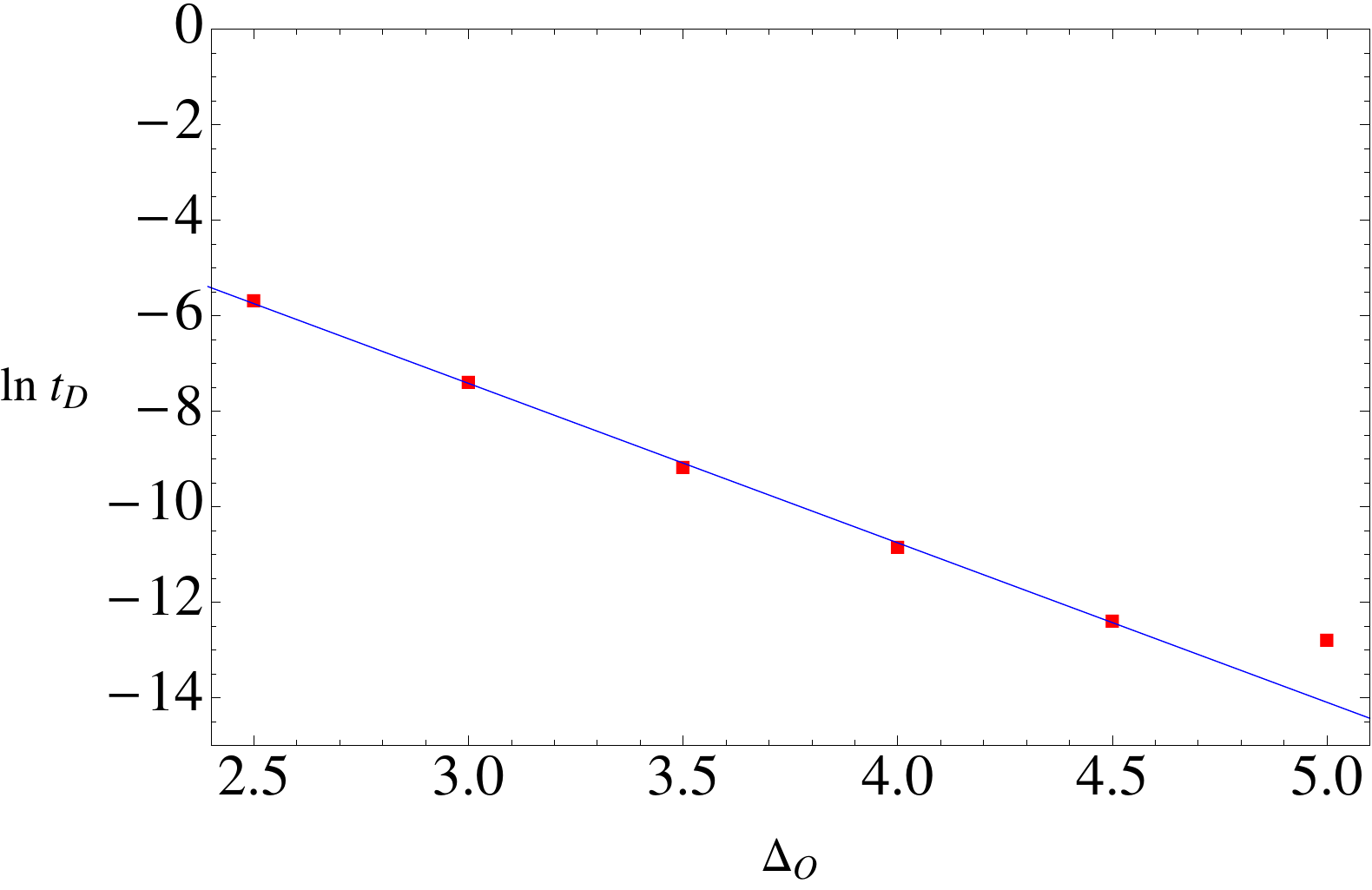}
\end{center}
\singlespace\caption{$\ln t_D$ v.s.  $\Delta_{\mathcal{O}}$ for
$2.5  \le  \Delta_{\mathcal{O}} \le 5$. The blue line is the
fitting function: $\ln\ t_D=2.26 - 3.34 \Delta_{\mathcal{O}}$.}
\label{tDvsD}
\end{figure}
\begin{center}
\begin{figure}[tbp]
\begin{tabular}{ll}
\begin{minipage}{70mm}
\begin{center}
\unitlength=1mm
\resizebox{!}{4.2cm}{
   \includegraphics{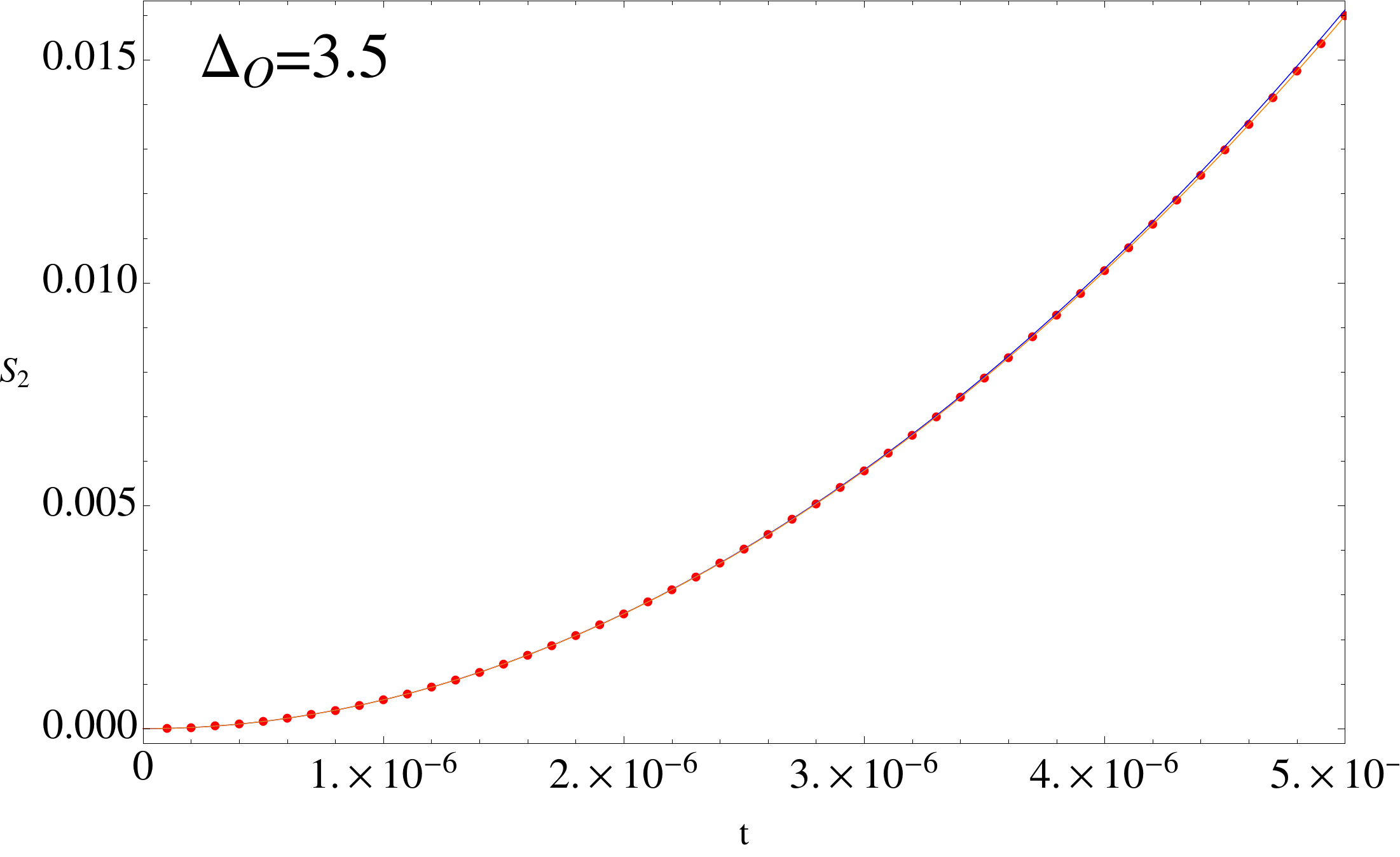}
                  }
\end{center}
\end{minipage}
&
\begin{minipage}{70mm}
\begin{center}
\unitlength=1mm
\resizebox{!}{4.2cm}{
   \includegraphics{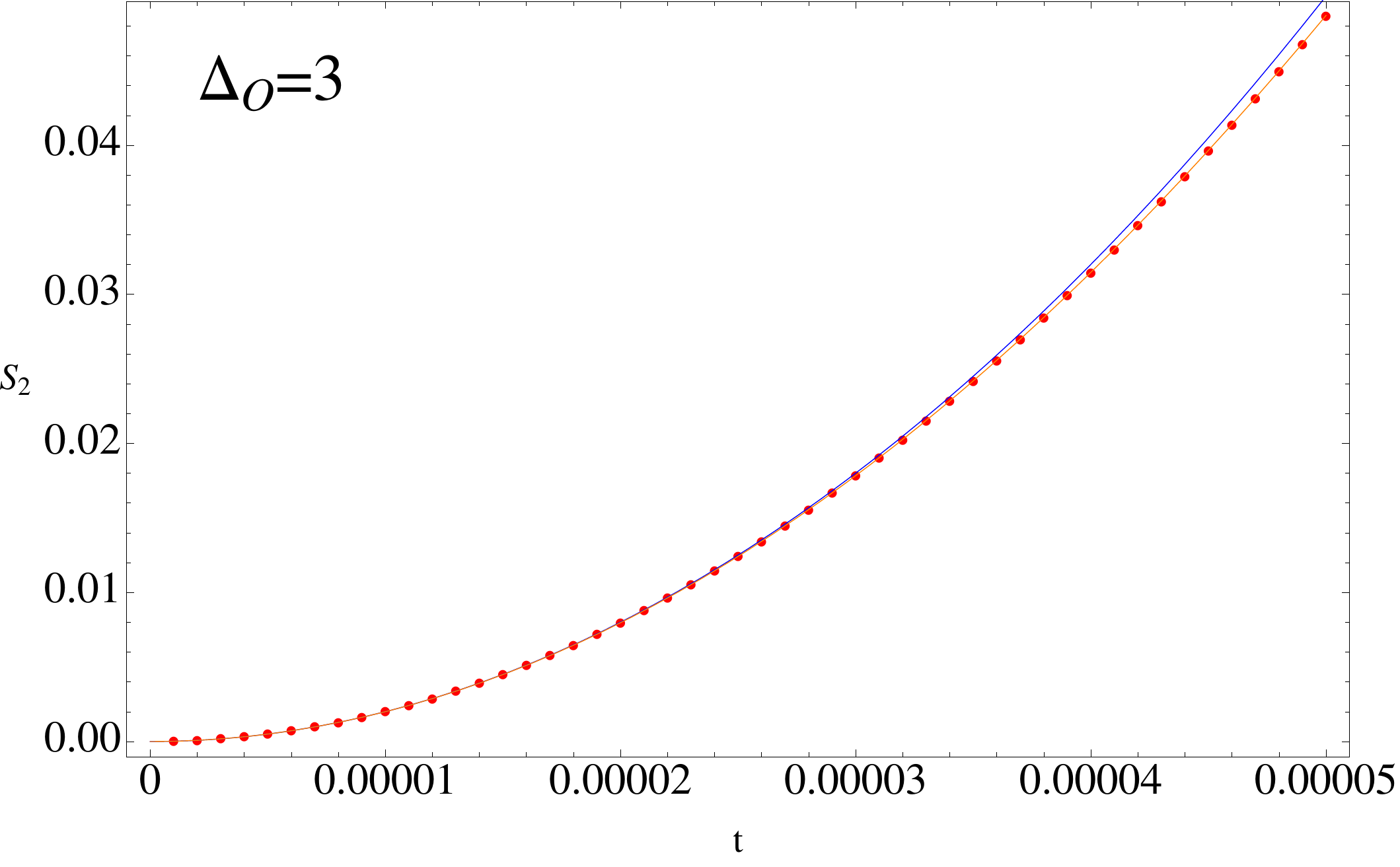}
                  }
\end{center}
\end{minipage}  \\

\begin{minipage}{70mm}
\begin{center}
\unitlength=1mm
\resizebox{!}{4.2cm}{
   \includegraphics{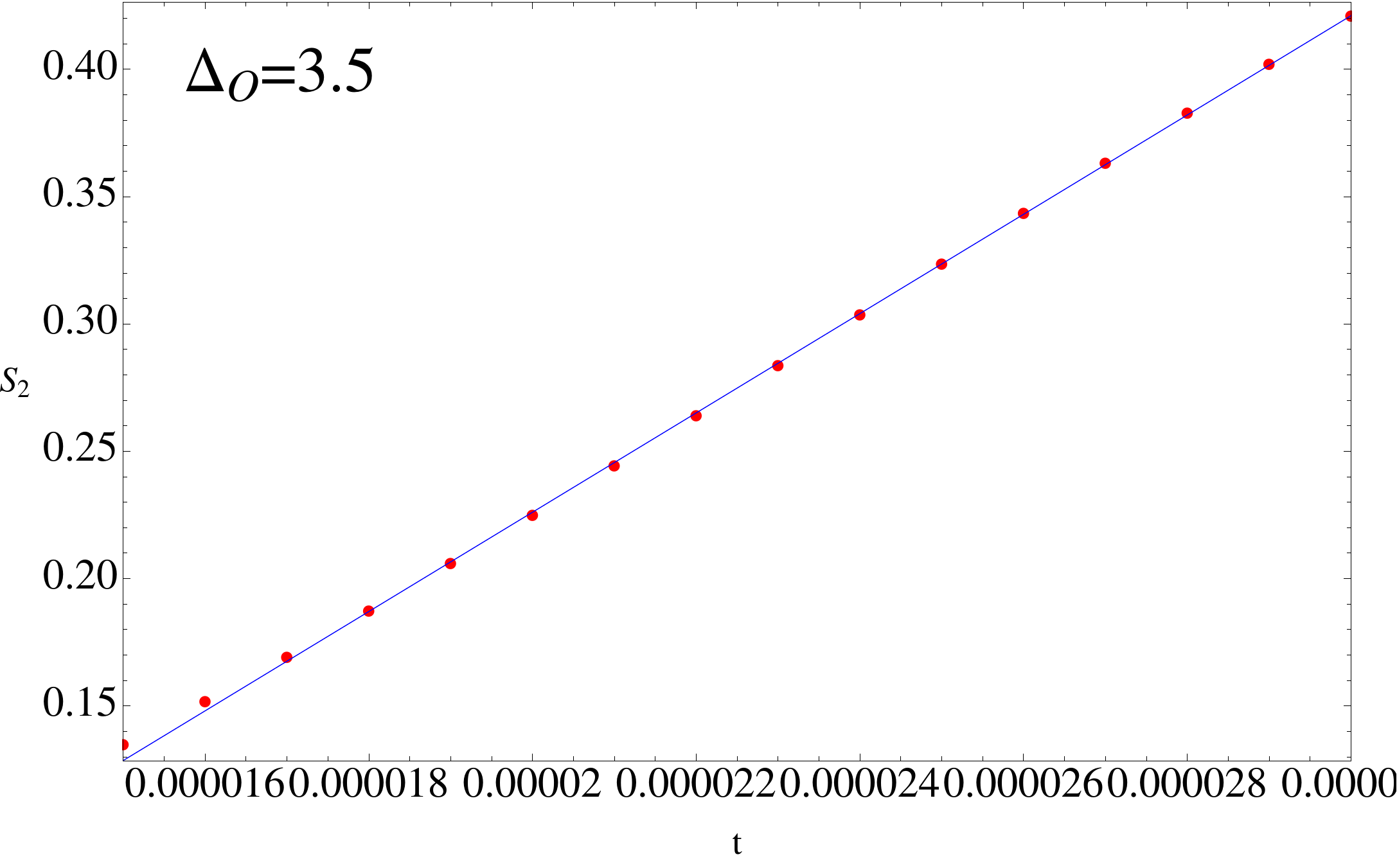}
                  }
\end{center}
\end{minipage}
&
\begin{minipage}{70mm}
\begin{center}
\unitlength=1mm
\resizebox{!}{4.2cm}{
   \includegraphics{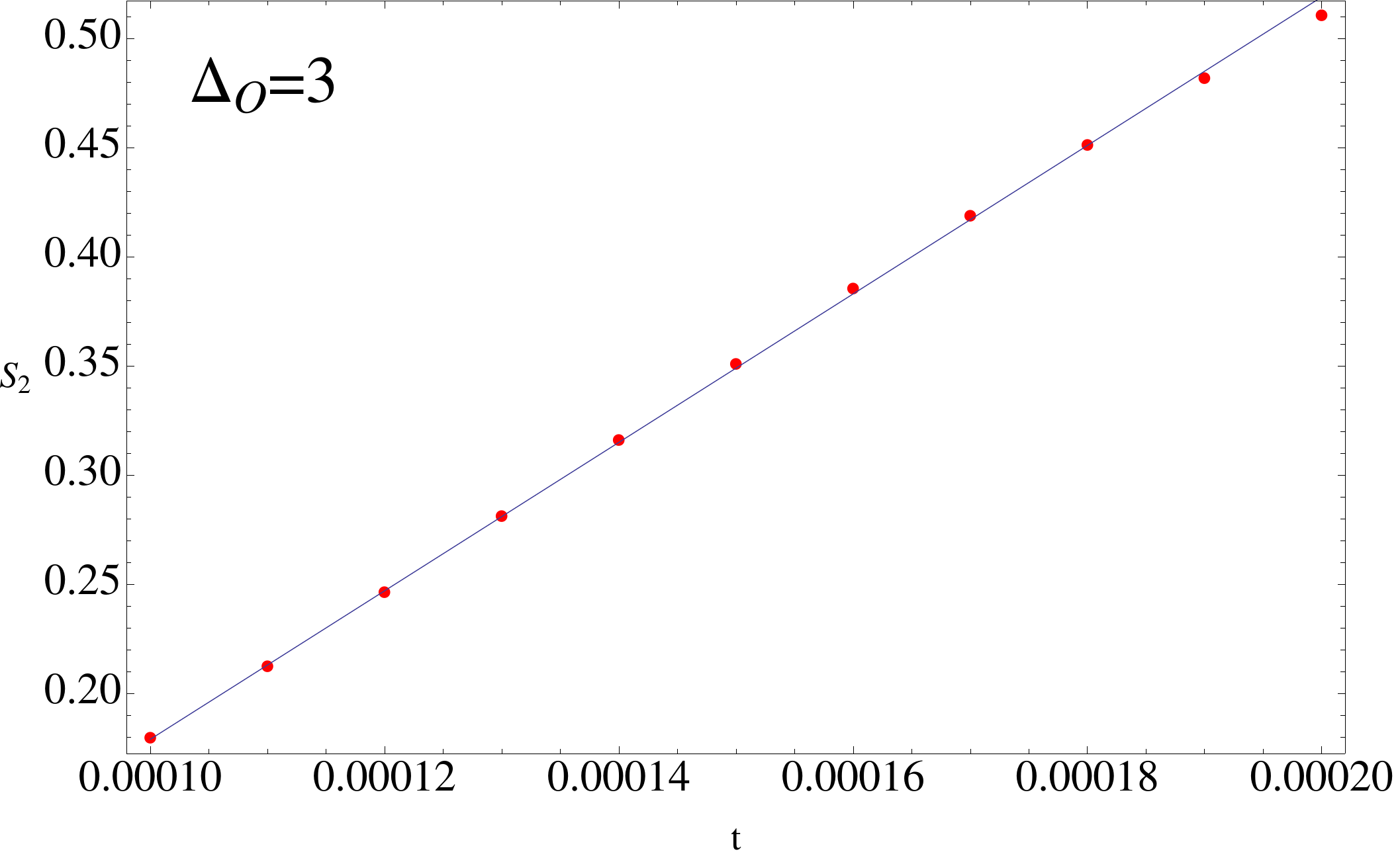}
                  }
\end{center}
\end{minipage}\\

\begin{minipage}{70mm}
\begin{center}
\unitlength=1mm
\resizebox{!}{4.2cm}{
   \includegraphics{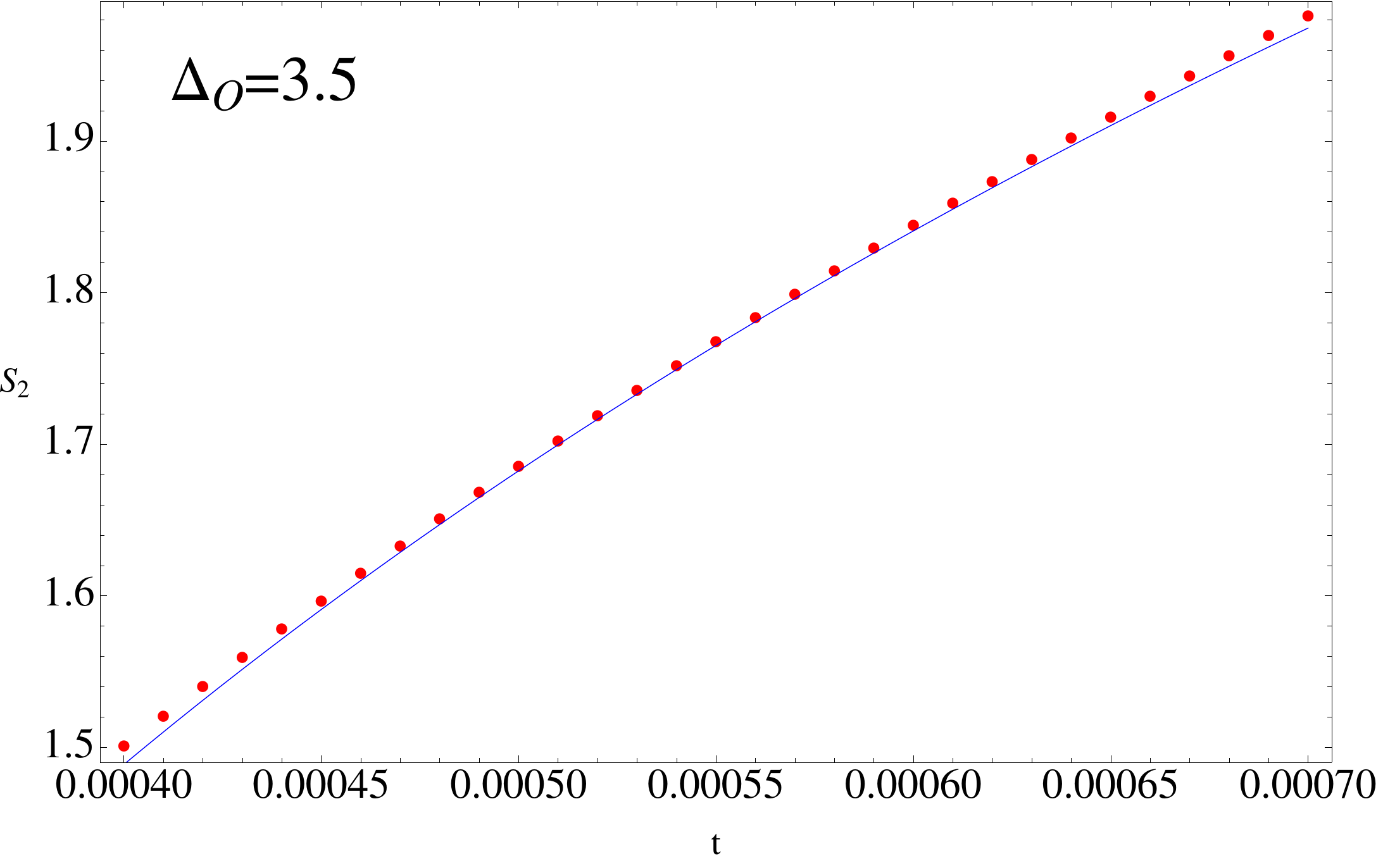}
                  }
\end{center}
\end{minipage}
&
\begin{minipage}{70mm}
\begin{center}
\unitlength=1mm
\resizebox{!}{4.2cm}{
   \includegraphics{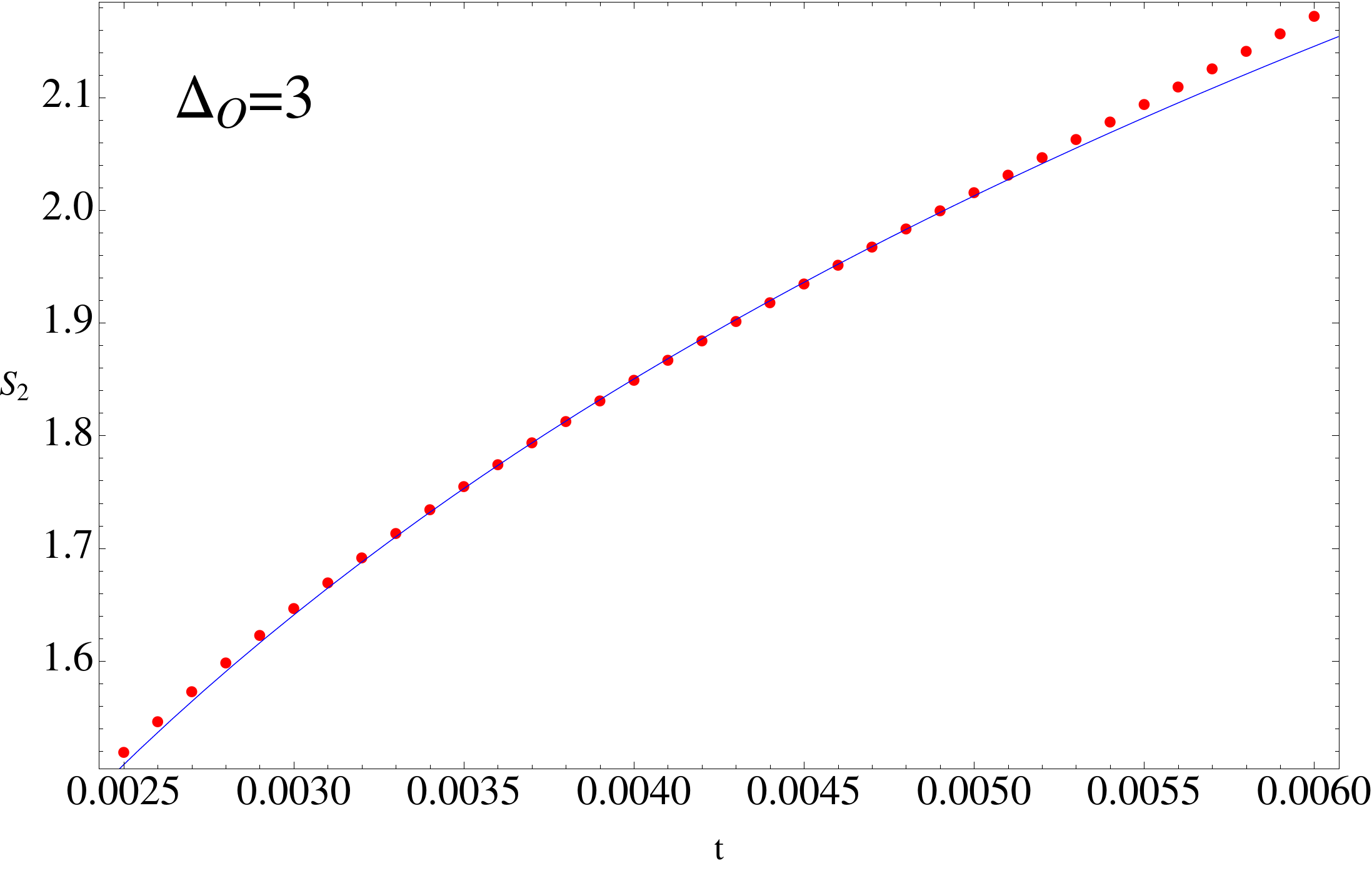}
                  }
\end{center}
\end{minipage}

\end{tabular}
\singlespace\caption{Scaling behaviors of $S_2$ at different stages of the growth shown in
the right panel of Fig. \ref{scalar_deco}. The three rows from up to down correspond to the three stages defined in the main text. Initial: The solid
(orange) lines correspond to $S_2 \sim  t^2 + \mathcal{O}(t^4)$ and fit better than the solid (blue) lines that have only $t^2$ term.
Intermediate: $S_2 \sim t + \mathcal{O}(1)$. Late: $S_2 \sim \ln t + \mathcal{O}(1)$.
}
\label{S2-scalar}
\end{figure}
\end{center}

\subsubsection{Scalar probe}

  Similar to the string probe case, we first present in Fig. \ref{scalar_deco} the evolution of
$\mathcal{A}_w(t)$ and $S_2(t)$ for operator $\mathcal{O}$ with different conformal dimensions.  The qualitative behavior is the same as the string probe case. Again, we find that the quench and  decoherence time scales are of the same order. In fact, since now the `environment' is at zero temperature, it is closer to the quantum quench setup. This provides more supporting evidence for the connection between quantum decoherence and  quantum quench.

Next, we fit the scaling behavior of $t_D$ w.r.t. $\Delta_{\mathcal{O}}$ and find (as shown in Fig. \ref{tDvsD}):
\be
\ln t_D \sim  - \; C_{\Delta} \Delta_{\mathcal{O}}
\ee
with $C_{\Delta}>0$. (This is to be compared with the scaling $t_D \sim  1/T$ in the previous case (shown in Fig. \ref{tDvsT-string}).) This means that the pure state of the superposition of
two Gaussian wave-packets has a lifetime of order $\frac{1}{\Delta_{\mathcal{O}}}$.
Namely, if we try to realize a qubit embedded in strongly-coupled environments of the type considered here, we should choose those with as small a  $\Delta_{\mathcal{O}}$ as possible, in order for the `system' to be more robust (against decoherence).

  Finally, using the fitting tool  `formulize' we determine the scaling behaviors of the second order R\'enyi entropy $S_2(t)$ at different stages for two
different  $\Delta_{\mathcal{O}}$'s and the results are shown in Fig. \ref{S2-scalar}. Although the total system is different from the previous case, we found the scaling behavior to be basically the same as  \eq{S2vst}.
(And all the constants $C$'s  in \eq{S2vst} increase with $\Delta_{\mathcal{O}}$.)  This suggests that the scaling behaviors of $S_2$ and the definition of the different stages
given in \eq{S2vst} might be universal for different probes and environments.\footnote{The $\ln t$
growing behavior in \eq{eeL} or \eq{S2vst} could be a feature of 2D CFT. Although
we consider the scalar probe in AdS$_5$ at very beginning, we only focus on the decoherence
behavior of its zero-mode, which can be thought as an effective $(1+1)$-dimensional problem.
Thus, we obtain the scaling behaviors \eq{S2vst} as expected.}   It is also very reassuring  to see all our results agree with the studies based on the (holographic) entanglement entropy for
the local quantum quench \cite{quench1,quench2,holo-quench,Nozaki:2013vta,quench}. It would be
interesting to explore this connection further.

\bigskip
\bigskip

\subsection{Initial state profile dependence of decoherence time}
  The decoherence time $t_{\textrm{D}}$ depends on how the `system' (the scalar $\phi$) interacts with the `environment' and on the properties of $\phi$ such as its mass or conformal dimension. The other relevant attribute of the `system' is the profile of its initial state, e.g. in our case of two Gaussian wave-packets the relevant data is the width ($\sigma$) of each packet and the `distance' ($2\phi_0$) between the two.

  Further, the similarity between \eq{S2vst} and \eq{eeL} implies a close relation between the quantum decoherence and the local quench once we identify the decoherence time $t_{\textrm{D}}$ in \eq{S2vst} with the effective length $L$ in \eq{eeL}.  Then once we determine $t_{\textrm{D}}$'s dependence  on the initial state profile, it immediately gives the relation between the initial state profile and the effective length.  This dependence is nontrivial due to the complexity of $\mathcal{A}_w(t)$ given in \eq{Nw}, and needs to be extracted by numerical methods.

\begin{figure}[h]
\begin{center}
\includegraphics[scale=0.45]{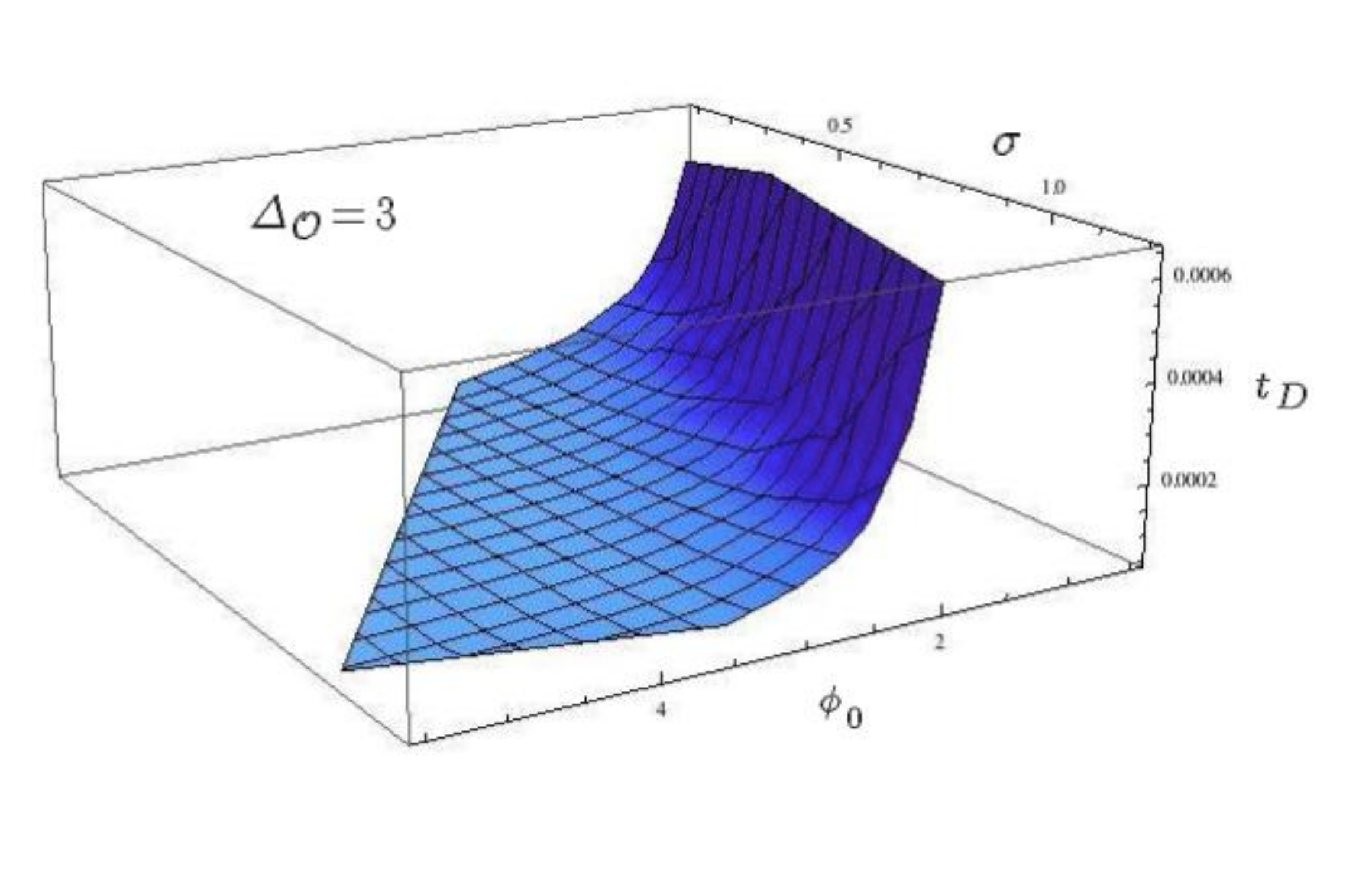}
\end{center}
\singlespace\caption{Dependence of decoherence time $t_{\textrm{D}}$ on the profile ($\{\sigma, \phi_0\}$) of initial state of the two Gaussian wave-packets, for scalar probe with $\Delta_{\mathcal{O}}=3$. All other parameters (except for $\sigma$ and $\phi_0$) are the same as used in Fig. \ref{scalar_deco}.}
\label{tDL}
\end{figure}

In Fig. \ref{tDL} we show the decoherence time $t_{\textrm{D}}$ as a function of the profile ($\{\sigma, \phi_0\}$) for the scalar probe with $\Delta_{\mathcal{O}}=3$. Fitting the data by constant $\sigma$ or $\phi_0$ slices, we find
\begin{equation}
t_D \sim \frac{c_1(\sigma)}{\phi_0+c_2(\sigma)} \qquad \textrm{for fixed $\sigma$}\,; \qquad \qquad t_D \sim c_3(\phi_0)+c_4(\phi_0)\sigma^3 \qquad \textrm{for fixed $\phi_0$}\,.
\end{equation}
This shows that the effective length ($\approx t_D$) is not proportional to $\phi_0$ (the distance between two packets) as naively expected for uniform probe. Instead, the relevant quantum information is only encoded in the properties of each individual Gaussian wave-packet, and the decoherence time $t_{\textrm{D}}$ or the effective length $L$ increase with $\sigma$. Thus, for fixed $\sigma$ a larger $\phi_0$ means a bigger zero-information region therefore lower concentration of nontrivial information, which accelerate the decoherence or the quench process. Therefore, to suppress decoherence, we should prepare initial states with small $\phi_{0}$. 

\subsection{Numerical studies for discrete time systems}

All the above results are based on exact solutions of $h(\tau)$ without any (numerical)
approximation. This is possible because the retarded Green's function $G_{\textrm{R}}$ is simple enough to allow $h(\tau)$ to be computed analytically via Laplace transform.  In general, $G_{\textrm{R}}$ is more complicated and solving \eq{hinverseLap} to obtain $h(\tau)$ requires  numerical methods. In order to generalize our proposed scheme in this paper and examine the
environment-induced decoherence scenario for more general situation, we compose a python code
\cite{python} to solve \eq{hinverseLap} and then evaluate $\mathcal{A}_w(t)$ and $S_2(t)$ for a given retarded Green's function.
\begin{figure}[h]
\begin{center}
\includegraphics[scale=0.3]{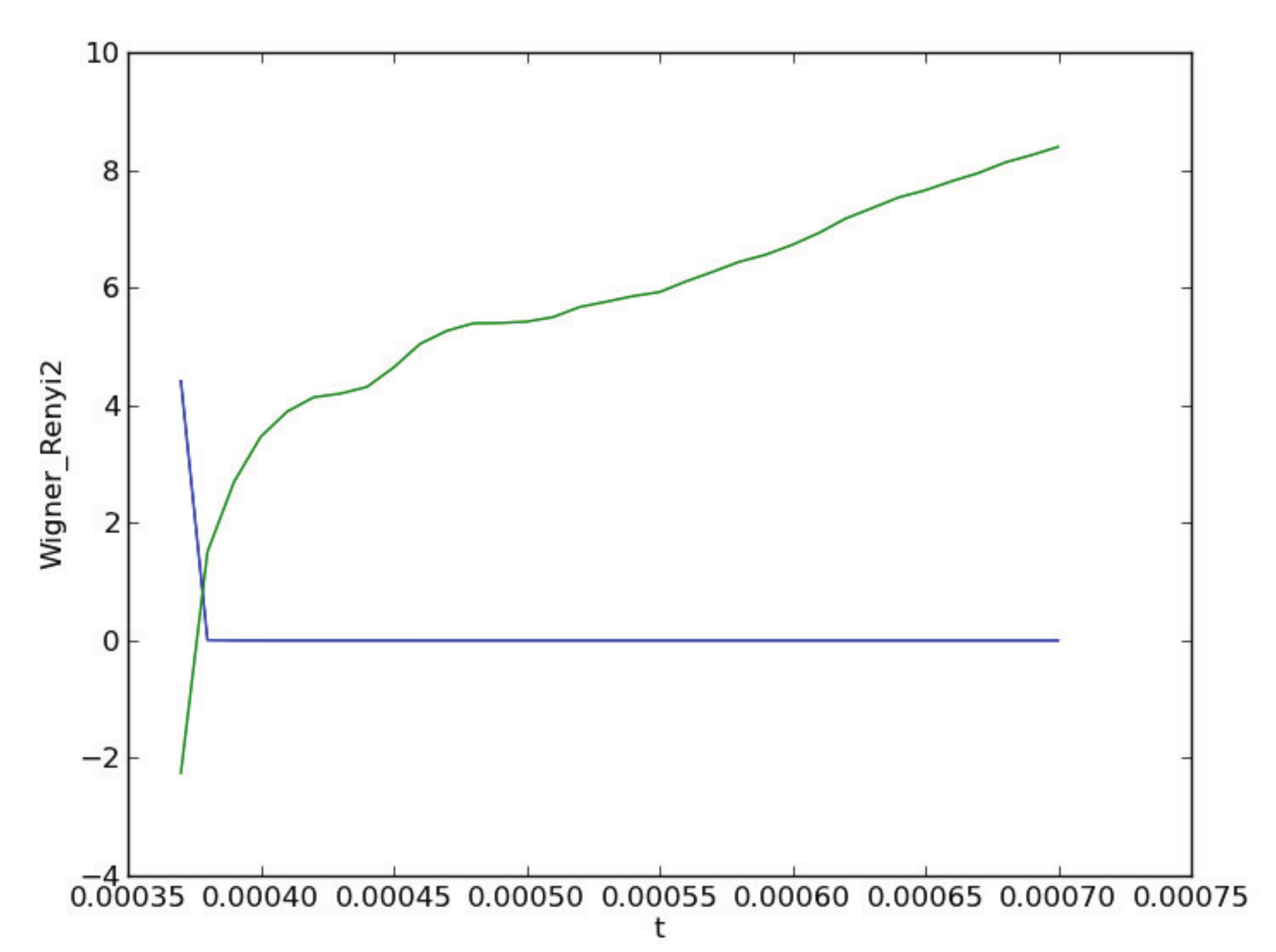}
\end{center}
\singlespace\caption{Evolution of  $\mathcal{A}_w$ (decreasing) and $S_2$ (increasing) for the two Gaussian wave packet of
scalar probe  for  $\Delta_{\mathcal{O}}=10/3$. The discretized time unit is $10^{-5}$ and $\Gamma_w=4000$.}
\label{python-AS2}
\end{figure}

   Here we show a typical example computed by our python code
\cite{python}: the retarded Green's function is
given by \eq{S01} with $\Delta_{\mathcal{O}}=\frac{10}{3}$. Note that  for this case \eq{hinverseLap} can only be solved numerically. The results for $A_w(t)$
and $S_2(t)$ are shown in Fig. \ref{python-AS2}. It again shows the decoherence behavior as
expected, however, the detailed behavior depends on the discretized time unit and $\Gamma_w$.\footnote{Despite the efficiency of this numerical code, we still need to discretize the time (with the main bottleneck being the integral-differential equation \eq{heom-main}),
which leads to a discretized retarded Green's function.
Therefore, the final results depend on the unit scale of the discretized time and would be different from the continuum time limit. }

The python code still needs to be improved, in particular, its sensitivity to the choice of the (discretized) time unit and of the window function needs to be further reduced. Once this is achieved, we could use it to search for the robust systems against generic quantum decoherence --- a necessary ingredient in designing real quantum computers.

\section{Conclusion}\label{sec6}

In this paper, based on the earlier observation \cite{Su:1987pi,Boyanovsky:2004dj} we elucidate
the connection between Feynman-Vernon and Schwinger-Keldysh formalisms: The influence functional
of Feynman-Vernon is the generating function of the Schwinger-Keldysh correlation functions. We
then apply this connection to study quantum decoherence when the `environment' is a holographic CFT, since the Schwinger-Keldysh retarded Green's function can be obtained holographically.  We consider
two cases for which the influence functional can be computed exactly, and derive the time evolution of the `system''s reduced density matrix, from which we then evaluate its Wigner function and R\'enyi entropy. We use the disappearance of the negative part of Wigner function to characterize the decoherence process and define the decoherence time. We then extract
the scaling behaviors of the decoherence time scale. Although we focus on these two particular examples, we expect the results to be rather universal for strongly-coupled CFT `environment'. The second example also proves that the quantum decoherence occurs even at zero temperature.

  The ability to determine the dynamics of the reduced density matrix allows us to directly compute the R\'enyi entropy which was used to characterize the quantum quench.  As explained earlier, the setup for the
quantum decoherence considered in this paper is very similar to that of the
local quantum quench. We demonstrate the close relation between these two quantum processes by
comparing their characteristic time scales and the scaling behaviors (with time) of their R\'enyi entropies. Our results indicate that the local quantum quench occurs when the local excitations decohere.
Moreover, we show that, for the quantum decoherence process, the long term scaling behavior of the R\'enyi entropy deviates from the logarithmic growth exhibited by the local quantum quench process. It would be interesting to further explore the relation between these two dynamical quantum phenomena.

Now a few words on the relevance of the present work to the design of realistic quantum computers. Although there are many different lab realizations of qubits, such as iron traps, liquid NMR, optical lattices, to name just a few, in order to have large number of qubits in the `system', solid-state based designs, such as semi-conductors and superconductors, are more promising candidates. Among solid-state structures in the lab, many are strongly-coupled and have (at least approximated) holographic duals. The method introduced in this paper can be applied to study the quantum decoherence in any strongly-coupled `environment' that has a holographic dual; therefore it is useful in the systematic study of the general behaviors of quantum decoherence and could hopefully provide some guidance in selecting the type of environments that are more coherence-friendly.



The most interesting open problem is how to incorporate the probe holographically. Answering this question would also solve the following technical problem. In this paper, we needed to implement numerical computations to determine the
time evolution of the reduced density matrix, which require solving integral-differential
equations and usually cause numerical artifacts. To bypass this technical
difficulty, one could try to find a method to directly compute the propagating function \eq{rho-dy} holographically, which is possible once we know how to describe the probe $\phi$ holographically.    We
hope to explore this possibility in the near future.

\medskip
\textbf{Acknowledgements.}
We thank Sung-Po Chao, Chung-Hsien Chou, Jen-Tsung Hsiang, Bei-Lok Hu, Matisse Tu and Wei-Min Zhang for very
helpful discussions. We also thank Ying-Jer Kao for providing valuable help in writing the
python code. FLL and BN are supported by NSC grant: 100-2112M-003-003-MY3, 101-2811-M-003-005,
102-2811-M-003-015.

\appendix

\section{Evaluating the propagating function: a review}\label{app-a}

In this appendix, we evaluate the propagating function $J$ by regarding it as a partition function \eq{Jrho-ra} with action \eq{seff-main}. This has been done before, e.g. in \cite{Grabert:1988yt,Hu:1991di}. Now we review the derivation to make the paper self-contained and to fix the notation.

First, we go to the momentum space, in which the effective action \eq{seff-main} becomes
\begin{equation}
\begin{aligned}
S_{\textrm{eff}} =& \sum_{\vec{k}} \int_0^t d\tau \left[\dot{\Sigma}_{\vec{k}} \dot{\Delta}_{\vec{k}} -
(\Omega^2-|\vec{k}|^2) \Sigma_{\vec{k}}\Delta_{\vec{k}}\right]
  \\ &-   g^2 \sum_{\vec{k}}\int_0^t d\tau  \int_0^t d\tau' \left[\Delta_{\vec{k}}
 (\tau)G_{\textrm{R}}^{(\vec{k})}(\tau-\tau')\Sigma_{\vec{k}}(\tau')-{i\over 2} \Delta_{\vec{k}}(\tau)
 G^{(\vec{k})}_{\textrm{sym}}(\tau-\tau') \Delta_{\vec{k}}(\tau') \right] \;.\label{seff0}
\end{aligned}
\end{equation}
As explained in the main text, now we will only consider the $\vec{k}=0$ mode to focus our attention on the time-dependent behavior. Then evaluating \eq{Jrho-ra} only takes two steps: first we expand the action \eq{seff0} around the
classical saddle points, then we integrate out the fluctuations to obtain the overall
normalization factor.  Expanding $\Sigma$ and $\Delta$ as follows:
\be\label{expan1}
\Sigma={\bf \Sigma}+q_1\;, \qquad \Delta(\tau)={\bf \Delta}+q_2
\ee
where ${\bf \Sigma}$ and ${\bf \Delta}$ are the classical solutions that solve the EOMs derived
from the real part of \eq{seff0}:
\bea\label{eom1}
&& \ddot{{\bf \Sigma}}(\tau)+\Omega^2 \,{\bf \Sigma}(\tau)+ g^2 \int_0^\tau d\tau' G_{\textrm{R}}(\tau-\tau') {\bf
\Sigma}(\tau')=0\;, \\ \label{eom2}
&& \ddot{{\bf \Delta}}(\tau)+\Omega^2 \,{\bf \Delta}(\tau)+ g^2 \int_\tau^t d\tau' G_{\textrm{R}}(\tau'-\tau) {\bf
\Delta}(\tau')=0
\eea
with the boundary conditions (in time)
\begin{equation}\label{initialfinal}
\begin{aligned}
{\bf \Sigma}(\tau=0)=\Sigma_0\equiv \tilde{\Sigma}\,,& \qquad {\bf \Sigma}(\tau=t)=\Sigma_1\equiv \bar{\Sigma}\;, \\
{\bf \Delta}(\tau=0)=\Delta_0 \equiv \tilde{\Delta}\,,& \qquad {\bf \Delta}(\tau=t)=\Delta_1 \equiv \bar{\Delta}\;.
\end{aligned}
\end{equation}
The time variable $\tau \in [0,t]$.  Note that for the ease of reading, in the derivation of this Appendix we use $\{\Sigma_{0,1},\Delta_{0,1}\}$ but will switch to the tilde and bar notation in the main text.

Now let us solve the boundary value problem \eq{eom1}-\eq{initialfinal}. First, note that \eq{eom1} and \eq{eom2} are related by a change of variable $\tau \rightarrow t-\tau$, therefore we only need to solve one of them. The matching of the boundary data can be done as follows. First we separate the boundary data out of $\{\mathbf{\Sigma},\mathbf{\Delta}\}$ by introducing $\{f_{0,1},g_{0,1}\}$:
\be
{\bf \Sigma}(\tau)= f_0(\tau) \Sigma_0+f_1(\tau) \Sigma_1  \qquad \mbox{and} \qquad
{\bf \Delta}(\tau)=g_0(\tau) \Delta_0+g_1(\tau) \Delta_1
\ee
then $\{f_{0,1}\}$ are two solutions of \eq{eom1} and $\{g_{0,1}\}$ of \eq{eom2}, and with the boundary conditions
\bea
\label{A18-2}
f_0(0)=1\,,\qquad f_0(t)=0\,; \qquad \qquad f_1(0)=0\,,\qquad f_1(t)=1\;, \\
\label{A18-3}
g_0(0)=0\,,\qquad g_0(t)=1\,; \qquad \qquad g_1(0)=1\,,\qquad g_1(t)=0\;.
\eea
Combining \eq{eom1}, \eq{eom2} with \eq{A18-2} and \eq{A18-3}, we see
\be
g_0(\tau)=f_1(t-\tau) \qquad  \mbox{and}  \qquad g_1(\tau)=f_0(t-\tau)\;.
\ee
Therefore we only need to solve for $f_i$'s.

$f_i$'s satisfy the integral equation
\begin{equation}\label{eomf}
\ddot{f}(\tau)+\Omega^2 \,f(\tau)+ g^2 \int_0^\tau d\tau' G_{\textrm{R}}(\tau-\tau') f(\tau')=0
\end{equation}
with initial and final values given by \eq{A18-2}. 
However a solution to \eq{eomf} can be uniquely determined either by fixing the initial and final values (boundary value problem) or the initial value plus its derivative (initial value problem); since a boundary value problem usually requires the shooting method therefore is rather difficult, we recast this boundary value problem into an initial value problem (by shifting the information at $\tau=t$ to $\tau=0$) following \cite{Grabert:1988yt}.

To convert \eq{eomf} into an initial value problem, let's first apply the Laplace transform on it, suggested by the convolution term.
The Laplace transform $\hat{f}\equiv \mathcal{L}(f)$ of $f$ satisfies
\begin{equation}\label{eomfLaplace}
\hat{f}(s)=\frac{1}{s^2+\Omega^2+g^2 \hat{G}_{\textrm{R}}(s)}\dot{f}(0)+\frac{s}{s^2+\Omega^2+g^2 \hat{G}_{\textrm{R}}(s)}f(0)\;.
\end{equation}
The original boundary value problem is now recast into an initial value problem: for given $f(0)$ and $\dot{f}(0)$, $f(\tau)$ is uniquely determined as the inverse Laplace transform of \eq{eomfLaplace}. 
It has two fundamental solutions:
\begin{equation}\label{hsolve}
\begin{cases}
h(0)=0\qquad \dot{h}(0)=1 \qquad &\Longrightarrow \qquad h(\tau)=\mathcal{L}^{-1}\left(\frac{1}{s^2+\Omega^2+g^2 \hat{G}_{\textrm{R}}(s)}\right)\\
H(0)=1\qquad \dot{H}(0)=0 \qquad &\Longrightarrow \qquad H(\tau)=\mathcal{L}^{-1}\left(\frac{s}{s^2+\Omega^2+g^2 \hat{G}_{\textrm{R}}(s)}\right)
\end{cases}\;.
\end{equation}
We see $H(\tau)=\dot{h}(\tau)$, since $\hat{H}(s)=s\hat{h}(s)$ and $h(0)=0$.

Since the integral equation \eq{eomf} is linear, all its solutions are linear combinations of $h(\tau)$ and $\dot{h}(\tau)$. The two solutions to the boundary value problem \eq{A18-2} are therefore easily determined to be:
\begin{equation}\label{f0f1}
f_0(\tau)=-\frac{\dot{h}(t)}{h(t)}h(\tau)+\dot{h}(\tau)\;, \qquad \qquad f_1(\tau)=\frac{1}{h(t)}h(\tau)
\end{equation}
where $h(\tau)$ is the solution to the initial value problem
\begin{equation}\label{eomh}
\ddot{h}(\tau)+\Omega^2 \,h(\tau)+ g^2 \int_0^\tau d\tau' G_{\textrm{R}}(\tau-\tau') h(\tau')=0\;, \qquad \qquad h(0)=0\;, \qquad \dot{h}(0)=1\;.
\end{equation}
For simple $G_{\textrm{R}}$, \eq{eomh} can be easily solved via inverse Laplace transform $h(\tau)=\mathcal{L}^{-1}\left(\frac{1}{s^2+\Omega^2+g^2 \hat{G}_{\textrm{R}}(s)}\right)$. Note that due to the boundary condition \eq{A18-2}, $\{f_{0},f_1\}$ are actually functions of both $\tau$ and $t$, with $t$ entering as the boundary point.

  Substituting \eq{expan1} into \eq{seff0} (with $\vec{k}=0$ mode being considered) and use the
equations of motion \eq{eom1} and \eq{eom2}, one obtain
\be
S_{\textrm{eff}}\equiv  S_{\textrm{cl}}[{\bf \Sigma},{\bf \Delta}]+S_{\textrm{q}}[q_1,q_2]
\ee
with
\be\label{scl}
S_{\textrm{cl}}[{\bf \Sigma},{\bf \Delta}]=\dot{{\bf \Sigma}}(\tau){\bf \Delta}(\tau) |_0^t +{i\over 2}g^2 \int_0^t d\tau \int_0^t d
\tau' {\bf \Delta}(\tau) G_{\textrm{sym}}(\tau-\tau'){\bf \Delta}(\tau')
\ee
and
\begin{equation}
\begin{aligned}
S_{\textrm{q}}[q_1,q_2] =&   \int_0^t d\tau \left[\dot{q_1}(\tau)  \dot{q_2} (\tau)-\Omega^2 q_1(\tau) q_2(\tau)\right]
  \\ &- g^2 \int_0^t d\tau  \int_0^t d\tau' \left[q_2(\tau)G_{\textrm{R}}(\tau-\tau') q_1(\tau')-{i\over 2}
q_2(\tau) G_{\textrm{sym}}(\tau-\tau') q_2(\tau') \right]
\end{aligned}
 \end{equation}
Note also that one can obtain the same results by using the complex saddle point.

One can factor out the boundary data in $S_{\textrm{cl}}$ and obtain
\be\label{Sclm}
S_{\textrm{cl}}=\sum_{i=0}^1 [\dot{f}_i(t) \Sigma_i \Delta_1 - \dot{f}_i(0) \Sigma_i  \Delta_0]+
{i\over 2} \sum_{i=0}^1\sum_{j=0}^1 a_{ij}(t)  \Delta_i \Delta_j
\ee
with
\be\label{aij}
a_{ij}(t)\equiv g^2 \int_0^t d\tau \int_0^t d\tau' g_i(\tau) G_{\textrm{sym}}(\tau-\tau') g_j(\tau')\;.
\ee
On the other hand, to carry out the integration over the fluctuation, it is convenient to
rewrite $S_{\textrm{q}}$ into the following form
\be\label{Sqq}
iS_{\textrm{q}}=-\frac{1}{2} \int_0^t d\tau \int_0^t d\tau'\; \mathbf{Q}^T(\tau)  \mathbf{O}(\tau-\tau')
\mathbf{Q}(\tau')
\ee
where $\mathbf{Q}^T(t)= \left(\begin{array}{cc}q_2(t), & q_1(t)\end{array}\right)$ and
\be \label{B08-2}
\mathbf{O}(t-t')=\left(\begin{array}{cc} g^2 G_{\textrm{sym}}(\tau-\tau') & i[(\partial_\tau^2+\Omega^2)\delta(\tau-\tau')+ g^2 G_{\textrm{R}}(\tau-\tau')] \\ i[(\partial_\tau^2+\Omega^2)\delta(\tau-\tau')+ g^2 G_{\textrm{R}}(\tau-\tau')] & 0 \end{array}\right)\;.
\ee
In the Appendix \ref{app-normal}, we show that the normalization factor
\be \label{E44}
A(t) \equiv \int D q_1  D q_2 \; e^{iS_q[q_1,q_2]} =[\det\mathbf{O}]^{-1/2}= {1\over 2\pi |h(t)|}
\ee
with $h(t)$ given in \eq{hsolve}.

Combining the above results, we obtain the propagating function $J$
\be\label{propf}
J[\Sigma_1,\Delta_1,t|\Sigma_0,\Delta_0,0]= A(t)\; e^{i \sum_{i=0}^1 [\dot{f}_i(t) \Sigma_i
\Delta_1 - \dot{f}_i(0) \Sigma_i  \Delta_0] - {1\over 2} \sum_{i=0}^1\sum_{j=0}^1 a_{ij}(t)
\Delta_i \Delta_j }
\ee
with $a_{ij}(t)$ given in \eq{aij} and $A(t)$ in \eq{E44}.

In summary, the propagating function \eq{propf} can be obtained by just solving $h(\tau)$ and
$H(\tau)$ for a given $G_{\textrm{R}}$ using
\eq{eomh}. Then, the time evolution of the reduced density matrix is given by \eq{rho-dy}.

\section{Derivation of the master equation for $\hat{\rho}_{\textrm{sys}}(t)$}\label{mastereq}

Now we derive the master equation of $\hat{\rho}_{\textrm{sys}}(t)$.
Start with the propagating function \eq{propf} and recall that the function $f_i(t)$  in \eq{propf} is actually
a shorthand for $f_i(t) \equiv f_i(\tau=t,t)$ where $\{f_{0}(\tau,t),f_{1}(\tau,t)\}$ are
the function $\{f_{0},f_1\}$ given in \eq{f0f1}: $f_1(\tau,t) = {h(\tau) \over h(t)}$ and
$f_0(\tau,t)= \dot{h}(\tau)- {h(\tau) \over h(t)} \dot{h}(t)$ and the dot refers to the
derivative w.r.t. the variable $\tau$. Therefore we have
\begin{equation} \label{C03}
\begin{aligned}
 \dot{f}_1(t) & \equiv {d \over d\tau} f_1(\tau)|_{\tau =t} = {\dot{h}(t) \over h(t)}\;,
 \qquad  \qquad \dot{f}_1(0)  = \frac{1}{h(t)}\;, \\
\dot{f}_0(t) & = \dot{H}(t) - {[H(t)]^2 \over h(t)}\;,
\qquad \qquad \dot{f}_0(0)= - {\dot{h}(t) \over h(t)} = - \dot{f}_1(t) \;.
\end{aligned}
\end{equation}
Now evaluate the  $t$-derivative of the propagating function \eq{propf}:
\begin{equation} \label{02}
\begin{aligned}
i{d\over dt} J(t)  =&  i A'(t) e^{iS_{eff}}   \\
& + i \left(i \Delta_1\Sigma_1 \dot{f}'_1(t)+  i \Delta_1 \Sigma_0  \dot{f}'_0(t)- i \Delta_0
\Sigma_1  \dot{f}'_1(0)-i \Delta_0 \Sigma_0  {d\over dt }\dot{f}_0(0)\right)J(t)    \\
 &  + i \left(-\frac{1}{2} \Delta_1^2  a'_{11}(t) -   \Delta_1 \Delta_0  \frac{a'_{10}(t)
 +a'_{01}(t)}{2}  -\frac{1}{2} \Delta_0^2 a'_{00}(t)  \right) J(t)
\end{aligned}
\end{equation}
where the prime denotes $t$-derivative, with
\begin{equation} \label{03}
\begin{aligned}
& \dot{f}'_1(t)  =\frac{d}{dt} \dot{f}_1(t) = \frac{d}{dt} \left[ {\dot{h}(t) \over h(t)}
 \right] =\ddot{f}_1(t) - [\dot{f}_1(t)]^2\;, 
\qquad \qquad \dot{f}'_1(0) = - \dot{f}_1(t) \dot{f}_1(0)\;,  \\
& \dot{f}'_0(t) = \ddot{f}_0(t) + \dot{f}_0(t) \dot{f}_0(0)\;, 
\qquad \qquad \dot{f}'_0(0)  = -\dot{f}_0(t) \dot{f}_0(0)\;, \\
& A'(t)  =  -{h'(t)\over h(t)^2} = -\dot{f}_1(t) A(t)\;.
\end{aligned}
\end{equation}

\bes \label{04}
Recall that $f_i(\tau,t)$ satisfy the equation of motion
\beq \label{04-1}
{d^2 \over d\tau^2} f(\tau) +\Omega^2 f(\tau)+ g^2 \int_0^{\tau} ds \ G_{\textrm{R}}(\tau-s)f(s)=0
\eeq
with boundary condition $f_1(\tau=0,t)=f_0(\tau=t,t)=0$ and $f_1(\tau=t,t)=f_0(\tau=0,t)=1$. Using this we compute the double derivatives in \eq{03}:
\begin{equation}
\label{04-2}
\begin{aligned}
\ddot{f}_1(t) &\equiv {d^2 \over d\tau^2} f_1(\tau)|_{\tau=t}
 =  -\Omega^2 - \int_0^t ds \ G_{\textrm{R}}(t-s)f_1(s) = -\Omega^2 - b_1(t)\;, \\
\ddot{f}_0(t) & \equiv{d^2 \over d\tau^2} f_0(\tau)|_{\tau=t}
 = - g^2 \int_0^t ds \ G_{\textrm{R}}(t-s)f_0(s)  = - b_0(t)
\end{aligned}
\end{equation}
where $b_i(t) \equiv g^2 \int^t_0 ds \ G_{\textrm{R}}(t-s)f_i(s)$.
\ees

Now let us derive some useful formulae:
\bes \label{14}
\beq
\label{14-1}
\frac{\partial}{\partial \Sigma_1} J  =&& [ i \dot{f}_1(t) \Delta_1 - i \dot{f}_1(0) \Delta_0 ]
J\;, \\
\label{14-2}
\frac{\partial}{\partial \Delta_1} J  = &&[ i \dot{f}_1(t) \Sigma_1 + i \dot{f}_0(t) \Sigma_0 -
a_{11} \Delta_1 -  \frac{a_{10}+a_{01}}{2} \Delta_0 ] J\;,  \\
\label{14-3}
\frac{\partial}{\partial \Sigma_1}\frac{\partial}{\partial \Delta_1} J  = && [ i \dot{f}_1(t)  -
[\dot{f}_1(t)]^2 \Delta_1 \Sigma_1 - \dot{f}_1(t) \dot{f}_0(t) \Delta_1 \Sigma_0 + \dot{f}_1(t)
\dot{f}_1(0) \Sigma_1 \Delta_0 + \dot{f}_0(t) \dot{f}_1(0) \Sigma_0\Delta_0 ] J \nonumber \\
&&+[ i a_{11} \Delta_1 ( \dot{f}_1(0) \Delta_0 - \dot{f}_1(t)\Delta_1) + \frac{i}{2} ( a_{10}
+a_{01} ) \Delta_0 (\dot{f}_1(0) \Delta_0 -\dot{f}_1(t)\Delta_1)]J \nonumber \\
 = && [ i \dot{f}_1(t)  - [\dot{f}_1(t)]^2 \Delta_1 \Sigma_1 - \dot{f}_1(t) \dot{f}_0(t)
\Delta_1 \Sigma_0 + \dot{f}_1(t)\dot{f}_1(0) \Sigma_1 \Delta_0 + \dot{f}_0(t) \dot{f}_1(0)
\Sigma_0\Delta_0 ] J \nonumber \\
&& - [\frac{i}{2} ( a_{10} +a_{01} )  \Delta_0 \frac{\partial}{\partial \Sigma_1}+a_{11}
\Delta_1 \frac{\partial}{\partial\Sigma_1} ] J\;.
\eeq
\ees
With (\ref{03}), (\ref{04}) and (\ref{14-3}), the first two lines in (\ref{02}) can be reduced into
\begin{equation} \label{15}
-\frac{\partial}{\partial \Sigma_1}\frac{\partial}{\partial \Delta_1}J + \Omega^2
\Delta_1\Sigma_1 J + b_1 \Sigma_1\Delta_1 J + b_0 \Sigma_0 \Delta_1 J- [\frac{1}{2} ( a_{10}
+a_{01} )  \Delta_0 \frac{\partial}{\partial \Sigma_1}+a_{11} \Delta_1 \frac{\partial}{\partial
\Sigma_1} ] J\;.
\end{equation}
Then using
\begin{equation} \label{03-1}
 a'_{22}(t)= g^2 \frac{d}{dt} \int^t \int^t ds ds' g_2(s) G_{sym} (s-s') g_2(s') = 2 g^2 g_2(t)
 \int^t G_{sym} (t-s)g_2(s) =0
\end{equation}
and (\ref{14-1}), we can rewrite the last line in (\ref{02}) into
\beq \label{17}
-i \left( {1 \over 2} a'_{11}+\frac{a'_{10}+a'_{01}}{2} \frac{\dot{f}_1(t)}{\dot{f}_1(0)}\right)
\Delta_1^2 J+ \frac{a'_{10}+a'_{01}}{2\dot{f}_1(0)} \Delta_1 \frac{\partial}{\partial \Sigma_1}J\;.
\eeq
Finally, combining (\ref{15}) and (\ref{17}) and using (\ref{14}) to replace all $\Sigma_0$ and
$\Delta_0$, we obtain
\beq \label{16}
i \frac{d}{dt} J && = \left[ -\frac{\partial}{\partial \Sigma_1}\frac{\partial}{\partial
\Delta_1}+ (\Omega^2 +\delta \Omega^2) \Delta_1\Sigma_1 \right]J  + \left(-i
\frac{b_0}{\dot{f}_0(t)} \right) \Delta_1 \frac{\partial}{\partial \Delta_1} J \nonumber \\
&& + \left[ \frac{\dot{a}_{10}+\dot{a}_{01}}{2\dot{f}_1(0)} - a_{11} -
\frac{2 \dot{f}_1(t)}{\dot{f}_1(0) (a_{10}+a_{01})}+
\frac{ b_0 (a_{10}+a_{01} )}{2\dot{f}_1(0) \dot{f}_0(t)} \right] \Delta_1
\frac{\partial}{\partial \Sigma_1} J \nonumber \\
&& + \left[ -i \left(  \frac{b_0 a_{11}}{\dot{f}_0(t)}+\frac{a_{10}\dot{f}_0(t)}{\dot{f}_1(t)
\dot{f}_1(0)} + \frac{1}{2} \dot{a}_{11} +\frac{\dot{a}_{10}+\dot{a}_{01}}{2}
\frac{\dot{f}_1(t)}{\dot{f}_1(0)}         \right)     \right]\Delta_1^2  J  \nonumber \\
&& -i \frac{a_{10}+a_{01}}{2\dot{f}_1(0)} \frac{\partial^2}{\partial \Sigma_1^2} J
\eeq
where $\delta\Omega^2 \equiv b_1 - b_0 \frac{\dot{f}_1(t)}{\dot{f}_0(t)}$.
Then it is straightforward to write down the master equation for the reduced density matrix
$\rho_{\textrm{sys}}(\Sigma_1,\Delta_1,t)$ from \eq{16} and \eq{rho-dy}.

\section{Normalization of $\hat{\rho}_{\textrm{sys}}$}\label{app-normal}

\subsection{Evaluating the normalization $A(t)$ using $\textrm{Tr}\hat{\rho}_{\textrm{sys}}(t)=1$}

We now use the propagating functional given by (\ref{propf}) and the condition
$\Tr \hat{\rho}_{\textrm{sys}}(t)=1$ to evaluate the normalization factor $A(t)$. Here we assume the
initial density matrix is a Gaussian wave packet.
\beq \label{App01}
\langle \tilde{\phi}_+|\hat{\rho}_{\textrm{sys}}(0)|\tilde{\phi}_- \rangle=\frac{1}{\sigma \sqrt{\pi}}
e^{-\frac{\tilde{\phi}_+^2+\tilde{\phi}_-^2}{2\sigma^2}}=\frac{1}{\sigma \sqrt{\pi}} e^{-
\frac{\Delta_0^2+4\Sigma_0^2}{4\sigma^2}}
\eeq
where the normalization factor $\sigma$ is fixed by the unitarity condition Tr$\hat{\rho}_{sys}(0)=1$.

The reduced density matrix is
\beq \label{App02}
\langle \bar{\phi}_+|\hat{\rho}_{\textrm{sys}}(t)|\bar{\phi}_-\rangle  = && \int d\Sigma_0\; d\Delta_0 \;
J[\Sigma_1,\Delta_1,t | \Sigma_0, \Delta_0, 0]   \langle \tilde{\phi}_+|\hat{\rho}_{sys}(0)|
\tilde{\phi}_- \rangle  \\
 =  &&A(t)  e^{-\frac{1}{2} a_{11}(t) \Delta_1^2+i\dot{f}_1(t)\Delta_1\Sigma_1-
 \frac{\sigma^2}{4} \dot{f}_0(t)^2 \Delta_1^2} \nonumber \\
&& \cdot \int d\Delta_0e^{-\left( \frac{\sigma^2}{4} \dot{f}_0(0)^2+\frac{1}{4\sigma}+
\frac{1}{2} a_{00}(t)\right)\Delta_0^2}  e^{\left(\frac{\sigma^2}{2}\dot{f}_0(t)\dot{f}_0(0)\Delta_1-\frac{1}{2} a_{10}(t) -
\frac{1}{2} a_{01}(t) \Delta_1-i\dot{f}_1(0)\Sigma_1\right)\Delta_0}   \nonumber \\
 = && \sqrt{\pi} C^{-1/2}A(t)  e^{-\frac{1}{2} a_{11}(t) \Delta_1^2+i\dot{f}_1(t)
\Delta_1\Sigma_1-\frac{\sigma^2}{4} \dot{f}_0(t)^2 \Delta_1^2} \nonumber \\
&& \times  \exp  \left[ \frac{\left(-i\dot{f}_1(0)\Sigma_1 + \frac{\sigma^2}{2}\dot{f}_0(t)
\dot{f}_0(0)\Delta_1-\frac{1}{2} (a_{10}(t)+a_{01}(t)) \Delta_1 \right)^2}{4C} \right] \nonumber
\\
=&& \sqrt{\pi} C^{-1/2}A(t) e^{-{1 \over L_{\Sigma}^2} \Sigma_1^2 - { i \over L_c^2}
\Sigma_1\Delta_1 - {1 \over L_{\Delta}^2} \Delta_1^2}
\eeq
where the characteristic length  $L_{\Sigma}$, $L_c$,  $L_{\Delta}$ and $C$ are defined in
\eq{TG07}.

Now using $\Tr \hat{\rho}_{\textrm{sys}}(t)$=1, we can derive the normalization factor $A(t)$.
\beq \label{App04}
1= \Tr\hat{\rho}_{\textrm{sys}}(t) && = \int d\phi d\phi' <\phi|\hat{\rho}_{sys}(t)|\phi'>
\delta(\phi - \phi' )\;, \\ \nonumber
&&= \int d\Sigma_1 d\Delta_1 \delta(\Delta_1) <\hat{\rho}_{sys}(t)>\;,  \\ \nonumber
&&= \int d\Sigma_1 \sqrt{\pi} A(t)  C^{-1/2} \exp
\left[ \frac{\left(-i\dot{f}_1(0)  \right)^2}{4C}\Sigma_1^2 \right]\;,   \\ \nonumber
&&= A \cdot 2\pi \frac{1}{|\dot{f}_1(0)|}\;.
\eeq

Using (\ref{C03}), we have
\beq \label{App05}
A(t)= \frac{1}{2\pi |h(t)|}.
\eeq

\subsection{A consistency check: Evaluating the normalization through functional determinant}
In this subsection, we compute the normalization factor $A(t)$ by evaluating the functional determinant in (\ref{E44}) directly.
We follow the formulation in
\cite{Alan}.

Suppose we want to compute the determinant of an operator $\hat{L}$
\beq \label{App06}
\hat{L}=\partial_t^2+P(t)\,,\qquad \quad \mbox{with}\quad t\in [ t_0, t_f]
\eeq
where $P(t)$ is a real function and the boundary condition for the eigenfunction of $\hat{L}$ is
$u(t_0)=u(t_f)=0$. The prescription takes only two steps.

First we write the boundary condition as
\bes \label{App07}
\beq \label{App07-1}
M\left(\begin{array}{c}u(t_0) \\\dot{u}(t_0)\end{array}\right)+N\left(\begin{array}{c}u(t_f) \\
\dot{u}(t_f)\end{array}\right)=\left(\begin{array}{c}0 \\ 0 \end{array}\right)
\eeq
where $M$ and $N$ are $2\times 2$ matrices and in our case
\beq \label{App07-2}
M=\left(\begin{array}{cc}1 & 0 \\0 & 0\end{array}\right) \mbox{ and } N=\left(\begin{array}{cc}0
& 0 \\1 & 0\end{array}\right)\;.
\eeq
Then we define matrices $\hat{H}(t)$ and $\hat{Y}(t_f)$ as
\be \label{App07-3}
\hat{H}(t) \equiv \left(\begin{array}{cc}y_1(t) & y_2(t) \\\dot{y}_1(t) & \dot{y}_2(t)\end{array}
\right)\;, \qquad \hat{Y}(t_f)\equiv \hat{H}(t_f) \hat{H}^{-1}(t_0)
\ee
where $y_1(t)$ and $y_2(t)$ are two independent solutions of the homogeneous differential
equation $\hat{L}\left(\begin{array}{c}y_1 \\y_2\end{array}\right)=0$.

The determinant $\det \hat{L}$ is certainly divergent, so the only meaningful
quantity is the renormalized determinant, i.e. the ratio (\ref{App07-5}). Usually we chose $
\bar{L}=\partial_t^2$ as in \cite{Kleinert}.
Finally we have the determinant ratio of $\hat{L}$ and some other (similar type of) operator
$\bar{L}$:
\beq \label{App07-5}
\frac{\det \hat{L}}{\det \bar{L}}= \frac{\det [M+N\hat{Y}(t_f)]}{\det [M+N\bar{Y}(t_f)]}\;.
\eeq
\ees
Here the matrices $M$ and $N$ are given in (\ref{App07-2}) for our case. The determinant is
\beq \label{App08}
\det [M+N\hat{Y}(t_f)]= \frac{y_1(t_0)y_2(t_f)-y_2(t_0)y_1(t_f)}{y_1(t_0)\dot{y}_2(t_0)-
y_2(t_0)\dot{y}_1(t_0)}\;.
\eeq
If we pick $y_1(t)$ being the solution whose initial condition is $y_1(t_0)=0$, the
determinant ratio (\ref{App07-5}) becomes
\beq \label{App09}
\frac{\det \hat{L}}{\det \bar{L}}= \frac{1}{\det \bar{L}}\frac{y_1(t_f)}{\dot{y}_1(t_0)}\;.
\eeq

Now the normalization factor obtained by integrating out the quadratic action \eq{Sqq} is
\beq \label{App12}
A(t)=[\det(\mathbf{O})]^{-1/2}={1\over |\det[(\partial_\tau^2+\Omega^2)\delta(\tau-\tau')
+G_{\textrm{R}}(\tau-\tau')]|}
\eeq
with $\mathbf{O}(t-t')$ defined in \eq{B08-2}.

 It is obvious that the solution $y(t)$ in (\ref{App08}) is given by the solution $f(t)$ in
(\ref{f0f1}). Hence the normalization factor (\ref{App12}) becomes
\beq \label{App14}
A(t)^{-1}=  \frac{|f_1(t)|}{|\dot{f}_1(0)|} =\left(\frac{|\dot{h}(0)|}{|h(t)|}\right)^{-1}=
\left(\frac{1}{2\pi }\frac{1}{|h(t)|}\right)^{-1}
\eeq
which agrees with (\ref{App05}).

\subsection{Another consistency check: The behavior of $J(t)$ and $\hat{\rho}_{\textrm{sys}}(t)$ as
$t \rightarrow 0$}

As a second consistency check, we compute the behavior of the propagating function at $t\rightarrow 0$  and check that it indeed recovers a Dirac delta function. Using (\ref{C03}) we have
\begin{equation}
\label{App15}
\begin{aligned}
 &\lim_{t=\epsilon \to 0} \dot{f}_1(0,t)|= \frac{1}{\epsilon} \;.
 \qquad \qquad \lim_{t=\epsilon \to 0} \dot{f}_1(t,t)  = \frac{1}{\epsilon}\;, \\
 &\lim_{t=\epsilon \to 0} \dot{f}_0(0,t)   = -\frac{1}{\epsilon} \;,
 \qquad \qquad \lim_{t=\epsilon \to 0} \dot{f}_0(t,t)  = -\frac{1}{\epsilon}\;,  \\
  &\lim_{t=\epsilon \to 0} A(t)  = \frac{1}{2\pi \epsilon}\;,
\end{aligned}
\end{equation}
and $a_{ij}(t \rightarrow 0) \rightarrow 0$. The propagatinf functional (\ref{propf})
becomes
\begin{equation} \label{A16}
\begin{aligned}
J(t \to 0)  \approx
\lim_{t=\epsilon \to 0} \frac{e^{i \left( \frac{1}{\epsilon} (\Delta_1
- \Delta_0)(\Sigma_1 - \Sigma_0)   \right)}}{2\pi \epsilon}   &=  \lim_{t=\epsilon \to 0} \frac{e^{i \frac{1}{\epsilon}
(\bar{\phi}_+ -\tilde{\phi}_+)^2} }{\sqrt{2\pi i \epsilon}}   \cdot \frac{e^{-i
\frac{1}{\epsilon}(\bar{\phi}_- -\tilde{\phi}_-)^2}}{\sqrt{2\pi (-i) \epsilon}}    \\
& = \delta(\bar{\phi}_+ -\tilde{\phi}_+) \cdot \delta(\bar{\phi}_- -\tilde{\phi}_-)\;.
\end{aligned}
\end{equation}
(\ref{A16}) again shows that our propagating functional is properly normalized.

Similarly, we expect the reduced density matrix to reduce to (\ref{App01}) when $t \to 0$. Using
(\ref{App15}) and $a_{ij}(t \to 0) \to 0$,  the exponent in (\ref{App02}) is
\begin{equation} \label{App17}
\begin{aligned}
- \frac{1}{L_{\Sigma}^2} \Sigma_1^2 - {i \over L_c^2} \Sigma_1\Delta_1 -
{1 \over L_{\Delta}^2} \Delta_1^2
&\longrightarrow  \frac{1}{4C} \left( -\frac{1}{\epsilon^2} \Sigma_1^2 - i \sigma^2
\frac{1}{\epsilon^3} \Sigma_1\Delta_1 - \frac{1}{4} \frac{1}{\epsilon^2} \Delta_1^2 +4i C
\frac{1}{\epsilon} \right)   \\
&\qquad = \frac{-1}{\sigma^2} \left( \Sigma_1^2 + \frac{1}{4} \Delta_1^2 \right)
\end{aligned}
\end{equation}
where we used $C \rightarrow \frac{\sigma^2}{4} \frac{1}{\epsilon^2}$. And the normalization factor in (\ref{App02}) takes the form
\beq \label{App18}
\sqrt{\pi} C^{-1/2}A(t) \rightarrow \left( \frac{\sigma^2}{4} \frac{1}{\epsilon^2}\right)^{-1/2}
\frac{1}{2 \pi \epsilon} = \frac{1}{\sigma \sqrt{\pi}}\;.
\eeq
Combining (\ref{App17}) and (\ref{App18}), we reproduce (\ref{App01}).


\begin{thebibliography}{99}



\bibitem{Zurek}
W. H. Zurek, Decoherence and the transition from quantum to classical, Physics Today 44 (10) 36-44  (1991)

W. H. Zurek, Decoherence and the transition from quantum to classical - Revisited, Los Alamos Science 27, 86-109 (2002)

W. H. Zurek, Decoherence, einselection, and the quantum origins of the classical, Reviews of Modern Physics, 75, 715-765 (2003)



\bibitem{Bohr}
N.~Bohr, Nature {\bf 121}, 580 (1928).

\bibitem{many_worlds}
 H.~Everett,
  ``Relative state formulation of quantum mechanics,''
  Rev.\ Mod.\ Phys.\  {\bf 29}, 454 (1957).

 J.~A.~Wheeler,
  ``Assessment of Everett's 'Relative State' Formulation of Quantum Theory,''
  Rev.\ Mod.\ Phys.\  {\bf 29}, 463 (1957).


\bibitem{Brune:1996zz}
  M.~Brune, E.~Hagley, J.~Dreyer, X.~Maitre, A.~Maali, C.~Wunderlich, J.~M.~Raimond and S.~Haroche,
  ``Observing the Progressive Decoherence of the 'Meter' in a Quantum Measurement,''
  Phys.\ Rev.\ Lett.\  {\bf 77}, 4887 (1996).



\bibitem{Feynman:1963fq}
  R.~P.~Feynman and F.~L.~Vernon, Jr.,
  ``The theory of a general quantum system interacting with a linear dissipative system,''
  Annals Phys.\  {\bf 24}, 118 (1963)

\bibitem{Caldeira:1982iu}
  A.~O.~Caldeira and A.~J.~Leggett,
  ``Path integral approach to quantum Brownian motion,''
  Physica {\bf 121A}, 587 (1983).

\bibitem{Grabert:1988yt}
  H.~Grabert, P.~Schramm, G.~L.~Ingold,
  ``Quantum Brownian motion: The functional integral approach,''
  Phys.\ Rept.\  {\bf 168}, 115 (1988).


\bibitem{Hu:1991di}
  B.~L.~Hu, J.~P.~Paz and Y.~-h.~Zhang,
  ``Quantum Brownian motion in a general environment: 1. Exact master equation with nonlocal dissipation and colored noise,''
  Phys.\ Rev.\ D {\bf 45}, 2843 (1992).

\bibitem{Weiss}

U.~Weiss, ``Quantum Dissipative Systems", 3rd Edition by World Scientific Publishing Co. Pte. Ltd., Singapore 2008

\bibitem{Kitaev:1997wr}
  A.~Y.~Kitaev,
  ``Fault tolerant quantum computation by anyons,''
  Annals Phys.\  {\bf 303}, 2 (2003).
  A.~Kitaev,
  ``Anyons in an exactly solved model and beyond,''
  Annals Phys.\  {\bf 321}, no. 1, 2 (2006).

\bibitem{Nayak:2008zza}
  C.~Nayak, S.~H.~Simon, A.~Stern, M.~Freedman and S.~Das Sarma,
  ``Non-Abelian anyons and topological quantum computation,''
  Rev.\ Mod.\ Phys.\  {\bf 80}, 1083 (2008).

\bibitem{TO} Xiao-Gang Wen,
``Quantum Field Theory of Many Body Systems - From the Origin of Sound to an Origin of Light and Electrons", Oxford Univ. Press, Oxford, 2004.

\bibitem{Majorana}
F. Wilczek. Majorana Returns - 2009. Nature Phys.,5,614

\bibitem{Ho&Lin}
S.~-H.~Ho and F.~-L.~Lin,
 ``Anti-de Sitter Space as Topological Insulator and Holography,''
  arXiv:1205.4185 [hep-th].

S.~-H.~Ho, F.~-L.~Lin and X.~-G.~Wen,
  ``Majorana Zero-modes and Topological Phases of Multi-flavored Jackiw-Rebbi model,''
  JHEP {\bf 1212}, 074 (2012).


\bibitem{Environment}
W.~H.~Zurek,
  ``Pointer Basis of Quantum Apparatus: Into What Mixture Does the Wave Packet Collapse?,''
  Phys.\ Rev.\ D {\bf 24}, 1516 (1981).

W.~H.~Zurek,
  ``Environment induced superselection rules,''
  Phys.\ Rev.\ D {\bf 26}, 1862 (1982).






\bibitem{Hu:1993vs}
  B.~L.~Hu, J.~P.~Paz and Y.~Zhang,
  ``Quantum Brownian motion in a general environment. 2: Nonlinear coupling and perturbative approach,''
  Phys.\ Rev.\ D {\bf 47}, 1576 (1993).


\bibitem{fermion-1}
C. Anastopoulos, B.~L.~Hu,
``Two-level atom-field interaction: Exact master equations for non-Markovian dynamics,
decoherence, and relaxation", Phys.\ Rev.\ A {\bf 62}, 033821 (2000). 

S.~Shresta, C.~Anastopoulos, A.~Dragulescu, B.~L.~Hu,
``Non-Markovian qubit dynamics in a thermal field bath: Relaxation, decoherence and entanglement"
Phys.\ Rev.\ A {\bf 71}, 022109 (2005). 

\bibitem{fermion-2}
M.~W.-Y.~Tu and W.-M. Zhang,
``Non-Markovian decoherence theory for a double-dot charge qubit",
Phys.\ Rev.\ B {\bf 78}, 235311 (2008). 

J.~S.~Jin, M.~W.-Y.~Tu, W.-M. Zhang and Y.~J. Yan,
``Non-equilibrium quantum theory for nanodevices based on the Feynman-Vernon influence
functional", New J.\ Phys.\ 12 083013 (2010). 


\bibitem{fermion-3}
A.~Ghosh, S.~ S.~ Sinha and D.~S.~Ray,
``Fermionic oscillator in a fermionic bath",
Phys.\ Rev.\ E {\bf 86}, 011138 (2012).

\bibitem{NonMarkov}
C.~U.~Lei and W.-M.~Zhang,
``Decoherence suppression of open quantum systems through a strong coupling
to non-Markovian reservoirs", Phys.\ Rev.\ A {\bf 84}, 052116 (2011).

W.-M.~Zhang, P.-Y.~Lo, H.-N. Xiong, M.~W-Y.~Tu, F.~Nori,
``General non-Markovian dynamics of open quantum systems", Phys.\ Rev.\ Lett.\ {\bf 109}, 170402 (2012). 


\bibitem{Schwinger:1960qe}
  J.~S.~Schwinger,
  ``Brownian motion of a quantum oscillator,''
  J.\ Math.\ Phys.\  {\bf 2}, 407 (1961).



\bibitem{Keldysh:1964ud}
  L.~V.~Keldysh,
  ``Diagram technique for nonequilibrium processes,''
  Zh.\ Eksp.\ Teor.\ Fiz.\  {\bf 47}, 1515 (1964)
  [Sov.\ Phys.\ JETP {\bf 20}, 1018 (1965)].



\bibitem{Su:1987pi}
  Z.~-b.~Su, L.~-y.~Chen, X.~-t.~Yu and K.~-c.~Chou,
  ``Influence functional and closed-time-path Green's function,''
  Phys.\ Rev.\ B {\bf 37}, 9810 (1988).

\bibitem{Matsumoto:1982ry}
  H.~Matsumoto, Y.~Nakano, H.~Umezawa, F.~Mancini and M.~Marinaro,
  ``A Causal Formulation of Multipoint Functions at Finite Temperature,''
  Prog.\ Theor.\ Phys.\  {\bf 70}, 599 (1983).

\bibitem{Herzog:2002pc}
  C.~P.~Herzog, D.~T.~Son,
  ``Schwinger-Keldysh propagators from AdS/CFT correspondence,''
  JHEP {\bf 0303}, 046 (2003).

\bibitem{Boyanovsky:2004dj}
  D.~Boyanovsky, K.~Davey and C.~M.~Ho,
  ``Particle abundance in a thermal plasma: Quantum kinetics vs. Boltzmann equation,''
  Phys.\ Rev.\ D {\bf 71}, 023523 (2005).


\bibitem{KCChou}
  K.~-c.~Chou, Z.~-b.~Su, B.~-l.~Hao and L.~Yu,
  Phys.\ Rept.\  {\bf 118}, 1 (1985).

\bibitem{Son:2002sd}
  D.~T.~Son and A.~O.~Starinets,
  ``Minkowski space correlators in AdS / CFT correspondence: Recipe and applications,''
  JHEP {\bf 0209}, 042 (2002).

  \bibitem{SpinBath}
  Nikolay Prokof'ev and Philip Stamp, ``Theory of the spin bath," Rep.\ Prog.\ Phys.\ {\bf63}, 669 (2000). \\
  Nikolay Prokof'ev and Philip Stamp, ``Spin bath-mediated decoherence in superconductors," [cond-mat/0006054].



\bibitem{Martin:1959jp}
  R.~Kubo,
  ``Statistical mechanical theory of irreversible processes. 1. General theory and simple applications in magnetic and conduction problems,''
  J.\ Phys.\ Soc.\ Jap.\  {\bf 12}, 570 (1957).

  P.~C.~Martin and J.~S.~Schwinger,
  ``Theory of many particle systems. 1.,''
  Phys.\ Rev.\  {\bf 115}, 1342 (1959).




\bibitem{GKPW}
  S.~S.~Gubser, I.~R.~Klebanov and A.~M.~Polyakov,
  ``Gauge theory correlators from noncritical string theory,''
  Phys.\ Lett.\ B {\bf 428}, 105 (1998).

E.~Witten,
  ``Anti-de Sitter space and holography,''
  Adv.\ Theor.\ Math.\ Phys.\  {\bf 2}, 253 (1998).




\bibitem{Iqbal:2008by}
  N.~Iqbal and H.~Liu,
  ``Universality of the hydrodynamic limit in AdS/CFT and the membrane paradigm,''
  Phys.\ Rev.\ D {\bf 79}, 025023 (2009).





\bibitem{Israel:1976ur}
  W.~Israel,
  ``Thermo field dynamics of black holes,''
  Phys.\ Lett.\ A {\bf 57}, 107 (1976).


\bibitem{Maldacena:2001kr}
  J.~M.~Maldacena,
  ``Eternal black holes in anti-de Sitter,''
  JHEP {\bf 0304}, 021 (2003).

\bibitem{deBoer:2008gu}
  J.~de Boer, V.~E.~Hubeny, M.~Rangamani, M.~Shigemori,
  ``Brownian motion in AdS/CFT,''
  JHEP {\bf 0907}, 094 (2009).


\bibitem{Son:2009vu}
  D.~T.~Son and D.~Teaney,
  ``Thermal Noise and Stochastic Strings in AdS/CFT,''
  JHEP {\bf 0907}, 021 (2009).

\bibitem{quench1}
P.~Calabrese and J.~L.~Cardy,
 ``Evolution of entanglement entropy in one-dimensional systems,''
  J.\ Stat.\ Mech.\  {\bf 0504}, P04010 (2005).

 P.~Calabrese and J.~L.~Cardy,
 ``Time-dependence of correlation functions following a quantum quench,''
  Phys.\ Rev.\ Lett.\  {\bf 96}, 136801 (2006).


P. Calabrese and J. L. Cardy, ``Entanglement and correlation functions following a local quench:
a conformal field theory approach," J. Stat. Mech. 10 (2007) P10004.

\bibitem{zurek1993}
W.~Zurek,
  ``Preferred states predictability classicality and the environment-induced decoherence",
  Prog.\ Theor.\ Phys.\  {\bf 89}, 281 (1993).

\bibitem{quench2}
V. Eisler, I. Peschel, ``Evolution of entanglement after a local quench," J. Stat. Mech. (2007)
P06005.

V. Eisler, D. Karevski, T. Platini, I. Peschel, ``Entanglement evolution after connecting finite
to infinite quantum chains," J. Stat. Mech. (2008) P01023.


\bibitem{quench}
 H.~Liu and S.~J.~Suh,
  ``Entanglement Tsunami: Universal Scaling in Holographic Thermalization,''
  arXiv:1305.7244 [hep-th].


\bibitem{holo-quench}
V.~E.~Hubeny, M.~Rangamani and T.~Takayanagi,
  ``A Covariant holographic entanglement entropy proposal,''
  JHEP {\bf 0707}, 062 (2007).

J.~Abajo-Arrastia, J.~Aparicio and E.~Lopez,
  ``Holographic Evolution of Entanglement Entropy,''
  JHEP {\bf 1011}, 149 (2010).

T.~Albash and C.~V.~Johnson,
  ``Evolution of Holographic Entanglement Entropy after Thermal and Electromagnetic Quenches,''
  New J.\ Phys.\  {\bf 13}, 045017 (2011).


 \bibitem{Nozaki:2013vta}
  M.~Nozaki, T.~Numasawa, A.~Prudenziati and T.~Takayanagi,
  ``Dynamics of Entanglement Entropy from Einstein Equation,''
  arXiv:1304.7100 [hep-th].

 M.~Nozaki, T.~Numasawa and T.~Takayanagi,
  ``Holographic Local Quenches and Entanglement Density,''
  JHEP {\bf 1305}, 080 (2013).

 J.~Bhattacharya and T.~Takayanagi,
  ``Entropic Counterpart of Perturbative Einstein Equation,''
  arXiv:1308.3792 [hep-th].

\bibitem{python}
The python code can be found at the following website:

 http://phy.ntnu.edu.tw/\%7Elinfengli/QISp/deco-Wigner-Renyi-string.py


\bibitem{formulize}
Eureqa formulize, http://formulize.nutonian.com

\bibitem{Alan}
  Alan~J.~McKane and Matin~B.~Tarlie,
  ``Regularization of functional determinants using boundary perturbations,''
  J. Phys. A {\bf 28}, 6931 (1995)

\bibitem{Kleinert}
H. Kleinert. World Scientific. Path Integrals in Quantum Mechanics, Statistics, Polymer Physics, and Financial Markets Singapore, 2006.
















\end{thebibliography}
 \end{document}